\documentclass[aps,showpacs,preprintnumbers,amsmath,amssymb,nofootinbib,superscriptaddress,showkeys]{revtex4}

\usepackage{graphicx}
\usepackage{epsfig}
\usepackage{graphicx}
\usepackage{graphicx}
\usepackage{mathptmx}      
\usepackage{latexsym}

\def\slashchar#1{\setbox0=\hbox{$#1$}
   \dimen0=\wd0 \setbox1=\hbox{/} \dimen1=\wd1
   \ifdim\dimen0>\dimen1 \rlap{\hbox to \dimen0{\hfil/\hfil}} #1
   \else  \rlap{\hbox to \dimen1{\hfil$#1$\hfil}} / \fi}

\begin{document}

\title{Large--$\mathbf{N_C}$ Properties of the $\mathbf{\rho}$ and
$\mathbf{f_0(600)}$ Mesons from Unitary
  Resonance Chiral Dynamics.}
\author{J.~Nieves}
\affiliation{Departamento de F\'\i sica Te\'orica, IFIC, Centro
  Mixto Universidad de Valencia -- CSIC,
\\
Edificio de Institutos de Investigaci\'on de Paterna, Aptdo. 22085, 46071, Valencia, Spain
}
\author{A.~Pich}
\affiliation{Departamento de F\'\i sica Te\'orica, IFIC, Centro
  Mixto Universidad de Valencia -- CSIC,
\\
Edificio de Institutos de Investigaci\'on de Paterna, Aptdo. 22085, 46071, Valencia, Spain
}

\affiliation{Physik-Department and TUM Institute for Advanced Study, Technische
Universit\"at M\"unchen,
D-85748 Garching, Germany}

\author{E. Ruiz
  Arriola}\affiliation{Departamento de
  F\'{\i}sica At\'omica, Molecular y Nuclear
 and Instituto Carlos I de F{\'\i}sica Te\'orica y Computacional,\\
  Universidad de Granada, E-18071 Granada, Spain.}


\date{\today}

\begin{abstract}
\rule{0ex}{3ex} We construct $\pi\pi$ amplitudes that fulfill exact
elastic unitarity, account for one loop Chiral Perturbation Theory
contributions and include all $1/N_C$ leading terms, with the only
limitation of considering just the lowest-lying nonet of exchanged
resonances. Within such scheme, the $N_C$ dependence of $\sigma$ and
$\rho$ masses and widths is discussed. Robust conclusions are drawn in
the case of the $\rho$ resonance, confirming that it is
a stable meson in the limit of a large number of QCD colors, $N_C$.

Less definitive conclusions are reached in the
scalar-isoscalar sector. With the present quality of data, we cannot
firmly conclude whether or not the $N_C=3$ $f_0(600)$ resonance
completely disappears at large $N_C$ or it has a sub-dominant component
in its structure, which would become dominant for a number of quark colors
sufficiently large.

\end{abstract}

\maketitle



\section{Introduction}

Light $J^P= 0^+$ scalar resonance properties are of great interest,
since they might help to unravel details of QCD chiral symmetry
breaking and confinement. Despite many theoretical efforts, the
current understanding of the microscopic structure of these resonances
is still far from being complete. The difficulty is triggered by the
fact that scalar mesons carry vacuum quantum numbers, and also because
strong final-state interactions hide their underlying nature when they
are produced. Among other resonances, the lightest one $f_0(600)$,
currently denoted as the $\sigma$ meson, is an essential ingredient of
the nuclear force, as anticipated long ago~\cite{PhysRev.98.783}. Its
contribution to the mid-range nuclear attraction provides saturation
and binding in atomic nuclei. During many years, there has been some
arbitrariness on the ``effective'' scalar meson mass and coupling
constant to the nucleon, partly stimulated by lack of other sources of
information. The existence of this broad low-lying state is by now out
of question; its mass and width have accurately been extracted from
data analysis incorporating a large body of theoretical and
experimental
constraints~\cite{Caprini:2005zr,Kaminski:2006qe,GarciaMartin:2011jx}. The
debate on the nature of the $\sigma$ meson is nonetheless not
completely over. Structures of the tetraquark or glueball type have
been proposed (see e.g. Ref.~\cite{Klempt:2007cp} for a recent review
and references therein). It is remarkable that such an accurately
determined state is so poorly understood from the more fundamental
point of view of the underlying QCD dynamics of $N_C \times N_F$
quarks and $N_C^2-1$ gluons, where $N_C=3$ is the number of color
species and $N_F$ is the number of flavors. To clarify the issue on
the nature of the $\sigma$ meson, it has been suggested to follow the
dependence on a variable number of colors $N_C \neq 3 $ of its mass
and width~\cite{Pelaez:2003dy} by assuming that hadronic properties
scale similarly as if $N_C$ was large. A prerequisite for this scaling
approach to work is that at least {\it all leading}--$N_C$ effects are
taken into account.

The limit of an infinite number of quark colors keeping $\alpha_s N_C
$ fixed ($\alpha_s$ is the strong coupling constant),
turns out to be a very useful and simplifying starting point to
understand many qualitative features of the strong
interaction~\cite{'tHooft:1973jz,Witten:1979kh}.
While keeping essential properties of Quantum Chromodynamics (QCD),
under the assumption of confinement, the large--$N_C$ limit provides a
weak coupling regime to perform quantitative QCD studies. In this work
we are interested in describing the large--$N_C$ scaling of $\pi\pi$
scattering and the induced large--$N_C$ behavior of two-pion
resonances.  At leading order in $1/N_C$, the meson-meson scattering
amplitudes are given by sums of tree diagrams induced by the exchange
of an infinite number of weakly interacting physical (stable) hadrons.
Indeed, meson and glueball masses scale as ${\cal O}
(N_C^0)$ whereas the widths do as ${\cal O}(1/N_C)$ and ${\cal
  O}(1/N_C^2)$, respectively. Crossing symmetry implies that this sum
is the tree-level approximation to some local effective
relativistically-invariant Lagrangian, which by assuming that
confinement still holds at large $N_C$, can be re-written in terms of
mesonic fields and hence complies with quark-hadron
duality. Higher-order $1/N_C$ corrections correspond to hadronic loop
diagrams and effectively restore unitarity in the time-like region.

\begin{figure*}[tbh]
\begin{center}
\makebox[0pt]
{\hspace{-1cm}\includegraphics[height=2cm]{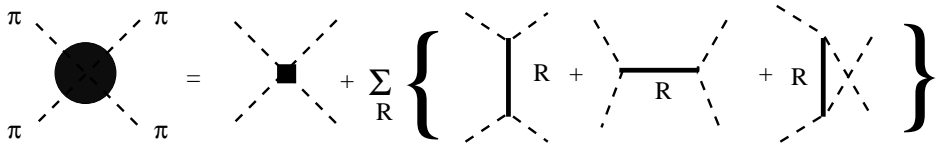}}
\end{center}
\caption{Leading large--$N_C$ $\pi\pi$ scattering diagrams made of
  contact (polynomial) and resonance-exchange (pole) terms
in $t$, $s$ and $u$ channels.}
\label{fig:pipi}
\end{figure*}

Resonance Chiral Theory (R$\chi$T)~\cite{Ecker:1988te,Ecker:1989yg}
includes the pseudo-Goldstone bosons and the resonances as dynamically
active degrees of freedom of the theory. The low-energy limit of
R$\chi$T must comply with low-energy theorems based on Chiral
Perturbation Theory (ChPT) \cite{Weinberg:1978kz,Gasser:1983yg,Gasser:1984gg},
and this property has been used to predict
systematically the Low Energy Constants (LECs) of ChPT in terms of
masses and couplings of the resonances, when integrating them out of
the action, at the chiral orders ${\cal O}(p^4)$~\cite{Ecker:1988te}
and ${\cal O}(p^6)$~\cite{Cirigliano:2006hb}. The ChPT Lagrangian
includes the octet of pseudo-Goldstone bosons, however, when extending
ChPT, R$\chi$T incorporates the resonances as active degrees of
freedom that are included in nonets, since octet and singlet of a
$SU(N_F =3)$ group merge into a nonet for $N_C \to \infty$. The ChPT
Lagrangian is built using the spontaneously-broken chiral symmetry of massless
QCD. The explicit symmetry breaking from non-zero quark masses and electromagnetic
interactions is
incorporated in exactly the same way as it happens in QCD. The nonets
of resonances are added requiring the general properties and
invariance under charge conjugation and parity, and the structure of
the operators is determined by chiral symmetry. At first order in the
$1/N_C$ expansion,  terms with more than one trace and loops
are suppressed. The first property permits to postpone some terms
allowed by the symmetries as subleading. However, the theory
determined by just symmetries does not share yet some of the known
properties of QCD at high energies or accepted from hadronic Regge
phenomenology\footnote{Regge behavior, while not directly deduced from
  QCD, works rather well and it has played a decisive role in the
  benchmark and unprecedented accurate determination of the $\sigma$
  meson mass and
  width~\cite{Caprini:2005zr,Kaminski:2006qe, GarciaMartin:2011jx}, by
  extending the high energy region above $\sqrt{s} > 1.4$ GeV.}.
Further constraints on the couplings arise by matching the
interpolating resonance theory at intermediate energies with
asymptotic QCD at the level of Green functions and/or form factors.
The application of these properties determines a series
of relations between the couplings of R$\chi$T, reducing the number
of couplings and enhancing the predictability. We remind here that QCD,
besides current quark masses, has only one dimensionful parameter,
$\Lambda_{\rm QCD}$. Of course, there are infinitely many such
short-distance constraints and hence a sufficiently large number of states
may eventually be needed to avoid contradictory
results~\cite{Bijnens:2003rc}.

A further and less trivial question is related to whether or not
purely contact terms should be regarded as independent of exchange
terms in the large--$N_C$ framework. Within $\pi\pi$ scattering, this
corresponds to distinguish between contact $4\pi$ vertices and those
where a resonance field propagates between $2\pi$ states (see
Fig.~\ref{fig:pipi}).  While the large--$N_C$ expansion is expected to
provide a better approximation to data in the space-like region where
unitarity does not play an active role, the approach is not
specifically related to a given energy range. This means that if the
tower of {\it all} infinitely many states is included, the question on
the duality between contact and exchange terms is pertinent.
However, if the exchanged resonance spectrum is truncated above a given energy,
contact terms must necessarily arise to encode the explicitly
disregarded high-energy contributions while still complying to the
short-distance constraints~\cite{Ecker:1989yg}.   
The explicit values of the contact LECs depend on the
functional parametrization adopted for the resonance fields.

In the Single-Resonance-Approximation (SRA) scheme, each infinite
resonance sum is just approximated by the contribution from the
lowest-lying meson multiplet with the given quantum numbers. This is
meaningful at low energies where the contributions from higher-mass
states are suppressed by their corresponding propagators. The SRA
corresponds to work with a low-energy Effective Field Theory (EFT)
below the scale of the second resonance multiplets.  In this work, we
will use the SRA of R$\chi$T to re-analyze how the $\sigma$ and $\rho$
properties depend on $N_C$, mainly based on the study of $\pi\pi$
scattering. An early investigation was proposed in
Refs.~\cite{Sannino:1995ik,Harada:1995dc} keeping the leading $1/N_C$
contributions but omitting the ChPT chiral logarithms. We impose
leading--$N_C$ short-distance constraints, as discussed in detail in
Refs.~\cite{Ecker:1989yg, Pich:2002xy}. This turns out to be of
capital importance because it leads to a clear distinction between
leading and subleading $N_C$ contributions to the $\pi\pi$
amplitudes. It also allows for a meaningful extension of the framework
to the $N_C > 3$ world. Besides, enforcing the short-distance
constraints reduces the number of free parameters to only the
subtraction constants, needed to restore exact elastic unitarity, and
the masses of the exchanged resonances. In addition, we check that a
direct analysis of $\pi\pi$ scattering at leading order in $1/N_C$ and
using high-energy constraints, based on forward dispersion relations
and Regge phenomenology, generates relations between the resonance
properties compatible with those already obtained in
\cite{Ecker:1989yg, Pich:2002xy} by looking at other processes.

 We will first construct $\pi\pi$ amplitudes that fulfill exact
 elastic unitarity, account for one-loop ChPT contributions, and
 include all $1/N_C$ leading terms in the SRA. Next, we will look for
 poles in the appropriate unphysical sheets of the amplitudes, and
 discuss their properties when $N_C$ deviates from its physical
 value,\footnote{We will focus here on the properties of the poles in
 the complex $s$ plane. The $N_C$ behavior of the Breit-Wigner
 resonance parameters has been discussed, in a model independent
 manner, at length in Ref.~\cite{Nieves:2009kh}, and we refer the
 reader to this latter work for further details.}  to learn details on
 their nature.  In this manner, we improve on previous
 analyses~\cite{Pelaez:2003dy,Sun:2005uk,Pelaez:2006nj,Nieves:2009ez}
 where leading $1/N_C$ terms, beyond a certain order in the chiral
 expansion, were neglected.

On the other hand, in the strict chiral and large--$N_C$ limits, the pseudo-scalar
singlet $\eta_1$ and the $\pi$ are
degenerate~\cite{DiVecchia:1980ve,Witten:1980sp,Rosenzweig:1979ay}. The
interplay between ChPT and large $N_C$ has been addressed in
Ref.~\cite{Kaiser:2000gs}. $U(3)$ meson-meson scattering has been
treated in \cite{Beisert:2003zd} with only contact ${\cal O}(p^2)$
chiral interactions.  In Ref.~\cite{Albaladejo:2010tj}, in addition to
the contact ${\cal O}(p^2)$, the leading--$N_C$ scalar resonance
exchange has been included, but leaving aside the vector meson
exchange. This latter mechanism not only contributes to the P-waves,
but also to other S-wave channels via the left-cut
contribution. Recently, a full-fledged one-loop unitarized couple-channel
$U(3)$ calculation, including both scalar and vector resonance
exchanges, has been undertaken in Ref.~\cite{Guo:2011pa}. Besides
achieving an excellent description of phase shifts until
center-of-mass energies of around 1.4~GeV, this reference also analyzes the $N_C$
behavior of the amplitudes. Among other results, it is explicitly
shown there that when $N_C$ increases, the mass of the lowest eigenstate of the
 $\eta_1$--$\eta_8$ mass matrix decreases, reaching values of around
twice the pion mass in the $N_C=30$ region. Thus, a natural question
arises here, namely, do the $\eta$--$\eta'$ degrees of freedom play a
relevant role to determine the $N_C$ trajectory of the $\sigma$ and
$\rho$ resonances? This is addressed also in Ref.~\cite{Guo:2011pa},
from where one might infer that this is certainly not the case. Those
degrees of freedom turn out to be much more relevant in the study of
the $N_C$ dependence of masses and widths of higher resonances, as for
instance the $f_0(980)$. We benefit here from this observation, and we
will neglect $\eta$--$\eta'$ effects in what follows.

We should also point out that some aspects of the $N_C\ne 3$ extension
undertaken in Ref.~\cite{Guo:2011pa} deserve discussion, and we
believe they can be improved along the lines followed in this
paper. In particular, the leading $1/N_C$ contributions were not
properly considered in Ref.~\cite{Guo:2011pa}. As a consequence,
we cannot firmly conclude a
scenario where the $\sigma$ moves far away in the complex plane for
large $N_C$, as obtained in \cite{Guo:2011pa}. We will give some more
details below Eq.~(\ref{eq:eles-nc}). Note that crossing symmetry,
the possible absence of an exotic isotensor ($I=2$) state and the
well-established fact that the $\rho$ width decreases with $N_C$ imply
the existence of a narrow scalar-isoscalar resonance, in the large--$N_C$
limit, with a mass comparable to that of the
$\rho $meson~\cite{RuizdeElvira:2010cs}.

The paper is organized as follows. In Section~\ref{sec:general} we
introduce our notation in order to make the paper self-contained.  We
also review some important features of $\pi\pi$ forward dispersion
relations and crossing symmetry, which yield to two sum rules, keeping
an eye on the large--$N_C$ expansion. In Section~\ref{sec:largeN} we
discuss the $\pi\pi$ scattering amplitude in the SRA of R$\chi$T and
analyze high-energy conditions which yield to the short-distance
constraints. In Section~\ref{sec:oneloop}, the ${\cal O}(p^4)$ ChPT
contributions to the $\pi\pi$ scattering amplitude and its matching to the
leading $1/N_C$ piece are discussed. Actually, we manage to write an
amplitude which is correct to ${\cal O} (p^4)$ and includes, within
the SRA, the leading ${\cal O}(1/N_C)$ terms to all orders in the
chiral expansion. However, it still needs to be unitarized before
being confronted to scattering data. After unitarization, we fix our
parameters in Section~\ref{sec:phases} by fitting the $(I,J)=(0,0)$,
$(1,1)$ and $(2,0)$ phase shifts in the real $N_C=3$ world. This complies with
the determination of the R$\chi$T Lagrangian parameters at the leading--$N_C$
approximation.  Only then do we allow ourselves to analyze the
$N_C > 3 $ situation in Section~\ref{sec:results} and the emerging
picture for $\pi\pi$ resonances as the number of colors is
varied. Finally, in Section~\ref{sec:conclusions} we draw the main
conclusions of our work.

\section{General Features of $\pi\pi$ scattering}
\label{sec:general}

A comprehensive presentation of $\pi\pi$ scattering can be seen at the
textbook level \cite{martin1976pion} and more recently in
\cite{Yndurain:2002ud}.  We summarize here the relevant formulae to
fix our notation and to provide a proper perspective of our subsequent
analysis merging large $N_C$, ChPT and unitarity considerations.

\subsection{Kinematics}

The $\pi_a(p_1)+ \pi_b(p_2) \to \pi_c(p_1')+ \pi_d(p_2') $ scattering
amplitude is written as
\begin{eqnarray}
T_{ab;cd} = A(s,t,u) \delta_{ab} \delta_{cd}
+ A(t,s,u) \delta_{ac} \delta_{bd}
+ A(u,t,s) \delta_{ad} \delta_{bc}\, ,
\end{eqnarray}
with the standard choice of Mandelstam variables $s=(p_1+p_2)^2$,
$t=(p_1-p_1')^2$ and $u=(p_1-p_2')^2$, where $A(s,t,u)$ is the
$\pi^+\pi^-\to \pi^0 \pi^0$ amplitude, which is the only independent
amplitude thanks to isospin, crossing and Bose-Einstein
symmetries. That is, if $T_{I_s}(s,t,u)$ is the isospin combination with
total isospin $I$ (in the $s$-channel), one has,
\begin{eqnarray}
T_{I_s=0}(s,t,u) &=& \frac12 \left\{ 3A(s,t,u) + A(t,s,u)+
A(u,t,s)\right\}\, , \\
T_{I_s=1}(s,t,u) &=& \frac12 \left\{ A(t,s,u)-
A(u,t,s)\right\}\, ,\\
T_{I_s=2}(s,t,u) &=& \frac12 \left\{ A(t,s,u)+
A(u,t,s)\right\}\, .
\end{eqnarray}
For the normalization we will use here the conventions
from~\cite{Nieves:1999bx}.  The partial-wave decomposition in the
$s$-channel becomes
\begin{eqnarray}
T_{I}(s,t,u) = \sum_{J=0}^\infty (2J+1) T_{IJ}(s) P_J ( \cos \theta)\, ,
\end{eqnarray}
where $T_{IJ}(s)$ is the projection of the $\pi\pi$ elastic scattering
amplitude with given total isospin $I$ and angular momentum $J$:
\begin{eqnarray}
T_{IJ}(s) &=& \frac12 \int^{+1}_{-1}\, d\cos\theta\;
P_J(\cos\theta)\, T_I\left(s,t(s,\cos\theta), u(s,\cos\theta)\right)
\nonumber \\
&=&
-16\pi \, \left (\frac{\eta_{IJ} (s) e^{2{\rm i}\delta_{IJ}(s)}-1}{2{\rm i}\,\rho(s)}
\right)\, ,
\end{eqnarray}
with
\begin{equation}\label{eq:rho}
\rho(s) = \sqrt{1-\frac{4 m_\pi^2}{s}}\, ,
\end{equation}
being $m_\pi=139.57$ MeV the pion mass
and $P_J$ the Legendre polynomials. The in-elasticity $\eta_{IJ} (s)= 1
$ for $s< 16 m_\pi^2$ and $\eta_{IJ} (s) < 1 $ for $s> 16
m_\pi^2$. Besides, $\delta_{IJ}$ are the phase shifts and the
Mandelstam variables $t$ and $u$ depend on $s$ and on $\theta$, the
scattering angle in the center-of-mass frame (c.m.). The optical theorem reads
\begin{eqnarray}
\sigma_I (s) &=& - \frac{1}{s \,\rho(s)}\; {\rm Im} T_I(s,0,
4m_\pi^2-s) \; =\; \sum_{J} \,\sigma_{IJ}(s)\, ,
\end{eqnarray}
where the partial-wave total cross section is defined as
\begin{eqnarray}
\sigma_{IJ}(s) = 16 \pi\,\frac{(2J+1)}{s-4m_\pi^2}\;\left[
  \eta_{IJ}\sin^2 \delta_{IJ} + \frac12 (1
  -\eta_{IJ})\right]\, .
\end{eqnarray}
This value is bound by
\begin{eqnarray}
\sigma_{IJ}(s) \,\le\, 8 \pi (2J+1)\frac{1
+\eta_{IJ}}{s-4m_\pi^2}\, .
\end{eqnarray}
The contribution of a resonance state, with spin $J$ and
isospin $I$, to the partial cross section  in the narrow-width limit
and assuming $m_{IJ}^2 \gg 4 m_\pi^2$ reads,
\begin{eqnarray}
\sigma_{IJ} (s) = (2J+1) \,\frac{16 \pi^2 \Gamma_{IJ}}{m_{IJ}}\;
\delta (s-m_{IJ}^2) \label{eq:nra}\, .
\end{eqnarray}
Of course, one may think that such a limit does not apply to a broad
state as the $\sigma$. Within a Breit-Wigner model, the finite
width correction effectively corresponds to a reduction $\Gamma \to
\Gamma\, [1-\Gamma/(\pi m)]$, which even for the extreme case $\Gamma=m$
yields to a moderate $30\%$ correction.

\begin{figure}[tbh]
\begin{center}
\makebox[0pt]
{\hspace{-1cm}\includegraphics[height=8cm]{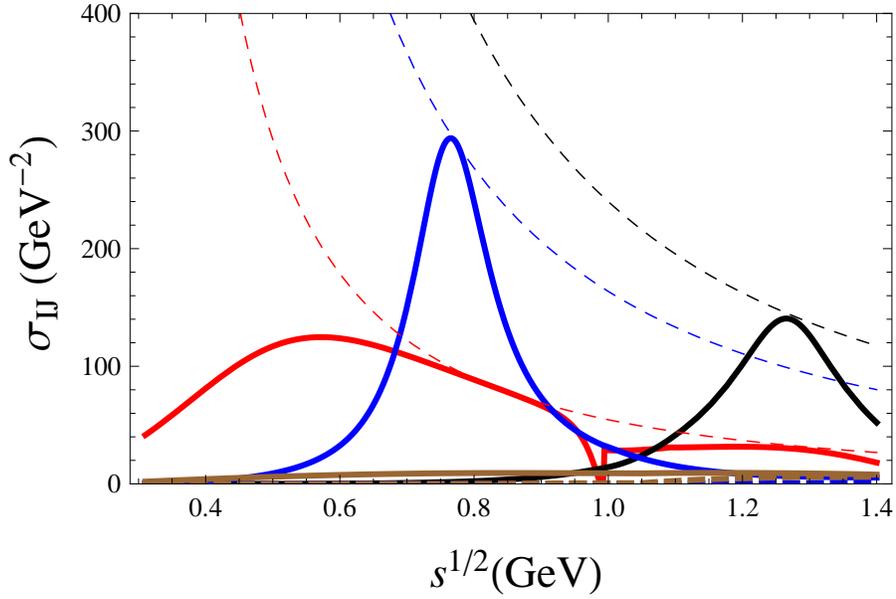}}
\end{center}
\caption{(Color online) The total partial-wave cross sections
  $\sigma_{IJ}(s)$ (in ${\rm GeV}^{-2}$), defined as a function of the
  cm energy variable $\sqrt{s}$ (in ${\rm GeV}$) for the S0
  (Solid, Red), D0 (Solid, Black), P (Solid, Blue), F (Dashed, Blue), S2
  (Solid, Brown) and D2 (Dashed, Brown) partial waves below $\sqrt{s} =
  1.42$ GeV, from the GKPRY parameterization with set UFD of
  parameters given in~\cite{Martin:2011cn}. We also draw the unitarity bounds
  (dashed lines) for S0 (red), P (blue) and D2 (black) waves, using
  $\eta_{00}=\eta_{11}=1$ and $\eta_{02}=0.88$ respectively.}
\label{fig:sigmaPW}
\end{figure}

Using the recent GKPRY parameterizations of Ref.~\cite{Martin:2011cn}
for the partial S, P, D and F waves, cross sections are presented in
Fig.~\ref{fig:sigmaPW} up to $\sqrt{s} \le 1.42$ GeV. As we see, the S0,
P and D0 waves play an outstanding role featuring the appearance of
the $f_0 (600)$, $\rho_1(770)$, $f_0 (980)$ and $f_2(1270)$
resonances. Below the $\bar K K$ production threshold, $s_{\bar K K} = 4
m_K^2 $, only S0 and P are essential.

\subsection{Crossing and forward dispersion relations}

In the (crossed) $t$-channel the amplitudes read
\begin{eqnarray}
\left(\begin{matrix}
 T_{I_t=0}(s,t,u) \\
T_{I_t=1}(s,t,u) \\
T_{I_t=2}(s,t,u)
\end{matrix}\right) =
\left(\begin{matrix}
\frac13 &  \, \, 1 & \, \frac53 \\
\frac13 & \, \frac12 & -\frac56  \\
\frac13 & -\frac12 &  \, \frac16
\end{matrix}\right)
\left(\begin{matrix}
 T_{I_s=0}(s,t,u) \\
T_{I_s=1}(s,t,u) \\
T_{I_s=2}(s,t,u)
\end{matrix}\right)\, ,
\label{eq:s-to-t}
\end{eqnarray}
where $I_t$ and $I_s$ are the corresponding isospins in the $t$- and
$s$-channels respectively.
\begin{figure}[tbh]
\begin{center}
\makebox[0pt]
{\hspace{0cm}\includegraphics[height=7cm]{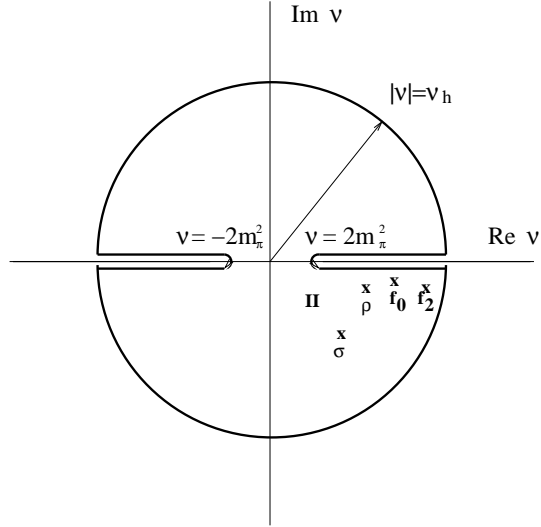}}
\end{center}
\caption{The contour in the complex crossing-odd $\nu$ plane used to
  derive FDRs for the $t$-channel isospin
  $\pi\pi$ scattering amplitude. For $t=0$ $\nu = s - 2 m_\pi^2$, so
  that $\nu= 2 m_\pi^2$ corresponds to the $\pi\pi$ threshold. The
  principal value of the forward scattering amplitude corresponds to
  $T_{I_t}(\nu + {\rm i}0^+$). Finite Energy Sum Rules arise for the
  Regge-subtracted amplitude for the finite circle $|\nu| \le \nu_h$.
  Low-lying resonances $f_0(600)\equiv \sigma,\,\rho_1(770), \, f_0(980)$ and
  $f_2(1275)$ are also marked by crosses in the Second Riemann Sheet,
  across the cut.}
\label{fig:FDR-contour}
\end{figure}

\begin{figure}[t]
\begin{center}
\makebox[0pt]
{\hspace{-1cm}\includegraphics[height=8cm]{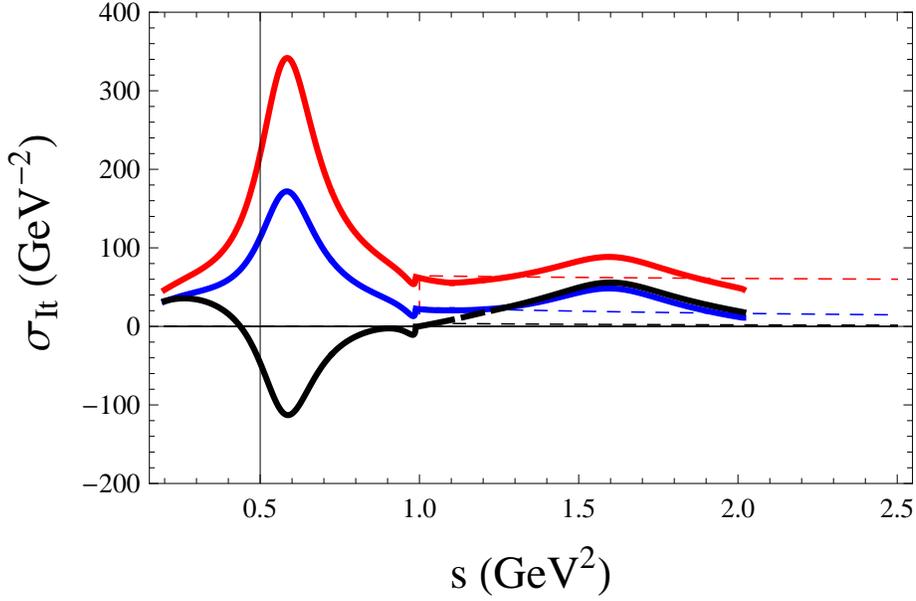}}
\end{center}
\caption{The $I_t=0,1,2$ total cross sections, as a function of the
  $s$ variable, including S, P, D and F waves below $\sqrt{s} = 1.42$ GeV
  (solid lines) and the corresponding Regge behavior starting at $
  \sqrt{s} = 1.42$ GeV (dashed lines) used in the solutions of
  Ref.~\cite{Martin:2011cn}. $I_t=0$ (Red), $I_t=1$ (Blue) and $I_t=2$
  (Black).}
\label{fig:sigmaIt}
\end{figure}

The rigorous Froissart bound from axiomatic field theory requires that
in the forward direction ($t=0$) these amplitudes, up to logarithmic
corrections,  are asymptotically
polynomially bound by a single power of $s$. Actually, in terms of the
crossing-odd variable $\nu=(s-u)/2$,  the amplitude $T_{I_t} (\nu , t ) \equiv
T_{I_t} ( \nu + 2m_\pi^2-t/2, t, -\nu+2 m_\pi^2-t/2) $ satisfies
$T_{I_t} (-\nu , t )= (-)^{I_t}T_{I_t} (\nu , t ) \to \nu^{n_{I_t}}$,
with $n_{I_t}\le 1$ when $\nu \to \infty$. This means that a forward
($t=0$) once-subtracted dispersion relation is fulfilled, by
considering a closed contour excluding the cuts $-\infty < \nu < - 2
m_\pi^2 $ and $ 2 m_\pi^2 < \nu < + \infty $ (see
Fig.~\ref{fig:FDR-contour}), where $T(\nu + i \, 0^+) - T(\nu - i\,
0^+) = 2 i \, {\rm Im} T(\nu + i \, 0^+) \equiv 2 i \, {\rm Im} T(\nu) $.
The subtraction constant can be fixed at low energies, and more
specifically at $\nu=0$ or equivalently $s = 2 m_\pi^2$, i.e. below
threshold.  Thus, at $t=0$,  we get
the Forward Dispersion Relations (FDRs):
\begin{eqnarray}
T_{I_t=0} (\nu, 0) &=& T_{I_t=0} (0, 0) + \frac{2 \nu^2}{\pi} \int_{2
  m_\pi^2}^\infty \frac{d\nu'}{\nu'} \frac{{\rm Im} T_{I_t=0} (\nu',
  0)}{\nu'^2-\nu^2} \, ,
\nonumber \\
\frac{T_{I_t=1} (\nu, 0)}{\nu} &=&
\lim_{\nu_0\to 0}\frac{T_{I_t=1} (\nu_0, 0)}{\nu_0}
+ \frac{2 \nu^2}{\pi} \int_{2 m_\pi^2}^\infty
\frac{d\nu'}{\nu'^2} \frac{{\rm Im} T_{I_t=1} (\nu', 0)}{\nu'^2-\nu^2}\, ,
\nonumber \\
T_{I_t=2} (\nu, 0) &=& T_{I_t=2} (0, 0)
+ \frac{2
  \nu^2}{\pi} \int_{2 m_\pi^2}^\infty \frac{d\nu'}{\nu'} \frac{{\rm Im}
  T_{I_t=2} (\nu', 0)}{\nu'^2-\nu^2}\, .
\label{eq:forw-disp}
\end{eqnarray}
The absorptive part of the amplitude can be written as an optical
theorem in the $\nu$ variable:
\begin{eqnarray}
{\rm Im} T_{I_t} (\nu,0)  = -\sqrt{\nu^2-4m_\pi^4}\,  \,  \sigma_{I_t}(\nu)\, ,
\end{eqnarray}
where one goes from $I_t$ to $I_s$ with the same $s$--$t$ crossing matrix
as for the amplitudes, see Eq.~(\ref{eq:s-to-t}). The FDRs converge,
provided $|{\rm Im} T_{I_t} (\nu,0)| < \nu^a$ with $a< 2$, for large
values of $\nu$.
Results for $\sigma_{I_t}$ are presented in Fig.~\ref{fig:sigmaIt}, up to $\sqrt{s}=1.42$~GeV,
using the UFD parameterization of Ref.~\cite{Martin:2011cn} for the partial
S, P, D and F waves

These general constraints are obeyed for any Quantum Field Theory and
by QCD in particular.  On the other hand, for $t \neq 0 $, and
sufficiently large $\nu$, say $\nu>\nu_h$, one has the phenomenologically
successful Regge behavior given by\footnote{ Our amplitude and that
  used ($ F(\nu,t)$) in Ref.~\cite{Martin:2011cn} are related by
 $T(\nu,t)= - 2  \pi^2 F(\nu,t)$.}
\begin{eqnarray}
{\rm Im} T_{I_t} (\nu,t) \to - 2\pi^2 \sum \beta_{I_t}(t)
\left[\frac{\nu}{\nu_0}\right]^{\alpha_{I_t}(t)}\, ,
\end{eqnarray}
where $\sum$ indicates summation over several Regge trajectories, which
for small $t$ are given by $\alpha_{I_t}(t)= \alpha_{I_t}(0) + t
\alpha_{I_t}' +\dots $
The slope parameter, $\alpha_{I_t}'(0) \sim
1/(2m_\rho^2) = 0.9 ~{\rm GeV}^{-2}$ is nearly universal and the
leading trajectory intercepts are $\alpha_0(0) \sim 1$, $\alpha_1(0)
\sim 1/2$ and $\alpha_2(0) \sim 0$ (for a review within a modern $\pi\pi$
context see e.g. Refs.~\cite{Pelaez:2003eh,Pelaez:2003ky} and
references therein). We also show in Fig.~\ref{fig:sigmaIt} the Regge
behavior used in the construction of the partial waves via FDRs,
Roy and GKPY equations.\footnote{For our purposes we need the following Regge
  tails~\cite{Martin:2011cn}, valid for $s > s_h = (1.42 {\rm ~GeV})^2$:
\begin{eqnarray}
 {\rm Im} T_{I_t=0}
        (\nu,0) &=& - 2\pi^2 \left[b_P (\nu/\nu_0) +  b_{P'}
        (\nu/\nu_0)^{a_{P'}}\right] \, ,\nonumber \\
{\rm Im} T_{I_t=1} (\nu,0) &=& -
        2\pi^2 b_1 (\nu/\nu_0)^{a_1} \, ,\nonumber \\
{\rm Im} T_{I_t=2} (\nu,0) &=& - 2\pi^2 b_2 (\nu/\nu_0)^{a_2}\, .
\end{eqnarray}
The numerical values of the parameters are given in section 8 of the
appendix A of Ref.~\cite{Martin:2011cn}.}
As we see, there is some mismatch between the partial waves and the
Regge behavior. This is somewhat expected, since Regge behavior
provides an average of the oscillating resonance contribution in the
high-energy region.  The region $\sqrt{s}< 1.42$ GeV is well
described by S, P, D and F waves, since the exchanged $\rho$ meson in the
$t$-channel corresponds to longest range $1/m_\rho$ and hence $J_{\rm
  max} \sim \sqrt{s} /m_\rho$.

\subsection{Finite Energy Sum Rules}

Separating the Regge tail in the dispersive
integral, corresponding to the integration region $\nu' > \nu_h  $, we
have for  $I_t=0,2$, in the limit $\nu \to + \infty$ and assuming
$\alpha_{I_t}(0) < 2$,
\begin{eqnarray}
T_{I_t} (\nu,0)|_{\rm Regge} & \equiv &  \frac{2 \nu^2}{\pi}
\int_{\nu_h}^\infty \frac{d\nu'}{\nu'} \frac{\left[-2 \pi^2 \sum
  \beta_{I_t}(0)
\left (\nu'/\nu_0\right)^{\alpha_{I_t}(0)}\right]}{\nu'^2-\nu^2 - {\rm
    i}\epsilon} \nonumber \\
& = & 2\pi^2 \sum \frac{ \beta_{I_t}(0) }{\sin \left(
  \frac{\alpha_{I_t}(0) \pi}2 \right)}
e^{-{\rm i} \alpha_{I_t}(0) \pi/2}\left(\frac{\nu}{\nu_0}\right)^{\alpha_{I_t}(0)}
- 4 \pi \sum \frac{ \beta_{I_t}(0)}{\alpha_{I_t}(0)}
\left(\frac{\nu_h}{\nu_0}\right)^{\alpha_{I_t}(0)} + {\cal O}
(\nu_0/\nu)
\label{eq:fesr02}
\end{eqnarray}
and similarly for $I_t=1$,
\begin{eqnarray}
\left.\frac{T_{I_t=1} (\nu,0)}{\nu}\right|_{\rm Regge} & \equiv&  \frac{2
\nu^2}{\pi} \int_{\nu_h}^\infty \frac{d\nu'}{\nu'^2} \frac{\left[-2 \pi^2 \sum
  \beta_{I_t=1}(0)
\left (\nu'/\nu_0\right)^{\alpha_{I_t=1}(0)}\right]}{\nu'^2-\nu^2 - {\rm i}\epsilon}
\nonumber \\
& = & \left(2\pi^2 \sum \frac{ \beta_{I_t=1}(0) }{\sin \left(
  \frac{\hat \alpha \pi}2 \right)}
e^{-{\rm i} \hat\alpha \pi/2}\left(\frac{\nu}{\nu_0}\right)^{\hat \alpha}
- 4 \pi \sum \frac{ \beta_{I_t=1}(0)}{\hat \alpha}
\left(\frac{\nu_h}{\nu_0}\right)^{\hat \alpha}\right)\frac{1}{\nu_0} +
  {\cal O} (1/\nu) \, ,
  \label{eq:fesr1}
\end{eqnarray}
with $\hat\alpha=\alpha_{I_t=1}(0)-1$. Note that the last term in both
Eqs.~(\ref{eq:fesr02}) and (\ref{eq:fesr1}) is a constant subleading
contribution. We will denote this constant term as
$-R_{I_t}$. Requiring
\begin{eqnarray}
T_{I_t} (\nu,0) &\to& 2\pi^2 \sum \frac{ \beta_{I_t}(0) }{\sin \left(
  \frac{\alpha_{I_t}(0) \pi}2 \right)} e^{-{\rm i} \alpha_{I_t}(0) \pi/2}
\left(\frac{\nu}{\nu_0}\right)^{\alpha_{I_t}(0)},\quad I_t=0,2\, , \\
\frac{T_{I_t=1} (\nu,0)}{\nu} &\to& \frac{2\pi^2}{\nu_0} \sum \frac{
\beta_{I_t=1}(0) }{\sin \left(
  \frac{\hat \alpha \pi}2 \right)}
e^{-{\rm i} \hat\alpha \pi/2}\left(\frac{\nu}{\nu_0}\right)^{\hat \alpha}\, ,
\end{eqnarray}
in the $\nu \to +\infty$ limit, we get the Finite Energy Sum Rules (FESRs)
\begin{eqnarray}
T_{I_t=0}(0)
&=&-\frac{2}{\pi}
\int_{2m_\pi^2}^{\nu_h} d \nu  \sqrt{1-\frac{4 m_\pi^4}{\nu^2}}
\, \sigma_{I_t=0} (\nu) + R_{I_t=0} \, ,  \nonumber \\
T_{I_t=1}'(0) &=& -\frac{2}{\pi}\int_{2m_\pi^2}^{\nu_h} \frac{d
  \nu}{\nu} \sqrt{1-\frac{4 m_\pi^4}{\nu^2}} \, \sigma_{I_t=1} (\nu) + R_{I_t=1}' \,
  , \nonumber \\
T_{I_t=2}(0)
&=&-\frac{2}{\pi}
\int_{2m_\pi^2}^{\nu_h} d \nu  \sqrt{1-\frac{4 m_\pi^4}{\nu^2}}
\, \sigma_{I_t=2} (\nu) + R_{I_t=2} \, , \label{eq:t00}
\end{eqnarray}
where
\begin{eqnarray}
R_{I_t=0} &=& 4 \pi b_P \left(\frac{\nu_h}{\nu_0}\right)+ 4 \pi
\frac{b_{P'}}{a_{P'}} \left(\frac{\nu_h}{\nu_0}\right)^{a_{P'}}\, ,
 \nonumber \\
R_{I_t=1}' &=&  \frac{4 \pi b_1}{a_1-1}
\frac1{\nu_h}\left(\frac{\nu_h}{\nu_0}\right)^{a_1} \, ,
\nonumber \\
R_{I_t=2} &=& 4 \pi \frac{b_{2}}{a_{2}} \left(\frac{\nu_h}{\nu_0}\right)^{a_{2}}\,
.
\label{eq:R's}
\end{eqnarray}
For $\sqrt{s_h} = \sqrt{\nu_h + 2 m_\pi^2}= 1.42$ GeV, we get the
values $T_{I_t=0}(0) = -121.8 + 64.3~(P) + 28.2~(P') =-29.3 $,
$T_{I_t=1}'(0) = -105.2- 19.4 ~(\rho) {\rm ~GeV}^{-2}= -124.6 {\rm
~GeV}^{-2} $ and $T_{I_t=2}(0) = -12.25 + R_{I_t=2}$. Indeed, in the
window $\sqrt{\nu_h} \in (1,1.42)$ GeV, we observe a smooth $\nu_h$
dependence of both the integral and the $R_{I_t}$
contributions.\footnote{We use the UFD parameterization of
Ref.~\cite{Martin:2011cn}.} Moreover, the total sum itself remains
fairly independent of $\nu_h$, as well. Being more quantitative the
right-hand side of Eqs.~(\ref{eq:t00}) changes at the level of 10\%
and 4\% for the $I_t=0$ and 1 cases, respectively, when $\sqrt{\nu_h}$
varies in the interval (1,1.42) GeV. For the case $I_t=1$,
$R_{I_t=1}'$ amounts, at energies as small as 1 GeV, to around 25\% of
the total, and decreases with increasing $\nu_h$, as deduced from its
Regge behavior.  For the case of $I_t=2$, the integral contribution
shows a more pronounced dependence on $\nu_h$, and it changes sign at
around half of the (1,1.42) GeV interval. However, this contribution
in size is just at maximum around one third of that of $R_{I_t=2}$,
which is almost constant, because of the smallness of the Regge
intercept, $a_2$. Furthermore, the sign of $R_{I_t=2}$ depends on the
sign of $a_2$ and actually the FESR becomes ambiguous for $a_2=0$,
since the amplitude would show a logarithmic growth $\sim
\log(-\nu^2/\nu_h^2)$ instead of a Regge behavior. For a
diverging/converging amplitude we have a positive/negative
contribution to the sum rule. These ambiguous signs appear in the CFD
and UFD solutions of Ref.~\cite{Martin:2011cn}, from where one gets
$R_{I_t=2}=-30 (30)$ and $R_{I_t=2}=20(40)$, respectively, both
results compatible with zero.

\subsection{Adler and $\sigma$ Sum rules}

If $R_{I_t=1}'$ and $R_{I_t=2}$ are neglected in Eqs.~(\ref{eq:t00}),
as suggested by the numerical values obtained from the fits performed
in Ref.~\cite{Martin:2011cn}, and $T_{I_t=1}'(0)$ and $T_{I_t=2}(0)$
are approximated by the lowest-order result in ChPT,
see Eq.~(\ref{eq:chpt_It}) below,
 the so-called Adler and $\sigma$ sum rules (see
e.g. \cite{martin1976pion} for a discussion based on current algebra
and PCAC) are obtained,
\begin{eqnarray}
\frac{1}{f_\pi^2} &=& \frac{2}{\pi}\int_{2m_\pi^2}^\infty \frac{d
  \nu}{\nu} \sqrt{1-\frac{4 m_\pi^4}{\nu^2}}\;
\left[ \frac13 \sigma_0 (\nu) + \frac12 \sigma_1 (\nu) -
  \frac56\sigma_2 (\nu) \right]\, ,
\label{eq:adler-sr}
\\
\frac{m_\pi^2}{f_\pi^2} &=&\frac{2}{\pi}
\int_{2m_\pi^2}^\infty d \nu  \sqrt{1-\frac{4 m_\pi^4}{\nu^2}}
 \;\left[ \frac13 \sigma_0 (\nu) - \frac12
  \sigma_1 (\nu) +\frac16\sigma_2 (\nu)\right]\, ,
\label{eq:sigma-sr}
\end{eqnarray}
with $f_\pi \sim 93$ MeV, the pion decay constant. Using the GKPRY
parameterizations~\cite{Martin:2011cn}, one finds that the contribution
of the region $\nu > 1.42$ GeV, for which Regge behavior is assumed,
is quite small. This is because $\sigma_{I_t} (\nu) \to 4 \pi^2
\beta_{I_t} (0) \nu^{\alpha_{I_t}(0)-1}$. In addition, we would like
to point out
\begin{itemize}
\item $I_t=1$: Following the discussion of the previous section,
neglecting the contribution to the sum rule arising from
  the region $\sqrt{\nu} \in (1,1.4)$ GeV might induce variations
  of order 20\%, which we expect to be of the same order as those
  stemming from ${\cal O}(p^4)$ ChPT terms neglected in the left-hand side
    of the sum rule.

  The Adler sum rule is satisfied to $5\%$ (see also
  \cite{Adler:2007aw}) in the partial-wave plus Regge representation.

\item $I_t=2$: This sum rule converges if $\alpha_{I_t=2}(0)<0$
  and is the $\sigma$ sum rule derived in~\cite{martin1976pion} on the
  basis of PCAC. The $\sigma$ sum rule is already
  approximately satisfied for an upper limit of the integration
  around $\sqrt{s}< 1.26$ GeV. The contribution above this upper
  limit tends to cancel and shows an oscillating behavior as a
  function of the upper limit of the integration. This is closely
  linked with having a value of $b_2$ compatible with zero and a not
  well-defined sign for the ratio $b_2/a_2$ appearing in
  $R_{I_t=2}$. For an upper limit of $\sqrt{s}< 1.26$, the truncated
  Adler sum rule also provides a rather reasonable value of
  89 MeV for $f_\pi$. This result is certainly more reasonable if one bears in
  mind that the left-hand side of the sum rule has been computed at
  lowest order in ChPT only. This observation suggests a kind of
  super-convergent dispersion relation which will be important to set
  up our model below.

\end{itemize}

Given the above discussion, one finds reasons to saturate the sum
rules with the lowest-lying resonances below 1 GeV. In addition, we
already mentioned that the $I_t=0$ sum rule in Eq.~(\ref{eq:t00}) is
saturated also, with great accuracy, at 1 GeV (integral and $R$
contributions tend to cancel above 1 GeV). All this, gives support to
the scenarios that will be presented below, in which we will take
into account only the lowest-lying resonance degrees of freedom.  Yet,
we will make use of the Adler and $\sigma$ sum rules, saturated at
energies of about 1 GeV, to find out short-distance constraints that
will make more predictive the R$\chi$T approach adopted here.

\section{Large--$\mathbf{N_C}$ aspects of $\mathbf{\pi\pi}$ scattering}
\label{sec:largeN}

In this section we discuss some aspects of the large--$N_C$ limit which
will provide some useful guidance in our analysis of $\pi\pi$
scattering.

\subsection{$\pi\pi$ elastic scattering amplitude in the SRA}

From the lowest-order R$\chi$T Lagrangian~\cite{Ecker:1988te,Ecker:1989yg},
describing
the couplings of the lowest-lying $V(1^{--})$ , $A(1^{++})$, $S(0^{++})$ and
$P(0^{-+})$
resonance nonet multiplets to the pions, we find\footnote{Here we use the
antisymmetric
field formulation where the
  $A$--$\pi$ mixing is absent. Note that the axial resonance does not
  contribute to the elastic $\pi\pi$ scattering amplitude.
  After proper incorporation of short-distance
  constraints, the Proca formulation using $g_V= G_V/f_\pi$ yields the same
  amplitude \cite{Ecker:1989yg}. $U(3)$ nonet resonance fields are generically
  parametrized as
  $R= \frac{1}{\sqrt{2}} \sum_{i=1}^8 R_i \lambda_i +
  \frac{1}{\sqrt{3}}R_0$ with $\lambda_i$ the standard Gell-Mann
  matrices.}
\begin{eqnarray}
A^{\rm SRA}
(s,t,u)&=&  \frac{m_\pi^2-s}{f_\pi^2}
+\frac{G_V^2 }{f_\pi^4} \left \{ \frac{t(s-u)}{t-m^2_V}
+\frac{u(s-t)}{u-m^2_V}\right \}  \nonumber\\
&+& \frac{2}{3f_\pi^4}\frac{\left [c_d(s-2m^2_\pi)+2m^2_\pi\ c_m
\right]^2}{s-m^2_{S_8}} \nonumber\\
&+&
 \frac{4}{f_\pi^4}\frac{\left [{\bar c}_d(s-2m^2_\pi)+
2m^2_\pi \ {\bar  c}_m \right]^2}{s-m^2_{S_1}}
\nonumber\\
&+&
\frac{8d_m^{\,2}}{f_\pi^4}\frac{m^4_\pi }{m^2_{P_8}-m_\pi^2}\, . \label{eq:asra}
\end{eqnarray}
In the large--$N_C$ limit,  $|\bar c_d|= |c_d|/\sqrt{3}$  and
$|\bar c_m|= |c_m|/\sqrt{3}$.
We have specified for clarity the contributions from
non-degenerate singlet, $S_1$, and  iso-singlet octet, $S_8$, fields.
Quite generally, $m_{S_8}-m_{S_1} = {\cal O} (1/N_C)$
and mixing effects have been analyzed in
Refs.~\cite{Black:1998wt,Cirigliano:2003yq}.\footnote{
Ref. \cite{Cirigliano:2003yq} finds sizeable mixing effects between the singlet and
octet scalar-isoscalar mesons and identifies two possible phenomenologically
acceptable scenarios for the tree-level mass eigenstates:
a) $M_L = 1.35$ GeV and $M_H = 1.47$ GeV,
or b) $M_L = 0.985$ GeV and $M_H = 1.74$ GeV.
The light solution was less preferential at it would correspond
to a case where $f_0(980)$ would not couple to pions. Therefore in
several studies the heavy solution has been adopted.}
 Glueball mixing
within R$\chi$T has been discussed in Ref.~\cite{Giacosa:2005zt}.

Taking $m_{S_8}=m_{S_1}=m_S$, $|\bar c_d|= |c_d|/\sqrt{3}$  and
$|\bar c_m|= |c_m|/\sqrt{3}$, we reproduce the expressions in
Ref.~\cite{Guo:2009hi}.
In principle, the couplings appearing in the scattering amplitude can be
determined by analyzing the decay processes $\rho \to 2 \pi $ and $S
\to 2 \pi $ with $S =\sqrt{2/3}\, S_1 + S_8 /\sqrt{3}$,
corresponding to a $(\bar u u + \bar d d)/\sqrt{2}$ flavor
composition in the $q\bar q$ picture,
which yield in the chiral limit (see also
Ref.~\cite{Guo:2007ff}),\footnote{The full expressions are $\Gamma_S =
\frac{3 m_S^3}{16 \pi f_\pi^4} \rho_S \left[ c_d + (c_m -c_d)
\frac{2m_\pi^2}{m_S^2}\right]^2$ and $\Gamma_V = \frac{G_V^2 m_V^3}{48
\pi f_\pi^4} \rho_V^3$, where $\rho_R = \sqrt{1-4m_\pi^2/m_R^2}$.}
\begin{eqnarray}
\Gamma_S &=& \frac{3 c_d^2 m_S^3}{16 \pi f_\pi^4} \, , \\ \Gamma_V &=&
\frac{G_V^2 m_V^3}{48 \pi f_\pi^4}\, .
\end{eqnarray}
The residues of the scalar-isoscalar and the vector-isovector poles in
the partial-wave amplitudes are
\begin{eqnarray}
g_S &=& \frac{c_d m_S^2}{f_\pi^2}\, , \nonumber\\
g_V &=& \frac{G_V m_V^2}{\sqrt{3}f_\pi^2}\, .
\label{eq:res-largeNc}
\end{eqnarray}
Note that the (large--$N_C$) relations $g_S^2 = 16 \pi \Gamma_S m_S /3$
and $g_V^2 = 16 \pi \Gamma_V m_V$ hold in the SRA, in the chiral limit.
The SRA amplitude in Eq.~(\ref{eq:asra}) contains too many parameters to
be analyzed in full detail. In the next subsections, we will discuss a
sensible way of reducing the number of couplings and masses.

After projecting onto partial waves we get for the $(I,J)=(0,0)$,
$(1,1)$ and $(2,0)$ channels the following asymptotic behavior at large values of
$s$:
\begin{eqnarray}
T_{IJ}^{\rm SRA} (s) = \lambda_{IJ} \;\frac{ 2 c_d^2 + 3G_V^2 - f_\pi^2}{f_\pi^4}\;
s + \dots\, , \qquad \lambda_{00} = 1\, , \quad \lambda_{11} = \frac16\, ,
\quad \lambda_{20} = -\frac12 \, .
\end{eqnarray}
The above behavior implies that subtractions would be
necessary to make convergent a dispersion relation.\footnote{
Requiring that the leading term proportional to $\lambda_{IJ}$ vanishes
would yield the relation $2 c_d^2 + 3G_V^2 = f_\pi^2$, advocated
in Ref.~\cite{Guo:2007ff}. This would imply $G_V \le f_\pi/\sqrt{3}$,
giving a 30\% too small $\rho \to 2 \pi $ decay width.
This relation has also been found as a necessary high-energy constraint in a
NLO R$\chi$T analysis of the vector form factor, incorporating
subleading $1/N_C$ corrections~\cite{Pich:2010sm}. In the $c_d=0$
limit, it gives $G_V = f_\pi/\sqrt{3}$, which was also found
in the study of one-meson radiative tau decays carried out
in~\cite{Guo:2010dv}.
Notice however that imposing this relation is not enough to make
subtractions unnecessary, because the partial waves would still grow
at large values of $s$ as
$T_{IJ} (s) \sim (m_V^2 G_V^2/f_\pi^4)\,\log{(s/m_V^2)}$.
}
We will add
subtraction constants {\it after} unitarization.

\subsection{Short-distance constraints}

The short-distance constraints encompass R$\chi$T with proper high-energy
behavior~\cite{Pich:2002xy}. In general they produce a set of
conditions which, for a limited set of resonances and in particular in
the SRA, reduce the number of independent couplings. These conditions
may be over-determined yielding at times to mutually inconsistent
values, a problem which can be side-stepped by introducing more
resonances. In the present case we think it of interest to pursue such
an analysis within $\pi\pi$ scattering.

In the chiral limit, the $t$-channel amplitudes, see
Eq.~(\ref{eq:s-to-t}), in the forward direction have the following
asymptotic behavior in terms of the crossing-odd variable $\nu =
(s-u)/2$:
\begin{eqnarray}
T_{I_t=0}^{\rm SRA} (\nu,0) &=& 2\, \frac{c_d^2 m_S^2 + \frac23 g_T^2 m_T^2 +  G_V^2
m_V^2}
{f_\pi^4} + {\cal O} (\nu^{-2}) \, ,\nonumber
\\
T_{I_t=1}^{\rm SRA} (\nu,0) &=& \frac{6 c_d^2  - 3 f_\pi^2 + 4 g_T^2  + 3 G_V^2 }
{3 f_\pi^4} \, \, \nu  + {\cal O} (\nu^{-1}) \, ,\nonumber
\\
T_{I_t=2}^{\rm SRA}  (\nu,0) &=& 2\, \frac{c_d^2 m_S^2 + \frac23 g_T^2 m_T^2 - G_V^2
m_V^2/2}
{f_\pi^4} + {\cal O} (\nu^{-2}) \, ,
\end{eqnarray}
where we have included momentarily the tensor meson coupling $g_T$ and
mass $m_T$, to be discussed below in more detail. This limit is
compatible with the Froissart bound, a specific merit of the
antisymmetric tensor
formulation~\cite{Ecker:1989yg}.\footnote{Physical results are
actually independent of the field representation. The naive exchange
of Proca fields does not satisfy the Froissart bound, but after
suitable polynomial subtractions to comply with the short-distance
constraints, one ends up with the same amplitude
\cite{Ecker:1989yg}. Fields remain a useful framework to incorporate
symmetries, see also the discussion in Ref.~\cite{Ecker:2007dj}.}

As already mentioned, it makes sense to divide the $\pi\pi$ scattering
amplitudes in Eq.~(\ref{eq:forw-disp}) into three pieces:
i) the low-energy part which takes the form of subtraction constants and is fixed
by chiral symmetry, ii) an intermediate-energy part, dominated by
resonance exchange, and iii) the high-energy remainder which we expect
to be responsible for the Regge behavior.  Therefore, if we impose a
behavior for the resonance contribution no worse than suggested by
Regge theory, we obtain the constraints
\begin{eqnarray}
f_\pi^2 &=&2 c_d^2   + \frac43 g_T^2  + G_V^2 \, ,\label{eq:argumentos-mseqmv-bis}
\\
0 &=& 6 c_d^2 m_S^2 + 4 g_T^2 m_T^2 -3 G_V^2 m_V^2 \, .\label{eq:argumentos-mseqmv}
\end{eqnarray}
These constraints correspond to require that
the $\nu$ and the  $\nu^0$ coefficients of $T_{I_t=1}^{\rm SRA}$ and
$T_{I_t=2}^{\rm SRA}$, respectively, vanish in
the large--$\nu$ regime. The second condition is less robust than the
first one, attending to what we have discussed on Regge phenomenology
above. Indeed,  Eqs.~(\ref{eq:argumentos-mseqmv-bis})
  and (\ref{eq:argumentos-mseqmv}) can be also obtained from the Adler and
    $\sigma$ sum rules, in the chiral limit, using the narrow
    resonance approximation of Eq.~(\ref{eq:nra}) to estimate the
    cross sections that appear in the right-hand sides of the sum
    rules. Note, that since we have not considered any exotic isotensor
    resonance, we are approximating $\sigma_2=0$.

In the absence of tensor couplings, $g_T=0$, these constraints imply $
c_d = \sin \phi f_\pi / \sqrt{2}$, $ G_V = \cos \phi f_\pi$
and $ m_V /m_S = \tan \phi $, where $\phi$ is a mixing
angle. The KSFR relation ($G_V = f_\pi/\sqrt{2}$) requires $\phi =
\pi/4$ and hence $m_S=m_V$, as well as $2 c_d= \sqrt{2} G_V= f_\pi $,
implying $\Gamma_S=9 \Gamma_V /2$.  These constraints have also been
found in the algebraic chiral-symmetry
approach~\cite{Gilman:1967qs,Weinberg:1969hw,Weinberg:1990xn} and
 can be re-written, in
terms of the decay widths, as
\begin{eqnarray}
1 &=&   \frac{\Gamma_S}{m_S} \frac{32 \pi f_\pi^2}{3 m_S^2}+
\frac92 \frac{\Gamma_V}{m_V} \frac{32 \pi f_\pi^2}{3 m_V^2}\, ,
\\
0 &=& \frac{\Gamma_S}{m_S} - \frac92 \frac{\Gamma_V}{m_V}\, ,
\end{eqnarray}
which yield the value of the scalar mass and width to be
\begin{eqnarray}
m_S = 660 {\rm ~MeV}\, , \qquad \Gamma_S = 570 {\rm ~MeV}\, ,
\end{eqnarray}
when phenomenological values for the mass and width of the $\rho$
meson are used. These numbers are quite sensitive to details. For
instance if the KSFR set of parameters is used, one gets instead
(taking as input the $\rho$ meson mass)
\begin{eqnarray}
m_S = m_V= 775 {\rm ~MeV}\, , \qquad \Gamma_S = \frac92 \Gamma_V = 805 {\rm ~MeV}\,
.
\end{eqnarray}
Of course, given the fact that the scalar turns out to be a broad
resonance, it is unclear what these estimates should be compared to,
since generally a resonance is characterized by the complex pole and
the complex residue of the scattering amplitude. The benchmark
calculation of the pole on the second Riemann sheet of the $\pi\pi$
scattering amplitude~\cite{Caprini:2005zr,Kaminski:2006qe}, when
written as $s_\sigma = m_\sigma^2 - {\rm i} m_\sigma \Gamma_\sigma $,
yields $ m_\sigma = 347(17)$ MeV and $ \Gamma_\sigma = 690(48)$ MeV.
On the other hand, the connection between the Breit-Wigner (BW) resonance
parameters, defined as $\delta(m_{\rm BW}^2)=\pi/2$ and $ \Gamma_{\rm
  BW} = 1/ (m_{\rm BW} \delta '(m_{\rm BW}^2))$, and the pole
resonance parameters has been discussed on the light of their $N_C$
behavior in a model-independent fashion~\cite{Nieves:2009kh},
suggesting that the large shift in the mass is ${\cal O} (1/N_C^2)$
and can be computed, yielding an acceptable extrapolation of $m_{\rm
  BW} \sim 700$ MeV. The recent $\pi\pi$--scattering analysis of
Ref.~\cite{Martin:2011cn} leads to the Breit-Wigner values $
\left [m_{\sigma,\ {\rm BW}}, \Gamma_{\sigma,\ {\rm BW}}\right ]=\left
[841(5) {\rm
  ~MeV}, 820 (20){\rm ~MeV}\right ]$.\footnote{The value of the pole is
  $\sqrt{s_\sigma}= 445(8) - {\rm i}\, 297(7)$ MeV, in agreement with
  Ref.~\cite{Caprini:2005zr}. On the other hand, the model-independent
  large--$N_C$-based extrapolation from the resonance pole mass to the
  BW pole mass \cite{Nieves:2009kh} yields $m_{\sigma,\ {\rm BW}} =
  670(20)$ MeV, when the UFD parameterization of Ref.~\cite{Martin:2011cn}
  is used.}
  The quoted errors above also account for the existing
differences when UFD and CFD parameterizations are used. This yields a
ratio $\Gamma_{\sigma,\ {\rm BW}} / m_{\sigma,\ {\rm BW}} \sim 5.0(1)\;
\Gamma_{\rho,\ {\rm BW}} / m_{\rho,\ {\rm BW}} $, which suggests a
$10\%$ accuracy of large $N_C$ in the SRA,
supporting as well  the identification of the large--$N_C$ parameters with the 
BW ones.

\subsection{Higher-energy resonances}

We have so far been limited to states below the $\bar K K$
threshold, $\sqrt{s} < 1 $ GeV. On the other hand, Regge
behavior works for $\sqrt{s} > 1.4$ GeV. So, it is interesting
to see the modifications induced by other resonances which may decay
into $2\pi$, in the mass range $1 {\rm ~GeV} < m_R < 1.4 {\rm ~GeV}$,
namely $f_0(980)$, $h_1 (1170)$, $b_1 (1235)$, $f_2(1275)$, $f_0
(1370)$ and $\rho_1 (1450)$.

After implementing the appropriate short-distance constraints, via the
Froissart bound, the inclusion of a $2^{++}$ tensor yields a resonance
amplitude~\cite{Ecker:2007us}
\begin{eqnarray}
A_T(s,t,u)= - \frac{2 g_T^2}{f_\pi^4} \,\frac{(t-u)^2-s^2/3}{m_T^2-s}
- \frac{4 g_T^2}{f_\pi^4} \,\frac{(s^2-t^2-u^2)}{m_T^2}\, ,
\end{eqnarray}
where the coupling is determined from the decay into $\pi\pi$ in a
relative D-wave yielding
\begin{eqnarray}
\Gamma_T = \frac{g_T^2 m_T^3}{40 \pi f_\pi^4}\; \rho_T^5\, .
\end{eqnarray}
The previous amplitude yields the following contributions to the
${\cal O}(p^4)$ ChPT couplings ~\cite{Ecker:2007us}: $L_1^T=L_2^T=0$ and
$L_3^T=g_T^2/(3m_T^2) \sim 0.16 \times 10^{-3}$ (see also
Ref.~\cite{Toublan:1995bk}).  The axial $1^{+-}$ mesons, such as $h_1
(1170)$ and $b_1 (1235)$, give just a purely polynomial contribution to
the $\pi\pi$ scattering amplitude (without $s$-channel propagator poles) which
cannot satisfy the Froissart bound, yielding to no contribution at all to
the LECs. We remind that the exchange of $J > 1$ resonances in the
$t$-channel naively violates the Froissart bound, a situation which has
been the standard motivation to rely on high-energy Regge behavior as
a way of introducing suitable cancellations.

If the tensor meson $f_2(1275)$ is considered, we might include $\rho'
\equiv \rho_1 (1450)$ and $f_0(980)$ as well, where the decay widths
into $\pi\pi$ are taken to be $\Gamma( f_2 \to \pi\pi)= 150$ MeV,
$\Gamma( f_0 \to \pi\pi)= 80$ MeV and $\Gamma( \rho' \to \pi\pi)= 250$
MeV (note the large inaccuracies).  This yields the extended sum rules
(we remain in the chiral limit)
\begin{eqnarray}
1 &=&   \sum_S \frac{\Gamma_S}{m_S} \frac{32 \pi f_\pi^2}{3 m_S^2}+
 \sum_V \frac92 \frac{\Gamma_V}{m_V} \frac{32 \pi f_\pi^2}{3 m_V^2}
+ \sum_T  5 \frac{\Gamma_T}{m_T} \frac{32 \pi f_\pi^2}{3m_T^2}\, ,
\nonumber \\
0 &=& \sum_S \frac{\Gamma_S}{m_S} +  \sum_T 5 \frac{\Gamma_T}{m_T}
- \sum_V \frac92 \frac{\Gamma_V}{m_V}\, .
\end{eqnarray}
Using PDG values~\cite{Nakamura:2010zzi}, the higher
resonances $f_0(980), f_2(1275)$ and $\rho_1(1450)$ produce
corrections of the order of $(0.06,-0.01)$ for the r.h.s of the first
and second sum rules respectively. This shows a trend to cancellation
which supports that higher states not only play a minor role at low
energies but also in the region of interest below 1 GeV, to leading
order in $N_C$. Therefore we will carry our analysis below with just
scalar $0^{++}$ and vector $1^{--}$ states.\footnote{Of course, it is
  intriguing to analyze the role of the $(I^G,J^{PC})=(2^+,0^{++})$
  exotic state, named X(1420) in the PDG~\cite{Nakamura:2010zzi} (and
  found also in the SU(6) study of Ref.~\cite{GarciaRecio:2010ki}),
  which decays into $2\pi$ in a S wave with estimated mass
  $M_X=1420(20)$ MeV and width $\Gamma_X = 160(20)$ MeV. The
  contributions of such a state to the Adler and $\sigma$ sum rules,
  Eqs.~(\ref{eq:adler-sr},\ref{eq:sigma-sr}), in the narrow-resonance
  approximation are $-80\pi \Gamma_X / 3 m_X^3 \sim -4.6 {\rm
    ~GeV}^{-2}$ and $16\pi \Gamma_X / 3 m_X \sim 1.9 $, respectively.
  As compared to the individual contributions from other mesons, they
  seem too small to provide a clear signal.}

\subsection{Other short-distance constraints}

Alternatively to the previous analysis, one may derive short-distance
constraints from other processes involving two- and three-point functions
\cite{Ecker:1989yg,Cirigliano:2006hb}.
Imposing the short-distance properties of the underlying QCD dynamics,
within the SRA, one gets~\cite{Ecker:1988te,Pich:2002xy}:
\begin{equation}
\sqrt{2} G_V = 2 c_d = 2 c_m = 2 \sqrt{3}\ {\bar c}_d = 2 \sqrt{3}\
     {\bar c}_m = 2\sqrt{2} d_m =  f_\pi\, .  \label{eq:rct-couplings}
\end{equation}
These constraints are obtained from a variety of processes, some of
them involving electroweak probes. It is remarkable that they turn out
to be totally compatible with those deduced here by looking to
$\pi\pi$ phenomenology at high energies, for a vanishing tensor-meson
contribution ($g_T=0$).  Neglecting the $f_2(1270)$ and higher-mass
resonance contributions leads to a realistic and simplified scenario
for the purpose of the present work.
On the other hand, from the $\pi\pi$ scattering amplitude, we have derived
the additional restriction $m_S=m_V$. Although we will explore the effects of
this constraint in one of the fits that will be presented
below, we should mention here that it relies on the assumption of
an asymptotic behavior for $T_{I_t=2} (\nu)$
more convergent than that of a constant ($\nu^0$). This is not a totally robust
result, though it is certainly plausible, given the accuracy of
current $\pi\pi$ analyses at high energies~\cite{Martin:2011cn}.

 For $m_V=775$ MeV, the conditions in
Eq.~(\ref{eq:rct-couplings}) lead to
\begin{eqnarray}
\Gamma_S &=& \frac{3 m_S^3}{64\pi f_\pi^2}\; \rho_S = 750 {\rm ~MeV}\, ,\\
\Gamma_V &=& \frac{m_V^3}{96\pi f_\pi^2}\; \rho_V^3 = 150 {\rm ~MeV}\, .
\end{eqnarray}
Besides, by requiring the two-point correlation functions of two
scalar or two pseudoscalar currents to be equal at high energies, up
to corrections of the order $\alpha_s f^4/t^2$, one
finds~\cite{Pich:2002xy} $m_{P_8}=\sqrt{2}m_S$. This relation
involves a small correction, of the order 5\%, which we neglect,
together with the tiny effects from light quark masses.  We will use
Eq.~(\ref{eq:rct-couplings}) to fix our parameters below, but allowing
$m_S$ to be different from $m_V$.

\section{One-Loop ChPT Corrections and Unitarization of the $\pi\pi$ elastic
scattering amplitude}
\label{sec:oneloop}

\subsection{$\pi\pi$ scattering at one loop in ChPT}

At ${\cal O}(p^4)$ in ChPT, the $\pi\pi$ elastic scattering amplitudes
can be written in the form \cite{Gasser:1983yg}:
\begin{eqnarray}
A^{\rm ChPT}(s,t,u) &=& A_2^{\rm ChPT}(s,t,u) + A_4^{\rm
    ChPT}(s,t,u) \, ,\\
 A_2^{\rm ChPT}(s,t,u) &=&
    \frac{m_\pi^2-s}{f_\pi^2}\, ,\\
A_4^{\rm ChPT}(s,t,u) &=&
    \frac{-1}{96\pi^2f_\pi^4}\Big \{ (2{\bar l}_1+{\bar l}_2-\frac72)s^2+
    ({\bar l}_2
-\frac56)(t-u)^2 + 4(3{\bar l}_4-2{\bar
    l}_1-\frac13) m_\pi^2 s-(3{\bar l}_3+12{\bar l}_4-8{\bar
    l}_1-\frac{13}{3})m^4_\pi \nonumber\\
  &+& 3(s^2-m_\pi^4){\bar
    J}(s)+ \left(t(t-u)-2m_\pi^2 t+4m_\pi^2 u-2m_\pi^4\right) {\bar
    J}(t)+ \left(u(u-t)-2m_\pi^2 u+4m_\pi^2
    t-2m_\pi^4\right) {\bar J}(u) \Big\} \, .\label{eq:a4chpt}
\end{eqnarray}
The lowest-order amplitude $A_2^{\rm ChPT}(s,t,u)$ is identical to the
first term in Eq.~(\ref{eq:asra}) (pion contribution) and only depends
on the pion mass and weak decay constant. The ${\cal O}(p^4)$
correction involves four SU(2) renormalization-scale-independent LECs:
${\bar l}_i$ ($i=1,2,3,4$).  In addition, $A_4^{\rm ChPT}(s,t,u)$
includes one-loop chiral corrections, which are suppressed by one
power of $1/N_C$; they are parameterized through the loop function
\begin{equation}
{\bar J}(s)= 2+\rho(s)\,\log \left[
\frac{\rho(s)-1}{\rho(s)+1} \right]\, .
\end{equation}
For the sake of completeness, we recall here, the relation between the
LECs ${\bar l}_i$ and the most common SU(3) parameters $L^r_i(\mu)$
\cite{Gasser:1983yg,Gasser:1984gg}:
\begin{eqnarray}
{\bar l}_i & = & \frac{32\pi^2}{\gamma_i} l^r_i(\mu)    - \log
(m_\pi^2/\mu^2)\, ,
\label{eq:lbar}
\end{eqnarray}
where
\begin{eqnarray}
\gamma_1=\frac13\, , \quad \gamma_2=\frac23\, , \quad
\gamma_3=-\frac12\, , \quad  \gamma_4=2
\end{eqnarray}
and
\begin{eqnarray}
l^r_1 &=& 4L_1^r+2L_3-\frac{\nu_K}{24}\, , \nonumber \\
l^r_2 &=& 4L_2^r-\frac{\nu_K}{12}\, , \nonumber \\
l^r_3 &=& -8L_4^r-4L_5^r+16L_6^r+8L_8^r-\frac{\nu_\eta}{18}\, , \nonumber \\
l^r_4 &=& 8L_4^r+4L_5^r-\frac{\nu_K}{2}\, ,
\label{eq:lrmu-lbar}
\end{eqnarray}
with $32\pi^2 \nu_{K,\eta} = 1+\log(\hat m^2_{K,\eta}/\mu^2)$, $\hat
m_\eta= 4 \hat m_K/3$ and $\hat m_K\sim 468$ MeV the kaon mass in
the limit $m_u=m_d=0$. The renormalized coupling constants $l^r_i(\mu)$ and
$L^r_i(\mu)$ depend logarithmically on the dimensional regularization
scale $\mu$:
\begin{eqnarray}
L_i^r (\mu_2) = L_i^r (\mu_1) +
\frac{\Gamma_i}{16\pi^2}\log(\frac{\mu_1}{\mu_2})\, ,
\end{eqnarray}
where
\begin{eqnarray}
\Gamma_1 = \frac3{32}\, , \quad
\Gamma_2 = \frac3{16}\, , \quad \Gamma_3=0\, , \quad
\Gamma_4= \frac1{8}\, , \quad\Gamma_5= \frac3{8}\, , \quad
\Gamma_6= \frac{11}{144}\, , \quad \Gamma_8= \frac5{48}\, .
\end{eqnarray}

The corresponding $t$-channel forward scattering amplitudes at $\nu=0$ are
given by:
\begin{eqnarray}
T_{I_t=0}(0) &=& \frac{m_\pi^2}{2 f_\pi^2}+
\frac{m_\pi^4 (-72 \bar l_1-48
  \bar l_2+ 45 \bar l_3 + 36 \bar l_4 + 39 \pi -101)}{576 \pi ^2 f_\pi^4} + {\cal O}
  (p^6)\, ,\nonumber
\\ T_{I_t=1}'(0) &=& -\frac{1}{f_\pi^2}+
\frac{(-48 \bar l_4+ 9 \pi -10) m_\pi^2
}{384 \pi ^2 f_\pi^4} + {\cal O} (p^6)\, , \nonumber \\
T_{I_t=2}(0)
&=& -\frac{m_\pi^2}{f_\pi^2}+
\frac{(-48 \bar l_2+18 \bar l_3 - 72 \bar l_4 + 21 \pi +10) m_\pi^4}{576 \pi ^2
f_\pi^4}+{\cal O} (p^6)\, .
\label{eq:chpt_It}
\end{eqnarray}
The ${\cal O} (p^4)$ corrections are at the $10-20\%$ level of the
lowest-order ones in the Adler and $\sigma$ sum
rules.

\subsection{${\cal O}(p^4)$ ChPT-improved SRA amplitudes and large--$\mathbf{N_C}$
counting rules.}

In the limit of a large number of colors~\cite{'tHooft:1973jz,Witten:1979kh},
$A^{\rm SRA}$ in Eq.~(\ref{eq:asra}) scales like $1/N_C$, since the pion
 weak decay constant behaves like ${\cal O} (\sqrt{N_C})$. Furthermore, $A^{\rm
 SRA}$
 provides the leading--$N_C$ prediction for the actual $\pi\pi$ scattering
 amplitude, with the only limitation of
 considering just the lowest-lying nonet of exchanged
 resonances~\cite{Ecker:1988te}. This
 latter approximation is justified as long as $s$, $t$ and $u$ are kept
 far from the second resonance region.

The lightest resonances have an important impact on the low-energy
dynamics of the pseudoscalar bosons. Below the resonance mass scale,
the singularity associated with the pole of a resonance propagator is
replaced by the corresponding momentum expansion; therefore, the
exchange of virtual resonances generates derivative Goldstone
couplings proportional to powers of $1/m^2_R$ . At lowest order in
derivatives, this gives the large--$N_C$ predictions for the ${\cal
O}(p^4)$ ChPT couplings~\cite{Ecker:1988te}.
At ${\cal O}(p^4)$ the amplitude $A^{\rm SRA}$ in Eq.~(\ref{eq:asra})
reduces to
\begin{eqnarray}
A_4^{\rm SRA}
(s,t,u)&=&  \frac{m_\pi^2-s}{f_\pi^2}
-\frac{G_V^2 }{f_\pi^4} \left \{ \frac{t(s-u)}{m^2_V}
+\frac{u(s-t)}{m^2_V}\right \} \nonumber \\
&-& \frac{2}{3f_\pi^4}\frac{\left [c_d(s-2m^2_\pi)+2m^2_\pi\ c_m
\right]^2}{m^2_{S_8}} \nonumber \\ &-&
 \frac{4}{f_\pi^4}\frac{\left [{\bar c}_d(s-2m^2_\pi) +
2m^2_\pi \ {\bar c}_m \right]^2}{m^2_{S_1}} \nonumber \\
&+&
\frac{8d_m^{\,2}}{f_\pi^4}\frac{m^4_\pi }{m^2_{P_8}}\, ,\label{eq:lsu3}
\end{eqnarray}
which constitutes the leading $1/N_C$ approximation to $A^{\rm ChPT}$.
The polynomial form can be re-written as
\begin{eqnarray}
 A^{\rm SRA}_4 (s,t,u)&=& \frac{m_\pi^2-s}{f_\pi^2} - \frac{4}{f_\pi^4} \,\Big\{ (
2L_1^{\rm SRA} + L_3^{\rm SRA})(s-2m_\pi^2)^2
+ L_2^{\rm SRA} \left[
  (t-2m_\pi^2)^2+(u-2m_\pi^2)^2 \right]  \Big\}
\nonumber \\
&-& \frac{8 m_\pi^2}{f_\pi^4}\,\Big\{
 (2 L_4^{\rm SRA} + L_5^{\rm SRA}) s + ( 4 L_6^{\rm SRA} + 2L_8^{\rm SRA} -4
L_4^{\rm SRA}-2 L_5^{\rm SRA})m_\pi^2 \Big\}  \, .
\label{eq:lsu3prim}
\end{eqnarray}
From Eqs.~(\ref{eq:lsu3}) and
    (\ref{eq:rct-couplings}), one trivially finds
\begin{eqnarray}
&& 2L_1^{\rm SRA}=L_2^{\rm SRA}= \frac{f_\pi^2}{8m_V^2}\, , \quad
L_3^{\rm SRA}= -\frac{3f_\pi^2}{8m^2_V}+\frac{f_\pi^2}{8m_S^2}\, ,
\nonumber \\
&& L_4^{\rm SRA}= 0\, , \quad
L_5^{\rm SRA}= \frac{f_\pi^2}{4m_S^2}\, , \quad L_6^{\rm SRA}= 0\, , \quad L_8^{\rm
SRA}=
\frac{3f_\pi^2}{32 m_S^2}\, ,
\label{eq:lsu3-sra}
\end{eqnarray}
in full agreement with Ref.~\cite{Pich:2002xy}.\footnote{If the
  relation $m_S=m_V$ is further assumed, one gets purely geometrical
  ratios for the non-vanishing LECs:
$$ 2L_1^{\rm SRA}=L_2^{\rm SRA}= -\frac12 L_3^{\rm SRA} = \frac12
  L_5^{\rm SRA} = \frac43 L_8^{\rm SRA} = \frac{f_\pi^2}{8m_V^2}\, .$$ In
  chiral quark models with meson dominance built-in one also has
  $m_S=m_V = \sqrt{24/N_C}\pi f \sim 781$ MeV, for $f=88$ MeV~\cite{Megias:2004uj},
  so that the above relations hold with
  $L_2 = N_C / 192 \pi^2 = 1.5 \times 10^{-3}$.}

Let us pay now some attention to the $N_C$ dependence of the one-loop ChPT
amplitude.  Note that from Eq.~(\ref{eq:a4chpt}), the logarithmic
contribution to $A_4^{\rm ChPT}$ scales as $1/N_C^2$, while the
polynomial piece behaves as $1/N_C$ in the $N_C\gg 1$ limit. This is
because, as shown in the relations (\ref{eq:lsu3-sra}),
 the LECs $L_{i}$ behave as ${\cal O}(N_C)$, with the
exceptions of $L_2-2 L_1$, $L_4$ and $L_6$ that scale as ${\cal
O}(N_C^0)$~\cite{Gasser:1984gg}. The renormalization
scale dependence of the LECs provides further subleading contributions
in the $1/N_C$ counting. Unfortunately, the measured values of the $L_i$
couplings cannot be phenomenologically split into their
large--$N_C$ leading and subleading parts. In
general, one has the scale-dependent relation
\begin{eqnarray}
  L_i^r (\mu) =  A_i \, N_C + B_i (\mu) \, ,\label{eq:eles-nc}
\end{eqnarray}
where
$A_i$ is scale independent. Note that only the $N_C=3$ combination is
experimentally accessible. However a meaningful extension of the
chiral amplitudes to an arbitrary number of colors requires some
knowledge of the coefficients $A_i$ and $B_i$. This
difficulty is precisely what $A_4^{\rm SRA}$ helps to overcome, and
thus the SRA predictions of Eq.~(\ref{eq:lsu3-sra}) can be used to
read off the $A_i$ coefficients.\footnote{Actually, the SU(2)
  scale-independent LECs defined by Eq.~(\ref{eq:lbar}) display the $N_C$
  separation in a scale-independent fashion, since $\bar l_i = a_i N_C
  + b$ where $b$ is {\it common} to all coefficients and stems from
  pion loops. This allows to build differences, $\bar l_i - \bar l_j=
  (a_i-a_j) N_C$, which can be used to extract the leading--$N_C$ contributions
  from
  data up to a constant. This is illustrated in quark model
  calculations~\cite{Megias:2004uj}.}

In the recent work of Ref.~\cite{Guo:2011pa}, the SRA parameters
$G_V,\, c_d,\, c_m,\, {\bar c}_d,\, {\bar c}_m,\, m_V,\, m_{S_8}$ and
$m_{S_1}$ are fitted to data. Afterwards and to extrapolate to $N_C>
3$, the resonance masses are kept constant, while all the couplings
are scaled by a $\sqrt{N_C/3}$ factor. This does not take into account
that singlet and octet scalar resonances become degenerate in the
large--$N_C$ limit. On the other hand, if one fits the resonance
couplings to data, the fitted values do not necessarily match their
leading--$N_C$ values in Eq.~(\ref{eq:rct-couplings}) and might
incorporate some significant $1/N_C$ subleading contributions, which
later on however, are scaled as if they were leading in the $N_C$
counting.  For instance, in Ref.~\cite{Guo:2011pa} a value of around
15 MeV is found for $c_d$, which is around a factor three smaller than
that of $f_\pi/2$ quoted in Eq.~(\ref{eq:rct-couplings}). A proper
extension of this parameter when $N_C$ deviates from 3 should be $2
c_d = f_\pi^{\scriptscriptstyle \, N_C=3} \times\sqrt{N_C/3} + (
30~{\rm MeV}-f_\pi^{\scriptscriptstyle \, N_C=3} ) $, instead of that
assumed in \cite{Guo:2011pa}. For other parameters, there exist also
large deviations between the fitted values found in \cite{Guo:2011pa}
and the leading--$N_C$ estimates given in
Eq.~(\ref{eq:rct-couplings}). These differences seem to indicate
$1/N_C$ corrections to the resonance parameters which are much larger
than expected.  One might wonder the underlying origin of these
disturbing large deviations.  The fitting strategy certainly might
play some role on this; for instance, the choice of the upper energy
limit or the choice of the unitarization procedure. This latter issue
has some relevance which we will address now. In the next section, we
will discuss our unitarization procedure, which is rather similar to
that used in Ref.~\cite{Guo:2011pa}. There appear independent
subtraction constants for each of the $(I,J)=(0,0),\, (1,1)$ and
$(2,0)$ sectors~\cite{Nieves:1998hp, Nieves:1999bx}. However, in
Ref.~\cite{Guo:2011pa}, all three subtraction constants were forced to
be equal. From the discussion in ~\cite{Nieves:1998hp, Nieves:1999bx},
this is somehow an arbitrary choice. The lack of flexibility of data
fits incorporating such constraint might influence the actual values
determined for the resonance couplings and their estimated scaling
with $N_C$.

Let us come back to our scheme. The amplitude $A^{\rm SRA}_4$ in
Eq.~(\ref{eq:lsu3prim}) contains neither pion loop terms, nor the
$1/N_C$ subleading contributions to the polynomial piece of the
one-loop ChPT amplitude. Thus, and to better describe the experimental
phase shifts for $N_C=3$, we propose to use
\begin{eqnarray}
A^{SRA + {\rm ChPT}} (s,t,u)&=& \underbrace{A^{\rm SRA}(s,t,u) }_{{\cal O}(1/N_C)}
+\underbrace{\left[ A^{\rm ChPT}(s,t,u)- A_4^{\rm
  SRA}(s,t,u)\right ]}_{{\cal O}(1/N_C^2)}\, .
\label{eq:modelforA}
\end{eqnarray}
In this way, by construction, we recover the one-loop ChPT
results, while at the same time, all terms in the amplitude that scale
like $1/N_C$ (leading) are also included, within the
SRA. Note, that in the $1/N_C$ counting, the correction $\left\{
A^{\rm ChPT}(s,t,u)- A_4^{\rm SRA}(s,t,u)\right \}$ is incomplete,
since it does not account for all existing subleading $1/N^2_C$
contributions to $A(s,t,u)$.
A complete $1/N_C^2$ calculation would require quantum corrections
stemming also from the low-lying resonances
\cite{Rosell:2004mn,Rosell:2006dt,Pich:2008jm,Pich:2010sm,Rosell:2005ai,Portoles:2006nr}.

\subsection{Unitarized amplitudes}

The ChPT loops incorporate the unitarity field theory constraints in a
perturbative manner, order by order in the chiral expansion. Though
subleading in the $1/N_C$ counting, their effect appears to be crucial
for a correct understanding of $S$-wave $\pi\pi$
phase shifts. Furthermore, resonances show up as  poles of the
unitarized amplitudes in unphysical sheets, which positions provide
their masses and widths. Thus, to discuss the nature of the $f_0(600)$
resonance, it is crucial to restore unitarity. Any unitarization
method re-sums a perturbative expansion of the scattering amplitude in
such a way that two-body elastic unitarity,
\begin{equation}
{\rm Im}T_{IJ}^{-1}(s) = \frac{\rho(s)}{16\pi}\, , \qquad s>
4m_\pi^2 \, ,\label{eq:unita}
\end{equation}
is implemented exactly. To restore
unitarity, we will make use here of a once-subtracted dispersive
representation of $T_{IJ}^{-1}(s)$ (see for instance Sect. 6 of
Ref.~\cite{Nieves:1999bx}). Let be ${\cal T}_{IJ}^{SRA+ {\rm
    ChPT}}(s)$ and $[T_2]_{IJ}(s)$, the $\pi\pi$ amplitudes, in the
$(I,J)$ sector, deduced from $A^{SRA + {\rm ChPT}}(s,t,u)$ and $A_2^{\rm
  ChPT}(s,t,u)$, respectively.  We define an unitarized amplitude
as~\cite{Nieves:1999bx}
\begin{eqnarray}
T^{-1}_{IJ} (s) &=& -C_{IJ}-\bar I_0(s) +V^{-1}_{IJ} (s) \, ,\label{eq:defT}\\
V_{IJ} (s) &=& {\cal T}_{IJ}^{SRA + {\rm ChPT}}(s)
-[T_2]_{IJ}(s)\left(\bar I_0(s)+C_{IJ}\right)[T_2]_{IJ}(s)\, ,
\end{eqnarray}
where $V_{IJ}$ stands for the two-particle irreducible amplitude, and the
subtraction constant and the loop function are given by
\begin{equation}
  C_{IJ}
= -T^{-1}_{IJ}(4m_\pi^2)+ V^{-1}_{IJ}(4m_\pi^2), \, \qquad
\bar I_0(s) =
\frac{1}{16\pi^2} \left[ 2-\bar J(s)  \right] \, .
\end{equation}
Note that {\it i)} the constants $C_{IJ}$ determine the scattering
length/volume in each sector and become undetermined free
parameters,  and {\it ii)} with the election of $V^{-1}_{IJ}(s)$, and
considering $A^{\rm SRA} (s,t,u)-A_4^{\rm SRA} (s,t,u)$
as ${\cal O}(p^6)$, we recover from $T_{IJ}(s)$
the ChPT series up to one loop, in all $(I,J)$ sectors.

Analytical expressions for the ChPT amplitudes projected onto the
different $(I,J)$ sectors can be found in the appendix B of
Ref.~\cite{Nieves:1999bx}.

A final remark concerning the $1/N_C$ counting rules is in order
here. The subtraction constants $C_{IJ}$ must scale as
\begin{equation}
C_{IJ} \sim {\cal O}(N_C^0) \, ,\label{eq:c's}
\end{equation}
because of their definition as the difference between the inverses of the
full and the two-particle-irreducible amplitudes at threshold for each
$(I,J)$ sector.\footnote{The full and the two-particle-irreducible
amplitudes differ by terms which always contains at least one
$s$-channel pion loop, which is subleading in the $1/N_C$ power
counting. Thus, the difference $T^{-1}-V^{-1}$ scales as ${\cal
O}(N_C^0)$, as one readily deduces by noting that both $TV$ and
$(T-V)$ scale as ${\cal O}(1/N_C^2)$ in the $N_C\gg 1$
limit.} Thus, the amplitude $T_{IJ}(s)$ reduces to the $1/N_C$
leading part of ${\cal T}_{IJ}^{SRA + {\rm ChPT}}(s)$ in the limit of
a large number of colors.

\section{$\mathbf{N_C=3}$ phase shifts and $\mathbf{\rho}$ and $\mathbf{f_0(600)}$
meson properties.}
\label{sec:phases}

In this section, we fit the $C_{IJ}$ parameters to phase-shift data
and show results for the poles found in the second Riemann sheet (SRS)
of the amplitudes. The SRS of the $T$ matrix is determined by the
definition of the loop function $\bar I_0(s)$ in the SRS. We follow
here Ref.~\cite{Nieves:2001wt} and use Eq.~(A13) of this latter
reference to compute $\bar I_0(s)$ in the SRS.\footnote{Note that the
  function $\bar J_0(s)$, defined in Eq.~(A8) of \cite{Nieves:2001wt},
  reduces to $\bar I_0(s)$ for equal masses.}

Masses and widths of the dynamically generated resonances are
determined from the positions of the poles, $s_R$, in the SRS of the
corresponding scattering amplitudes in the complex $s$ plane, namely
$s_R= M^2_R-{\rm i}\ M_R \Gamma_R$. For narrow resonances ($\Gamma_R
\ll M_R$), $\sqrt{s_R} \sim M_R - {\rm i}\Gamma_R/2 $ constitutes a
good approximation.\footnote{It has become customary to quote for the
broad $\sigma$ the number $\sqrt{s_\sigma}$ instead of $s_\sigma$.
Here, we will also quote $\sqrt{s_\sigma}$, but we will refrain from
identifying ${\rm Re} \sqrt{s_\sigma}$ with the mass and $-2 {\rm Im}
\sqrt{s_\sigma}$ with the width of the resonance. For the $\rho$
meson, the narrow-resonance approximation works within two standard
deviations.}

The coupling constants of each resonance to the pion pair
are obtained from the residues at the pole, by matching the amplitudes to
the expression
\begin{equation}
T^{IJ}_{SRS}(s)=\frac{g_R^2 }{(s-s_R)} \, , \label{eq:pole}
\end{equation}
for values of $s$ close to the pole. The couplings $g_R$ are
complex in general, and represent independent information from the
pole $s_R$.  In the narrow-resonance approximation, the extrapolation
of Eq.~(\ref{eq:pole}) to the real axis takes a Breit-Wigner form to
comply with unitarity and hence
$g_R^2 = 16\pi m_R \Gamma_R/\rho(m_R^2)$.

The first issue is to select the set of data points to be fitted. We
will consider the scalar--isoscalar, vector--isovector and
scalar--isotensor elastic $\pi\pi$ phase shifts, with a total of 107 data
points, as follows.
\begin{itemize}
\item $I=J=0$ sector: As our main input, we will use the Roy-equations
  results from Refs.~\cite{Ananthanarayan:2000ht,Yndurain:2007qm,
  Kaminski:2006qe} in the energy range $\sqrt{s} \le 750$ MeV. We take
  this upper c.m. energy cut to keep negligible coupled-channel $\bar
  K K$ effects, which give rise to the $f_0(980)$ resonance. We have
  considered the phase-shift determinations of
  Ref.~\cite{Ananthanarayan:2000ht} and that of Eq.~(4.8) of
  Ref.~\cite{Yndurain:2007qm}. For each value of $\sqrt{s}$, we have
  used as central value the average of both results, while the
  absolute difference between them is taken as the error for the
  $\chi^2$ fit. From threshold to the upper cut of 750 MeV, we have
  moved up in steps of 10 MeV, which amounts to a total of 48
  phase shifts to be fitted.

\item $I=J=1$ ($I=2, J=0$) sector: We fit to the phase shifts compiled
in Refs.~\cite{Estabrooks:1974vu,Protopopescu:1973sh}
$\big($\cite{Hoogland:1977kt,Losty:1973et}$\big)$ and consider an upper cut of
$\sqrt{s} \le 910$ MeV (1190 MeV), which comprises a total of 38 (21)
phase shifts.

\end{itemize}
An important point has been the selection of the fitting interval. Clearly
we should restrict the range to the elastic region and, in particular,
below the opening of the first inelastic $K \bar K$ channel. At lowest
order,  this effect corresponds to the two-step process $\pi\pi \to K
\bar K \to \pi\pi$, which in the sub-threshold region yields a real
contribution to the amplitude and is $1/N_C^2$ and
$s_{K\bar K}$--suppressed. The pure elastic re-scattering is just
$1/N_C^2$--suppressed. On the other hand, corrections to the SRA due to
heavier states with mass $M$ are of ${\cal O} (1/(N_C M^2)) $.
Therefore we expect
that, by restricting to low energies,  inelastic effects can be safely
included in $1/N_C$ corrections to the LECs. In other words, we may
allow $M \sim 2 M_K \sim 1 $ GeV without much trouble.

The second issue is to design the fit procedure. For three out of the
five fits examined here, we will fix
\begin{equation}
m_V=0.77~ {\rm GeV}, \qquad m_S=1 ~ {\rm GeV}.
\end{equation}
For the fit {\bf B.3}, we will force $m_V=m_S$, as
 suggested from our discussion of Eqs.~(\ref{eq:argumentos-mseqmv-bis})
 and (\ref{eq:argumentos-mseqmv}),
 and fit the common value to data.  Finally, in the fit {\bf B.1-2} we
 will fix $m_V$ to 0.77 GeV, while $m_S$ is considered as a free parameter.
We will always fit $C_{00}$,
 $C_{11}$ and $C_{20}$ to data; besides those parameters, $m_V$
 and $m_S=m_P/\sqrt{2}$, $A^{SRA + {\rm ChPT}}$ depends still on
 $L^r_{1,2,3,4,5,6,8}(\mu)$, once the relations in
 Eq.~(\ref{eq:rct-couplings}) are implemented.

We have considered two well-differentiated scenarios:
\begin{itemize}

\item {\bf A}: We just take the SU(3) Gasser-Leutwyler parameters
  $L^r_{i}(\mu)$, at  a certain scale $\mu$, from other
  phenomenological analyses. In this type of fits, only
the $C_{IJ}$ parameters are fitted to data.

\item {\bf B}: The contributions of the low-lying vector,
  axial--vector, scalar and pseudoscalar resonances to the $L_i$, and
  therefore to the effective chiral Lagrangian at order ${\cal
  O}(p^4)$, were given in Eq.~(\ref{eq:lsu3-sra}), and thus, the
  renormalized coupling constants  $L_i^r(\mu)$ can be written as a
  sum~\cite{Ecker:1988te}
\begin{equation}
L_i^r(\mu)= L_i^{\rm SRA}+ \hat L_i^r(\mu) \label{eq:Lremainder}
\end{equation}
of the resonance contributions, $L_i^{\rm SRA}$  and a remainder $\hat
L_i^r(\mu)$. The choice of the renormalization scale $\mu$ is
arbitrary. However, it is rather obvious that one can only expect the
resonances to dominate the $L_i^r(\mu)$ when $\mu$ is not too far away
from the resonance region. Therefore, it is common to adopt
$\mu=m_\rho$ as a reasonable choice.
However, one might take as a best fit parameter one scale, $\mu^{RS}$,
for which it occurs a complete resonance saturation of all the LECs
$L_i^r$, this  is  to say,
\begin{equation}
\hat L_i^r(\mu^{RS}) = 0\, .
\end{equation}
In other words, at this
privileged scale $\mu^{RS}$, there is no other contribution in addition
to the meson resonances.  In this type of fits, besides
the $C_{IJ}$ parameters, this scale should be fitted to data.

We have also considered a scenario where the complete resonance
saturation of the LECs $L_i^r$ occurs at two different scales,
$\mu^{RS}_V$ for $L_{1,2}^r$  and $\mu^{RS}_S$ for $L_{4,5,6,8}^r$,
depending whether the LEC is dominated by the vector  or the scalar
resonance contribution. Note that, $L_3$ is renormalization-scale
invariant.

\end{itemize}
\begin{figure*}[tbh]
\begin{center}
\makebox[0pt]{\hspace{-2cm}\includegraphics[height=6cm]{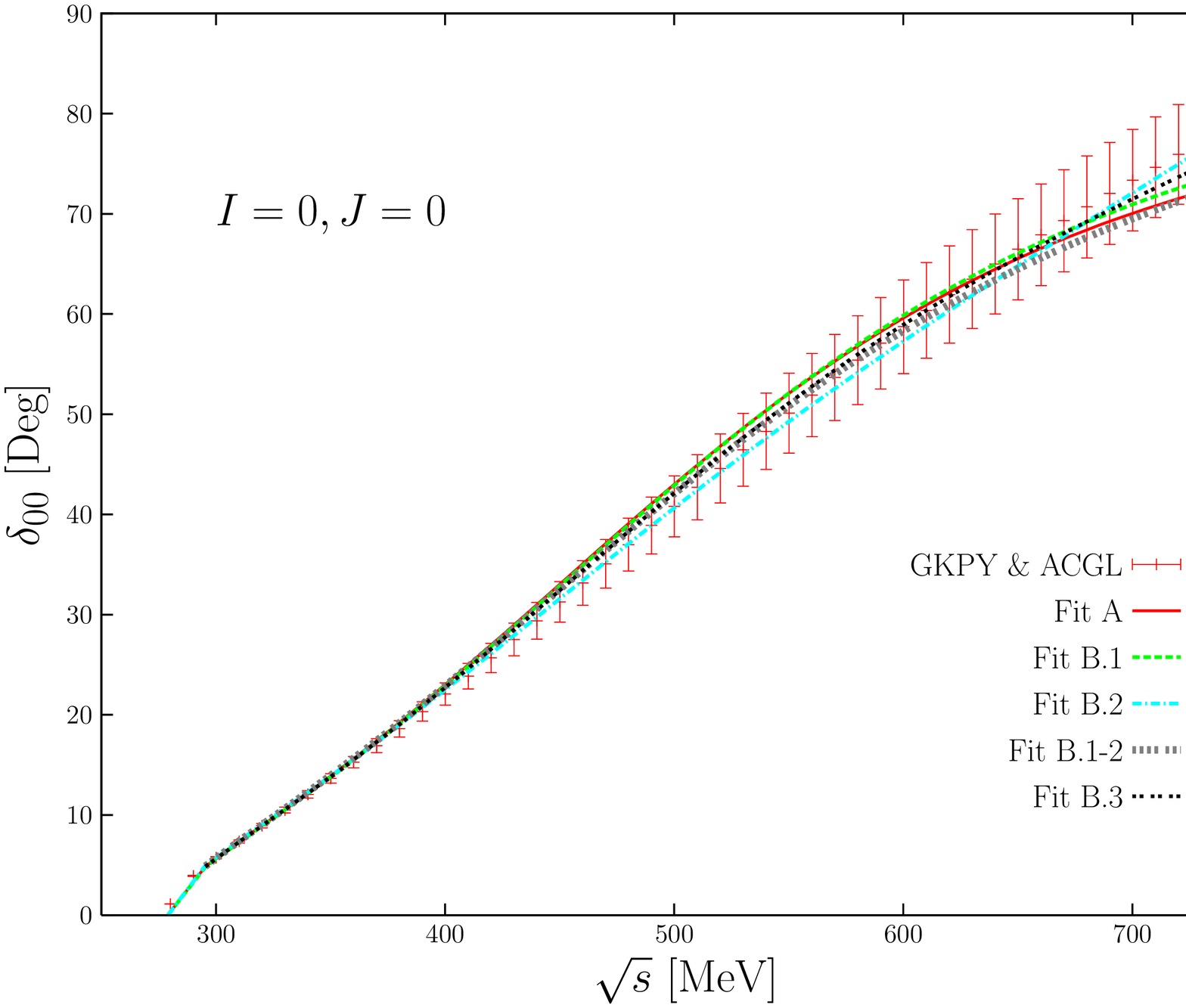}\hspace{1cm}\includegraphics[height=6cm]{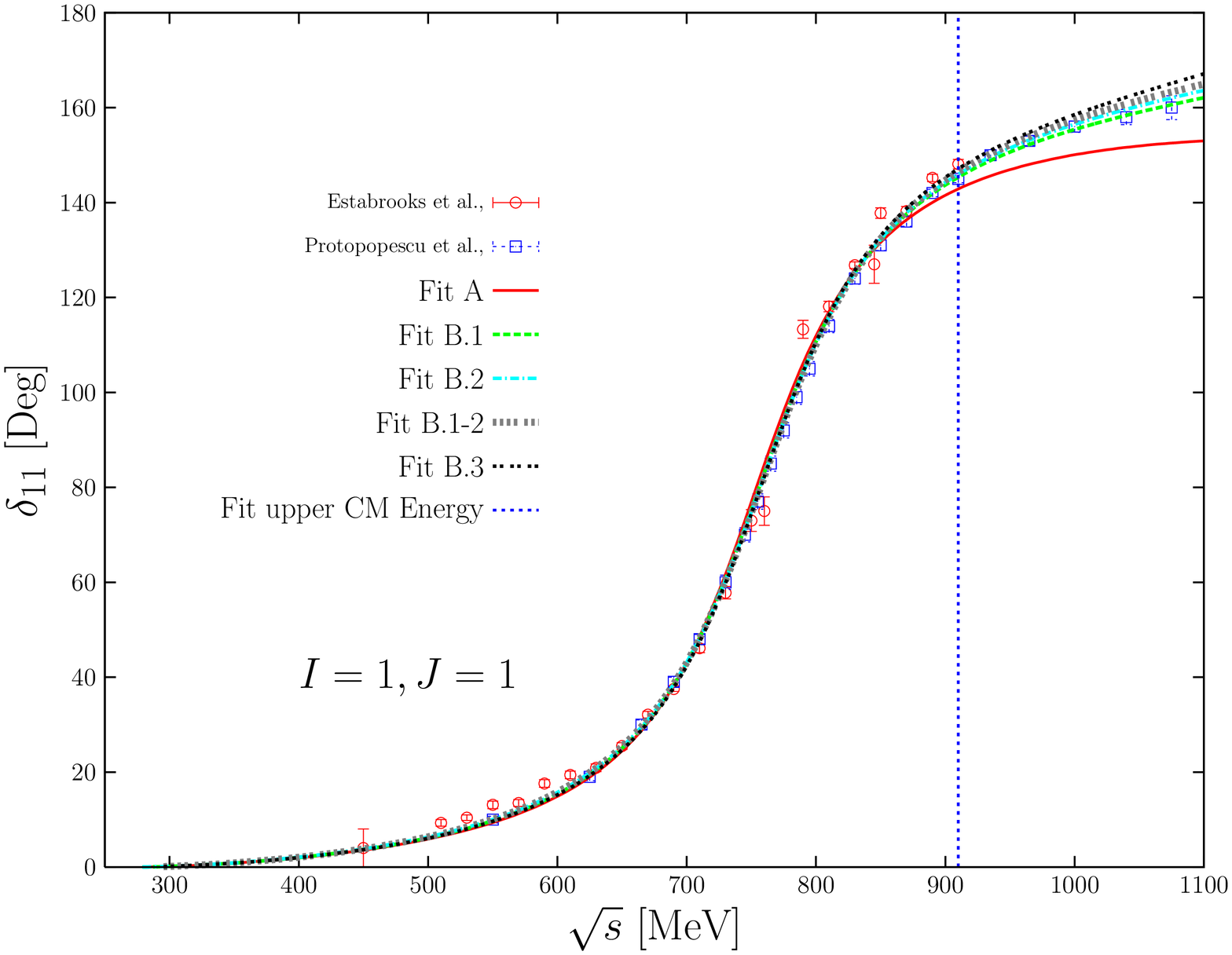}}\\\vspace{1cm}
\makebox[0pt]{\hspace{-2cm}\includegraphics[height=6cm]{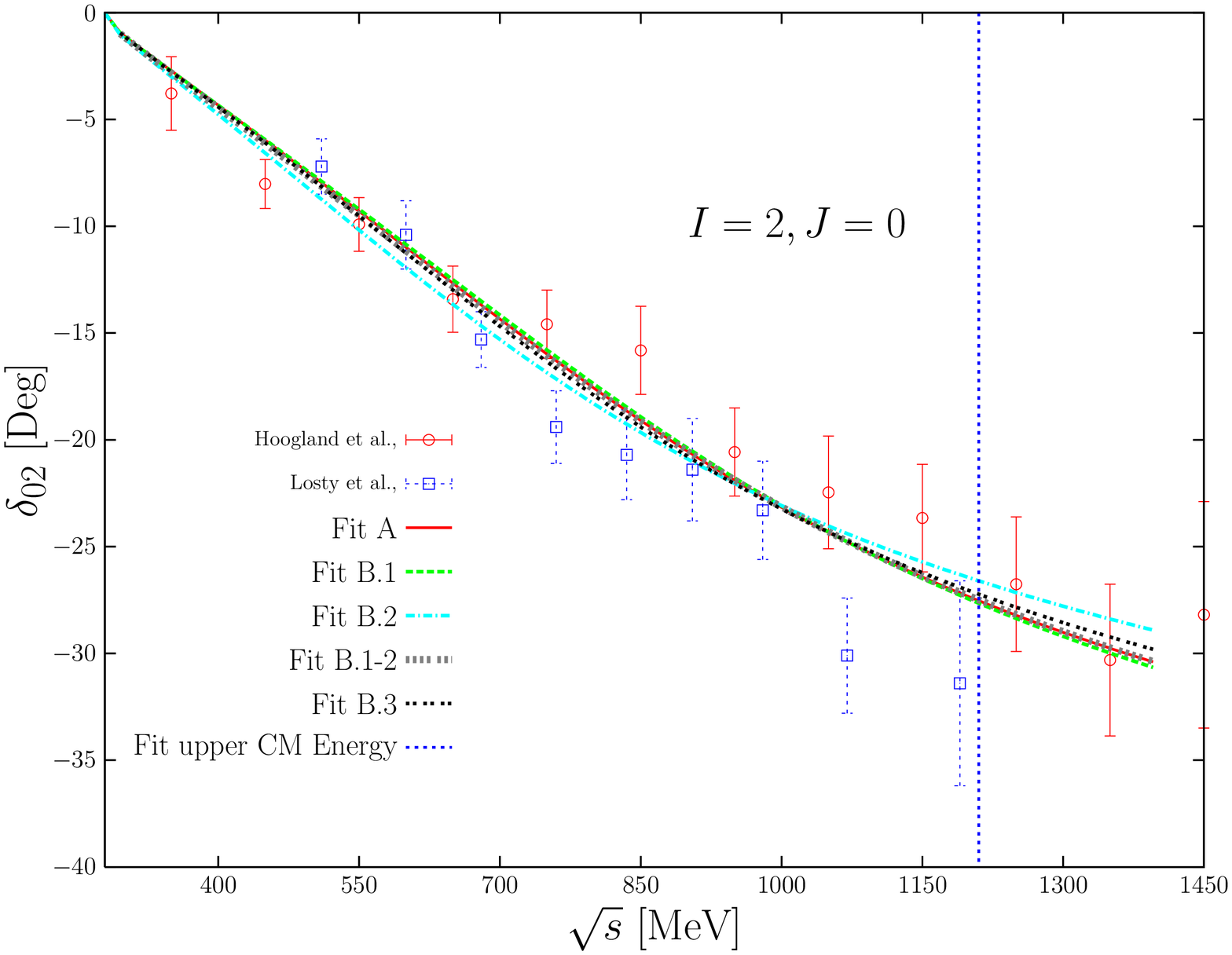}}
\end{center}
\caption{ Theoretical predictions for the phase shifts obtained from
  fits {\bf A}, {\bf B.1}, {\bf B.2}, {\bf B.1-2} and {\bf
  B.3}. Fitted data from Refs.
  ~\cite{Estabrooks:1974vu,Protopopescu:1973sh} ($I=1,J=1$) and
  \cite{Hoogland:1977kt,Losty:1973et} ($I=2,J=0$) are also
  displayed. In the isoscalar-scalar ($I=0,J=0$) channel, the used
  average of the results of
  Refs.~\cite{Ananthanarayan:2000ht,Yndurain:2007qm} is shown, as
  well.}
\label{fig:nc3}
\end{figure*}
Let us start discussing results obtained from five different fits to
the phase-shift data. Best fit parameters and pole properties are
compiled in Table~\ref{tab:1}, while predicted phase shifts and SRS
amplitudes are depicted in Figs.~\ref{fig:nc3},
\ref{fig:i0j0.SRS.nc3}, and~\ref{fig:i1j1.SRS.nc3}. In the first fit
({\bf A}), we take the SU(3) Gasser-Leutwyler parameters
$L^r_{i}(\mu=m_\rho)$ from the ${\cal O}(p^4)$ $K_{\ell 4}$ fit
compiled in Table 2 of Ref.~\cite{Amoros:2000mc}.\footnote{In units of
$10^{-3}$, $L_{1,2,3,4,5,6,8}^r(\mu=m_\rho)=0.46,\ 1.49,\ -3.18,\ 0.,\
1.46,\ 0.$ and 1.08, respectively. For simplicity in the analysis, we
have ignored the errors on these parameters, since they do not affect
the main conclusions of this work.}  A best fit to data using,
instead, the central values of the main ${\cal O}(p^6)$ fit of
Ref.~\cite{Amoros:2000mc} leads to a $\chi^2/dof$ more than twice
larger than that of fit {\bf A}. This is not entirely surprising,
since we are only considering here one-loop chiral logarithms, besides
those required to restore exact unitarity, in the $\pi\pi$ amplitudes.
In the four other fits ({\bf B.1}, {\bf B.2}, {\bf B.1-2} and {\bf
B.3}), we assume that complete resonance saturation occurs, the
corresponding scale (or scales) where it holds is fitted to data. In
the fit {\bf B.1-2}, we fix $\mu^{RS}_V$ to the value obtained in fit
{\bf B.1} and we fit $\mu^{RS}_S$ and $m_S$ to data. Besides in the fit {\bf
B.3}, we force $m_S=m_V$ in the SRA amplitude, and fit this common
mass to the data. We explore this possibility because it is suggested
from our previous discussion on short-distance conditions in
Eqs.~(\ref{eq:argumentos-mseqmv-bis}) and
(\ref{eq:argumentos-mseqmv}). The $m_S=m_V$ constraint was deduced in
Ref.~\cite{Nieves:2009ez} as well, when the one-loop SU(2) ChPT
amplitude, unitarized with the Inverse Amplitude Method (IAM), was
required to be consistent with the SRA.

Since in the SRA amplitudes we have
explicitly incorporated one vector an one scalar poles,
we expect at least one pole in each
of the $I=J=0$ and $I=J=1$ sectors. Because of the re-summation in
Eq.~(\ref{eq:defT}), the pole positions will change with respect to
those of the bare ones ($s=m_V^2$ and $s=m_S^2$) and the
resonances will acquire a width that accounts for their two-pion decay.
In addition, as we will see, some other poles  are generated as well
in the SRS of the scattering amplitudes.
\begin{table*}[htb]
{\small
\begin{center}
\begin{tabular}{|c|c|c|c|c|c|}  \hline
& {\bf A} & {\bf B.1} & {\bf B.2} & {\bf B.1-2} & {\bf B.3}   \\\hline
$C_{00}$ & $-$0.0210 (5)     & $-$0.0218 (9) & $-$0.0278 (18)  &
$-$0.0283 (18) &  $-$0.0144 (5) \\
$C_{11}$ & $-$0.02054 (11)   & $-$0.01996 (11) & $-$0.01979 (12)& 
$-$0.01968 (12) &
  $-$0.0108 (9)  \\
$C_{02}$ & $-$0.0594 (19)    & $-$0.0588 (20) & $-$0.0621 (17) &
$-$0.0593 (19) &  $-$0.0606 (18) \\
$\mu_V^{RS}$ (MeV) & 770 (fixed)    & 693 (26) & 474 (16) & 693 (fixed
  {\bf B.1} ) & 520 (21)  \\
$\mu_S^{RS}$ (MeV) & 770 (fixed)    & $\mu_V^{RS}$ & 1550 (180) & 1190
  (130) & $\mu_V^{RS}$  \\
$m_S$ (MeV)    & 1000 (fixed) & 1000 (fixed)  & 1000 (fixed) &1295 (40) & 738 (3)
  \\
$m_V$ (MeV)    & 770 (fixed) & 770 (fixed) & 770 (fixed)  & 770 (fixed) & $m_S$
  \\
$\chi^2/dof$    & 3.5    & 2.4     & 2.0 & 2.0 & 2.2\\
contrb. $\chi^2$ & 25/323/21   & 26/203/22 &  5/178/20 & 17/171/20 & 12/193/20
\\\hline
$r_{12}, r_{13},r_{14},r_{15}$ & 0, 0,-,- & $0.06$, $0.13$,
$0.80$, - & $0.04$, $0.05$, $0.43$, $0.20$ & $0.09$, $-0.01$, $0.47$, $-0.05$ & $0.20$, $0.08$,
$0.51$, $-0.20$  \\
$r_{23},r_{24},r_{25}$ & 0,-,- & $0.01$, $0.07$, -  & $-0.02$,
$-0.12$, $0.21$ & $-0.01$,
$0.29$, $0.27$ & $-0.02$, $-0.20$, $-0.99$ \\
$r_{34},r_{35}$ & -,- & $0.16$, -& $0.14$, $-0.10$ & $-0.03$, $-0.03$ & $0.15$, $0.02$  \\
$r_{45}$ & - & - & $-0.59$ & $0.60$ & $0.22$
\\\hline
$m_\rho$ (MeV) & 749.1 (4)  & 752.6 (5) & 754.0 (5) & 754.8 (6) & 754.6 (7) \\
$\Gamma_\rho$ (MeV) & 144.5 (3)  & 150.4 (4) & 152.0 (4) & 153.1 (5) & 149.5 (4)\\
$|g|_\rho$ (MeV) & 2404 (4)  & 2490 (5)  & 2515 (5) & 2534 (7) & 2504 (6)  \\
$\sqrt{s_\sigma}$ (MeV) & $(451 \pm 2, -i 234 \pm 1)$ & $(453 \pm 4, -i 238  \pm 4)$
& $(423 \pm 6, -i 267 \pm 3)$ & $(442 \pm 4, -i 248 \pm 2)$  & $(446 \pm 4, -i 246
\pm 4)$ \\
$|g|_\sigma$ (MeV) & 3005 (21)  & 3080 (70)  & 3070 (120) & 3100 (50)& 3070 (90) \\
$m_{scl}$ (MeV) & 1340 (40)  & $1600^{+0}_{-1200}$  & 772 (6) & $1020^{+70}_{-650}$ & 1030 (50)  \\
$\Gamma_{scl}$ (MeV) & $117^{+22}_{-0}$  & $300^{+300}_{-0}$ &
$580^{+120}_{-0}$ & $1070^{+240}_{-0}$ & $200^{+70}_{-0}$\\
$|g|_{scl}$ (MeV) & 2800 (300)    & $4700^{+0}_{-1700}$  & 2980 (100)& 2940 (120)
& $3300^{+500}_{-400}$ \\
\hline
\end{tabular}
\caption{Best fit parameters and pole properties (statistical
  uncertainties on these latter quantities define 68\%
  confidence-level regions, induced by the corresponding Gaussian
  correlated errors of the different fit parameters).  The three
  $C_{IJ}$ parameters are always fitted to data, and in addition
  $\mu_V=\mu_S$, $\mu_V$ and $ \mu_S$, $\mu_S$ and $m_S$, and
  $\mu_V=\mu_S$ and $m_V=m_S$ are also adjusted in the case of fits
  {\bf B.1}, {\bf B.2}, {\bf B.1-2} and {\bf B.3}, respectively.  In
  the row labeled as {\it contrb.}, the contributions to the $\chi^2$
  of the different $(I=J=0)$/$(I=J=1)$/$(I=2,J=0)$ sectors are
  displayed. Besides, $r_{ij}$ are Gaussian correlation coefficients
  between parameters $i$ and $j$. Note that the dispersive data
  analyses based in Roy~\cite{Caprini:2005zr} and
  GKPY~\cite{GarciaMartin:2011jx} equations predict for
  $\sqrt{s_\sigma}= (441^{+16}_{-8}, -i\, 272^{+9}_{-12})$ MeV and
  $(457^{+14}_{-13}, -i\, 279^{+11}_{-7})$ MeV respectively, while the
  Review of Particle Properties~\cite{Nakamura:2010zzi} quotes $m_\rho
  = 775.49 \pm 0.34$ MeV and $\Gamma_\rho = 149.1 \pm
  0.8$ MeV.\label{tab:1} }
\end{center}}
\end{table*}
The five fits have reasonable values of $\chi^2/dof$, though in the
$I=J=1$ channel {\bf B}-type fits lead to a better agreement with
data than fit {\bf A}. The major improvement occurs at the higher end
of the fitted region, and it is due to the the tail of a second
resonance located at around 1.4 GeV (see Fig.~\ref{fig:i1j1.SRS.nc3})
that it is generated in the {\bf B} schemes.  The relation $g^2_\rho =
m^4_\rho/(6f^2)$, deduced from the KSFR prediction $\Gamma_\rho =
m_\rho^3 / (96 \pi f_\pi^2)$, is satisfied within 3\% accuracy. This is
because the chiral logarithms are almost negligible in the $\rho$-meson
channel, and indeed at order ${\cal O}(p^4)$ they cancel out for the
SU(2) massless pion theory~\cite{Nieves:2009ez}.

Besides, it is also interesting to compare the leading--$N_C$ values
of the residues, as displayed by Eq.~(\ref{eq:res-largeNc}), with the
corresponding ones {\it after} unitarization, given in
Table~\ref{tab:1}.  Using the input values one gets $|g_V| \sim 2600$
MeV, in excellent agreement with the values for $|g_\rho|$ quoted in
the table that range in the interval $2400-2500$ MeV. For the scalar
resonance the rather stable values for $|g_\sigma| \sim 3000-3100$ MeV
can only be reproduced by Eq.~(\ref{eq:res-largeNc}) if $m_S=m_V$ (Fit
{\bf B.3}). Indeed, for the {\bf B.3} fitted $m_S=m_V$ mass, we find
$|g_S| \sim 3000 $ MeV.  For completeness, let us mention that the
residue of the $\sigma$ resonance has also been deduced from the
$\sigma \to \gamma \gamma$ decay~\cite{Bernabeu:2008wt}, yielding
$g_\sigma = 3204(28) + {\rm i}\, 1588(14)$ MeV, i.e. $|g_\sigma|=3580
(30)$ MeV. The result from Roy equations yields $|g_\sigma| = 3300
(300)$ MeV~\cite{Leutwyler:2008xd}.  Using the UFD and CFD
parameterizations from~\cite{Martin:2011cn} yields $g_S = 3735 (61) +
{\rm i}\, 874 (3)$ MeV and $g_\sigma = 3742 (60) + {\rm i}\, 874 (6)$
MeV or, equivalently, $|g_\sigma| = 3836 (85)$ MeV and $|g_\sigma| =
3843 (84)$ MeV respectively, while in the most recent analysis based
in GKPY equations, a value of $|g_\sigma| = 3590 (120)$ MeV is
quoted~\cite{GarciaMartin:2011jx}.

Moreover, the scalar-isoscalar phase shifts are significantly better
described when the complete resonance saturation of the $L_i^r$ occurs
at two different scales fitted to data (fit {\bf B.2}). In this latter
case, mass and width of the $f_0(600)$ or $\sigma$ resonance compare
also well with the results of Caprini et
al.~\cite{Caprini:2005zr}.\footnote{Bear in mind that in the $I=J=0$
sector we do not fit directly the Roy equation results of
Ref.~\cite{Ananthanarayan:2000ht}, used in the work of Caprini et al.,
but rather we take an average of these results with those obtained by
Yndurain and collaborators in Ref.~\cite{Yndurain:2007qm}. The recent
re-analysis~\cite{Martin:2011cn} provides errors fully compatible with
this assumption.}  The properties of the $\sigma$ resonance are
strongly influenced by chiral logarithms~\cite{Nieves:2009ez}.  In
this scalar-isoscalar channel, we will keep track of a second
resonance, which cannot be identified with the $f_0(600)$ and that we
will label as $scl$. There exist two types of scenarios: i) for fits
{{\bf A}, {\bf B.1} and {\bf B.3}, this second pole appears well above
$m_S$ (1 GeV for the first two fits and 0.738 GeV for the last one)
and it is relatively narrow, and ii) for fits {\bf B.2} and {\bf
B.1-2}, it is placed below $m_S$ (1 GeV and 1.295 GeV, respectively)
and it is quite wide ($\Gamma \ge $ 600 MeV). In the case of the fit
{\bf B.2}, the effects of this resonance ($scl$) on the phase shifts,
in the higher end (600-750 MeV) of the fitting range, are appreciable
and considerably improve the achieved description (see
Figs.~\ref{fig:nc3} and~\ref{fig:i0j0.SRS.nc3}). A different question
is whether or not such a wide state, below 1 GeV, does have a
correspondence with any physical state or it is just an artifact of
the fitting procedure. We should note that a state with these features
has not been reported neither in the Roy equation analysis of
Ref.~\cite{Leutwyler:2008xd}, nor in the most recent work based in the
GKPY equations of Ref.~\cite{GarciaMartin:2011jx}. As we shall see in
the next subsection, the $N_C \gg 1$ behaviour of the $\sigma-$pole
obtained from the {\bf B.2} fit is radically different to that
inferred from the {\bf A, B.1} and {\bf B.3} schemes. That is the
reason why we have proposed the fit {\bf B.1-2}, with the intention of
testing to what extent the dependence on $N_C$ of the
$\sigma-$resonance properties is determined by the existence of this
possible artifact\footnote{The minimization procedure, involving also
the election of the parameters which are fitted to data, is not
completely unique, and there exist obviously different local
minima. Some of them might not have a proper physical
interpretation. Bearing this in mind, we can not discard the existence of
artifacts.}.  In the fit {\bf B.1-2}, where the scale
$\mu_V^{RS}$ is fixed to the result of fit {\bf B.1} and the value of
$m_S$ is adjusted to the data, the second scalar resonance shows up
above 1 GeV, and as we will discuss below, it leads to qualitatively
the same $N_C$ dependence of the $f_0(600)$ mass and width as the fit
{\bf B.2} does.
\begin{figure*}[tbh]
\begin{center}
\makebox[0pt]{\hspace{-2cm}\includegraphics[height=5.5cm]{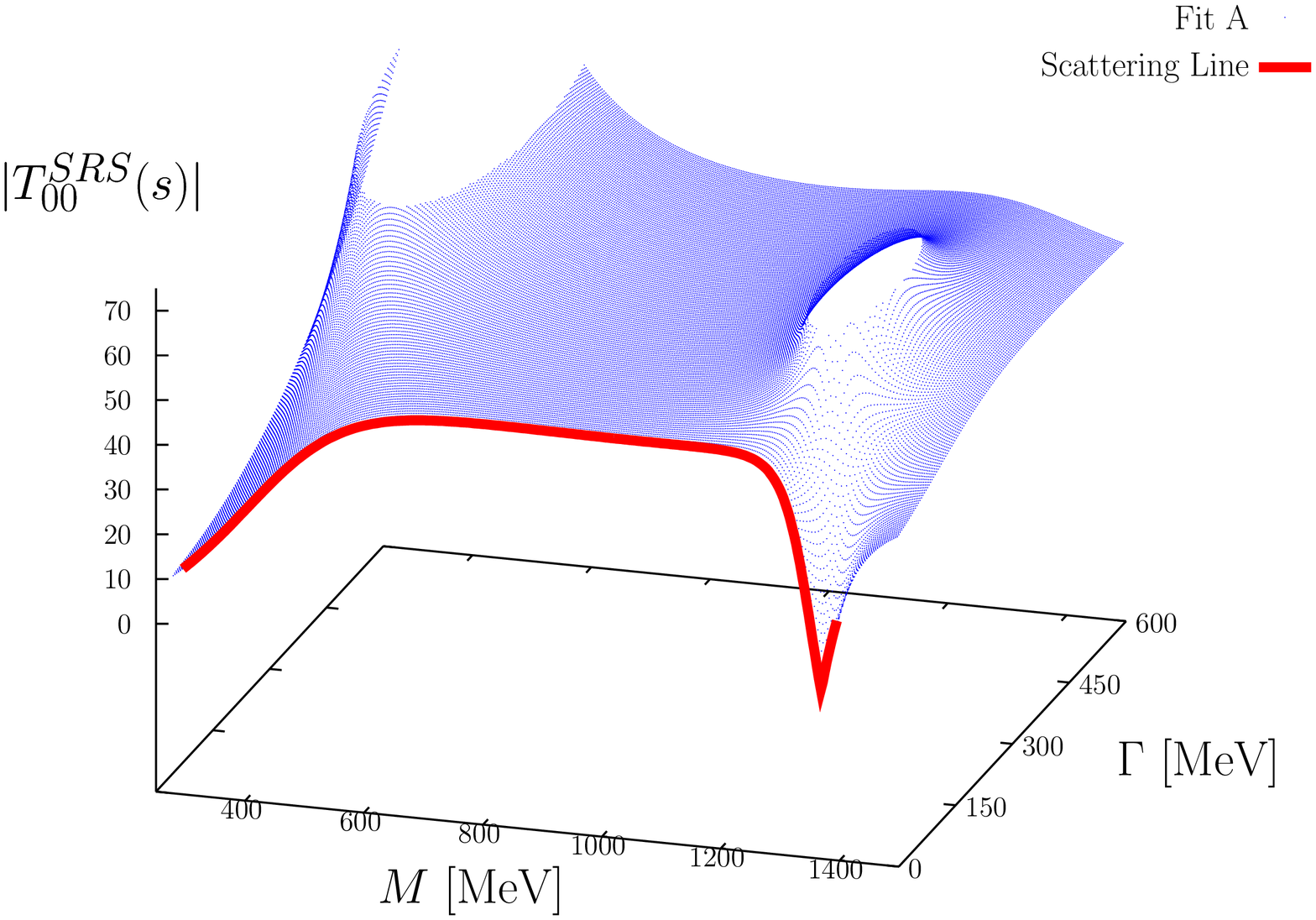}\hspace{1cm}\includegraphics[height=5.5cm]{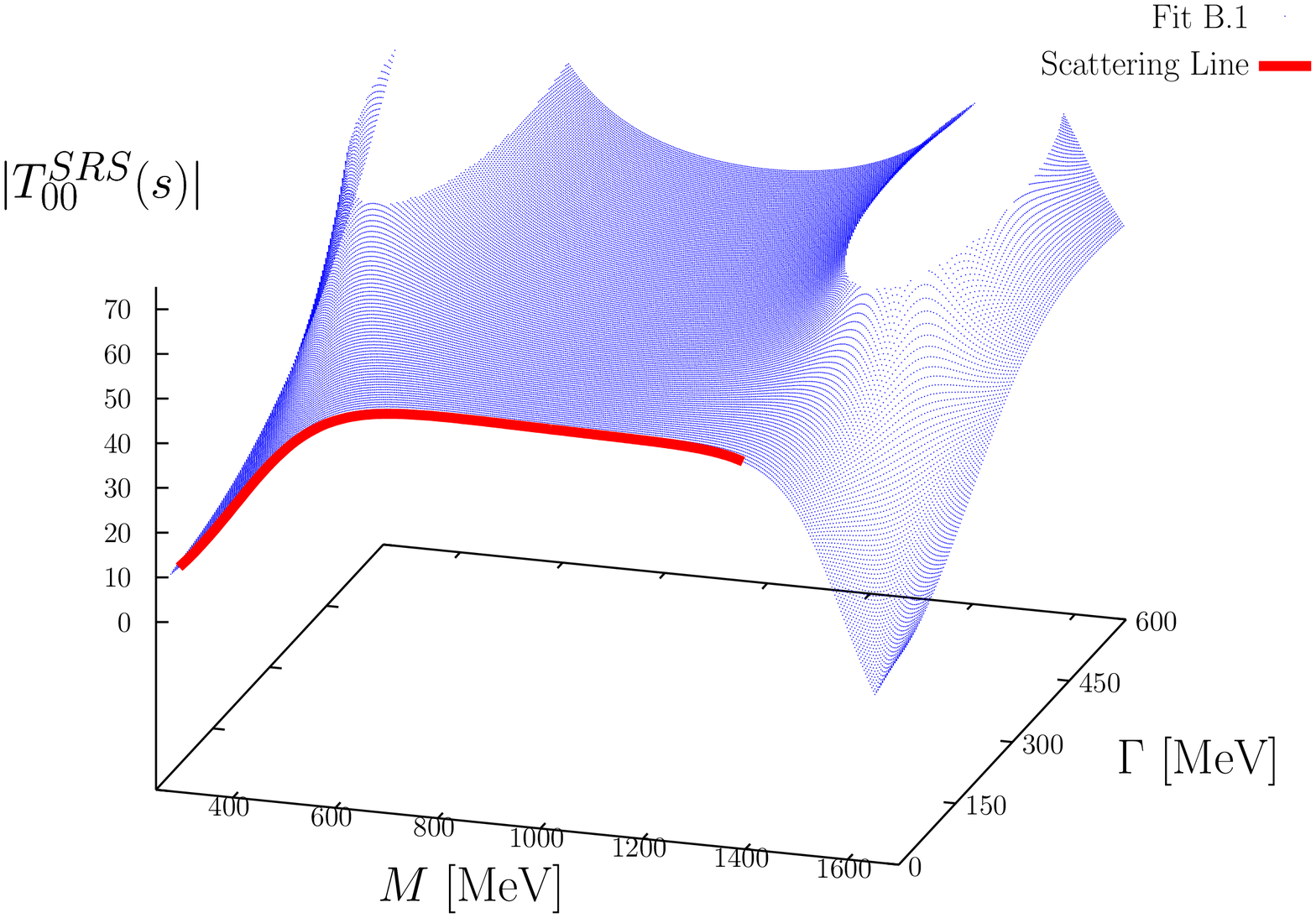}}\\\vspace{1cm}
\makebox[0pt]{\hspace{-2cm}\includegraphics[height=5.5cm]{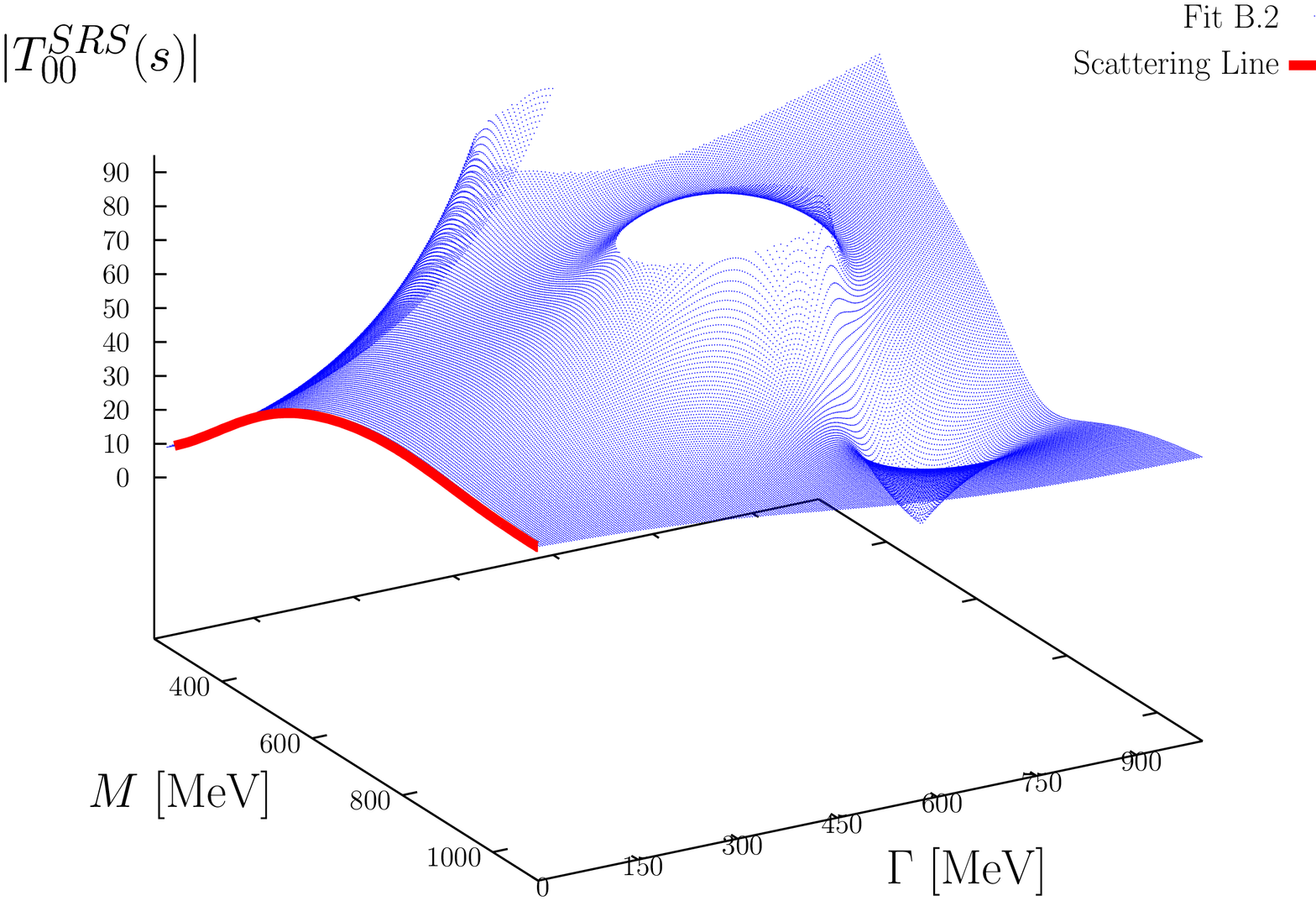}\hspace{1cm}\includegraphics[height=5.5cm]{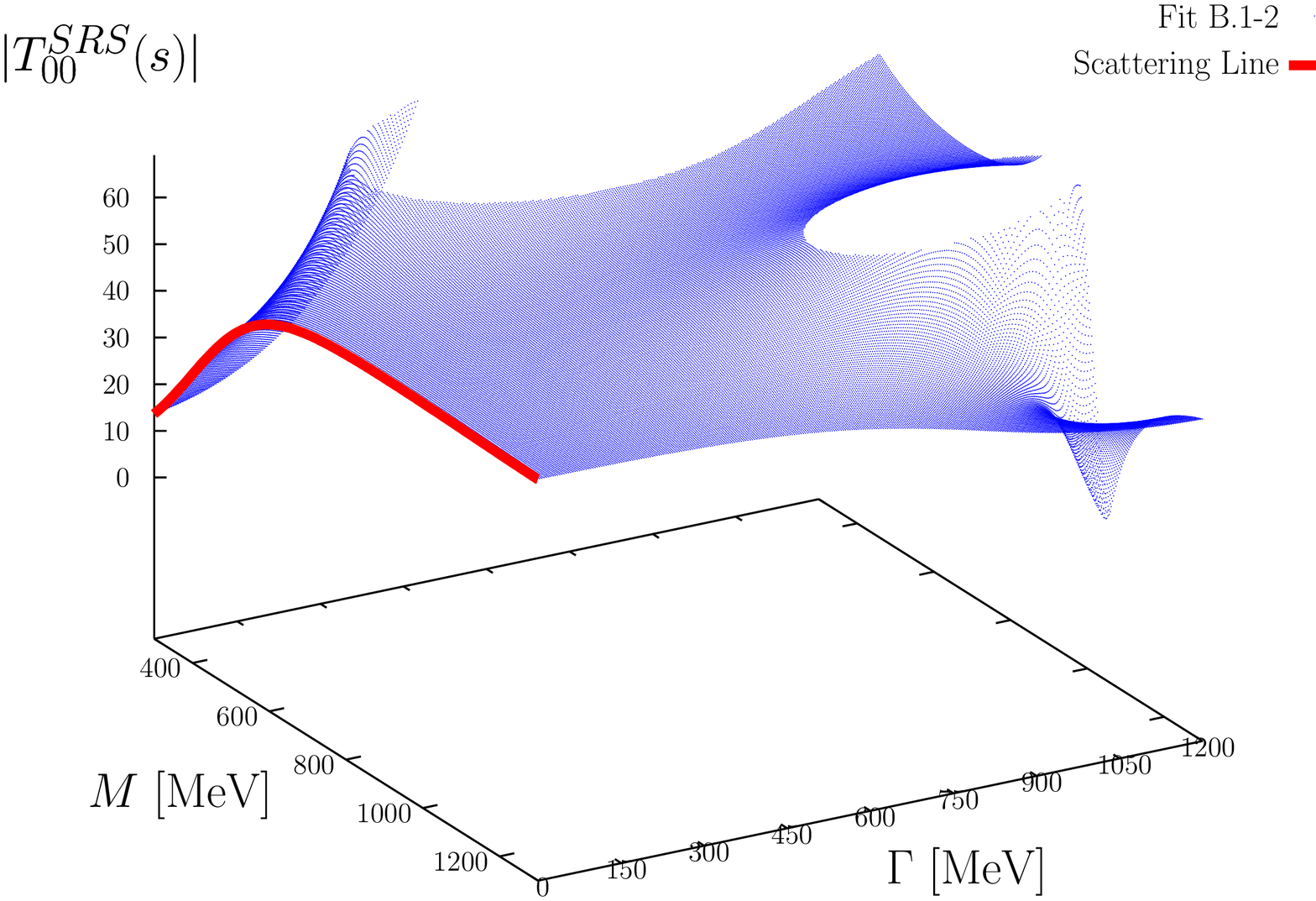}}\\\vspace{1cm}
\makebox[0pt]{\includegraphics[height=5.5cm]{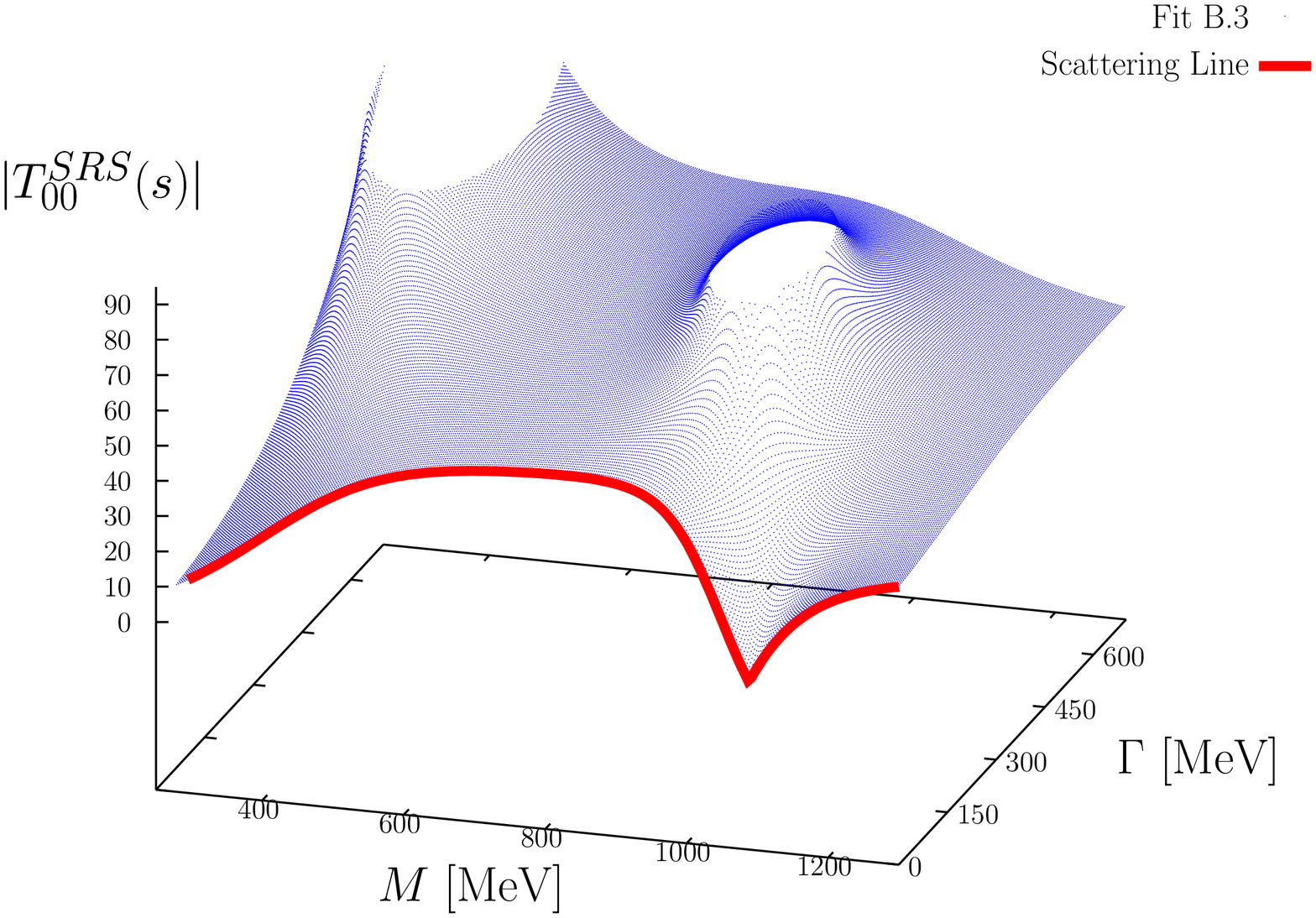}}
\end{center}
\caption{ Modulus of the SRS $T_{00}(s)$ amplitude, as a function of
complex $s$ ($s=M^2-{\rm i} M \Gamma$) taken in the fourth quadrant,
for fits {\bf A}, {\bf B.1}, {\bf B.2},{\bf B.1-2} and {\bf B.3}. 
In all plots the physical
scattering line ($\Gamma=0$) is also depicted.}
\label{fig:i0j0.SRS.nc3}
\end{figure*}
\begin{figure*}[tbh]
\begin{center}
\makebox[0pt]{\hspace{-2cm}\includegraphics[height=5.5cm]{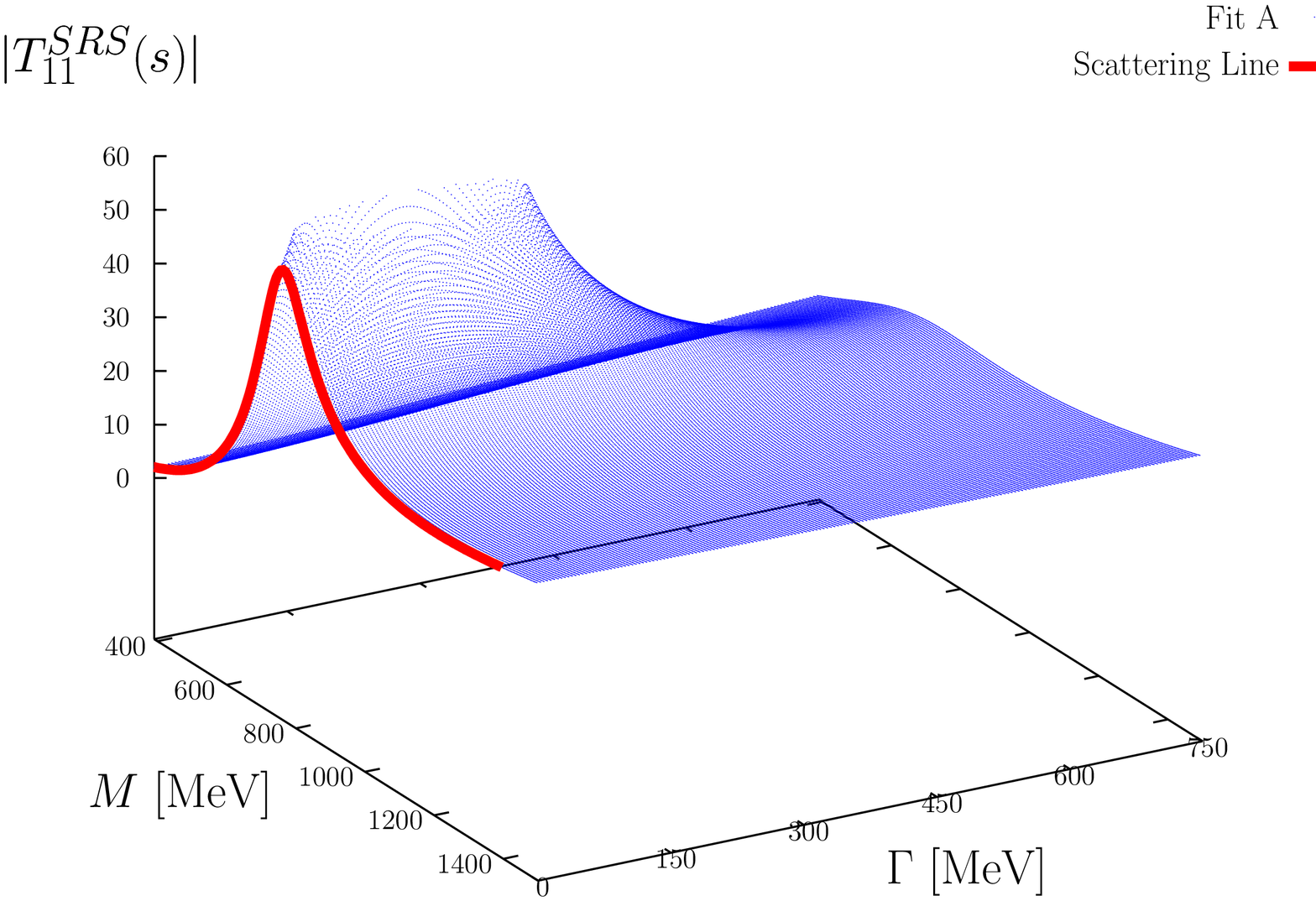}\hspace{1cm}\includegraphics[height=5.5cm]{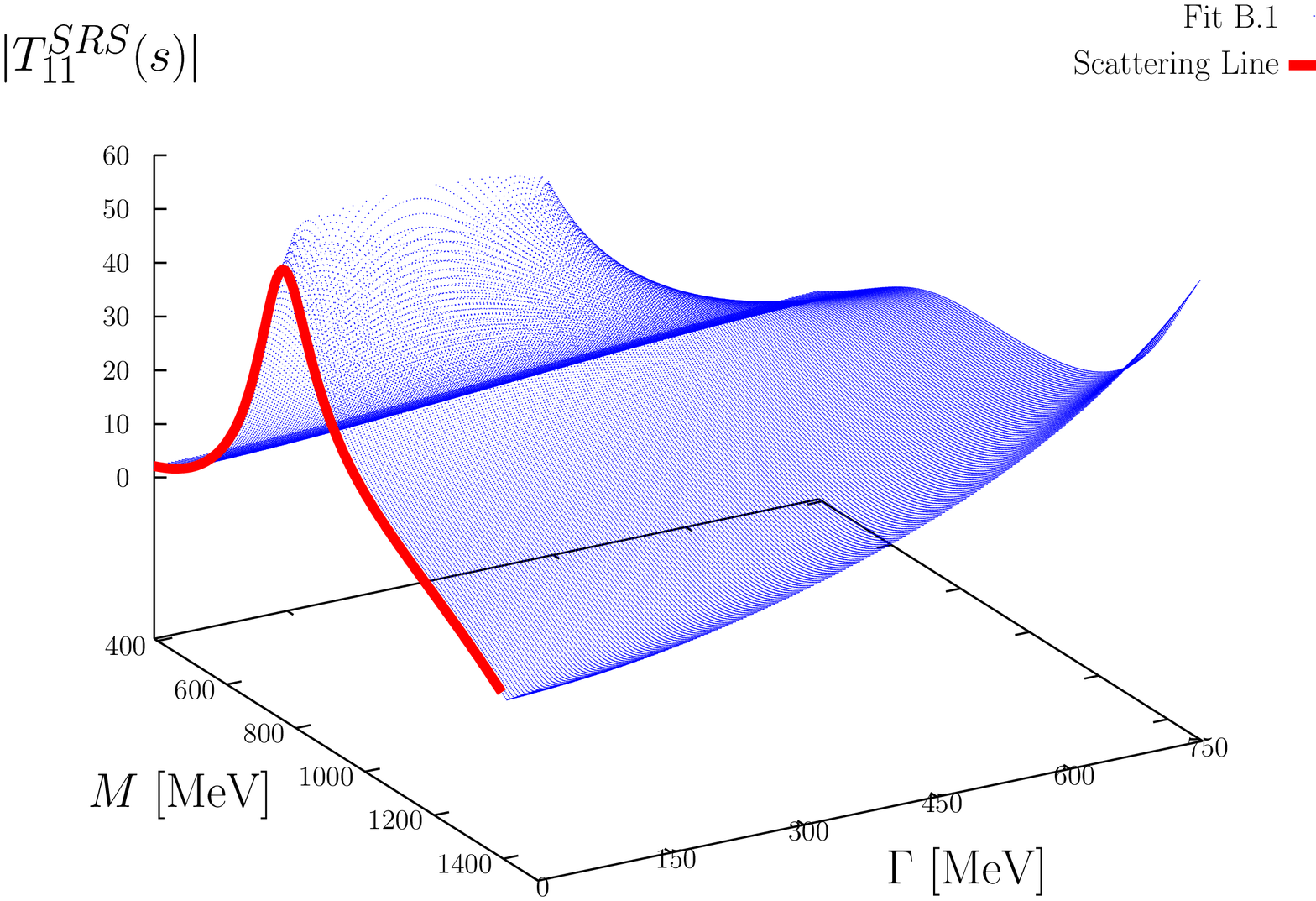}}\\\vspace{1cm}
\makebox[0pt]{\hspace{-2cm}\includegraphics[height=5.5cm]{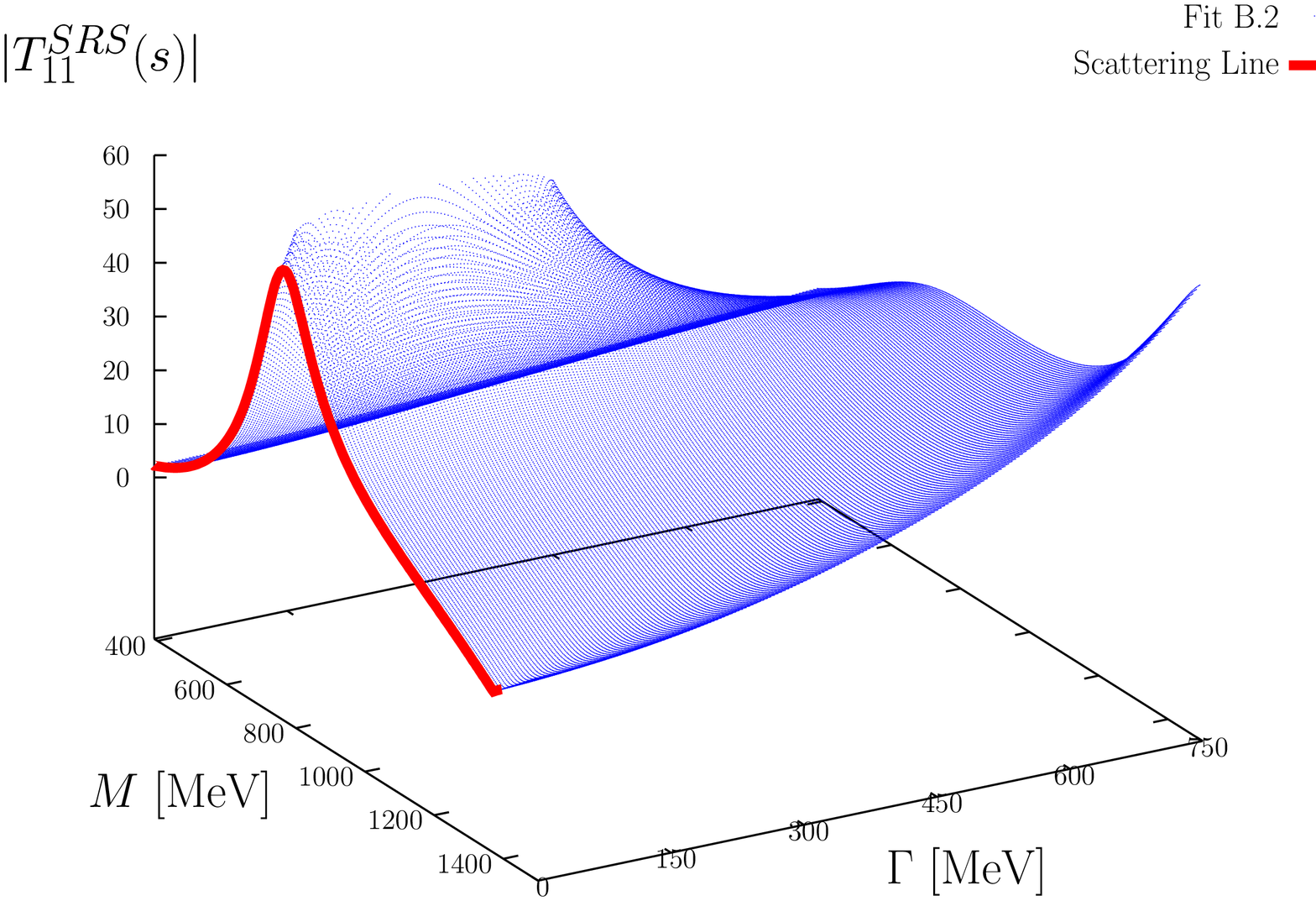}\hspace{1cm}\includegraphics[height=5.5cm]{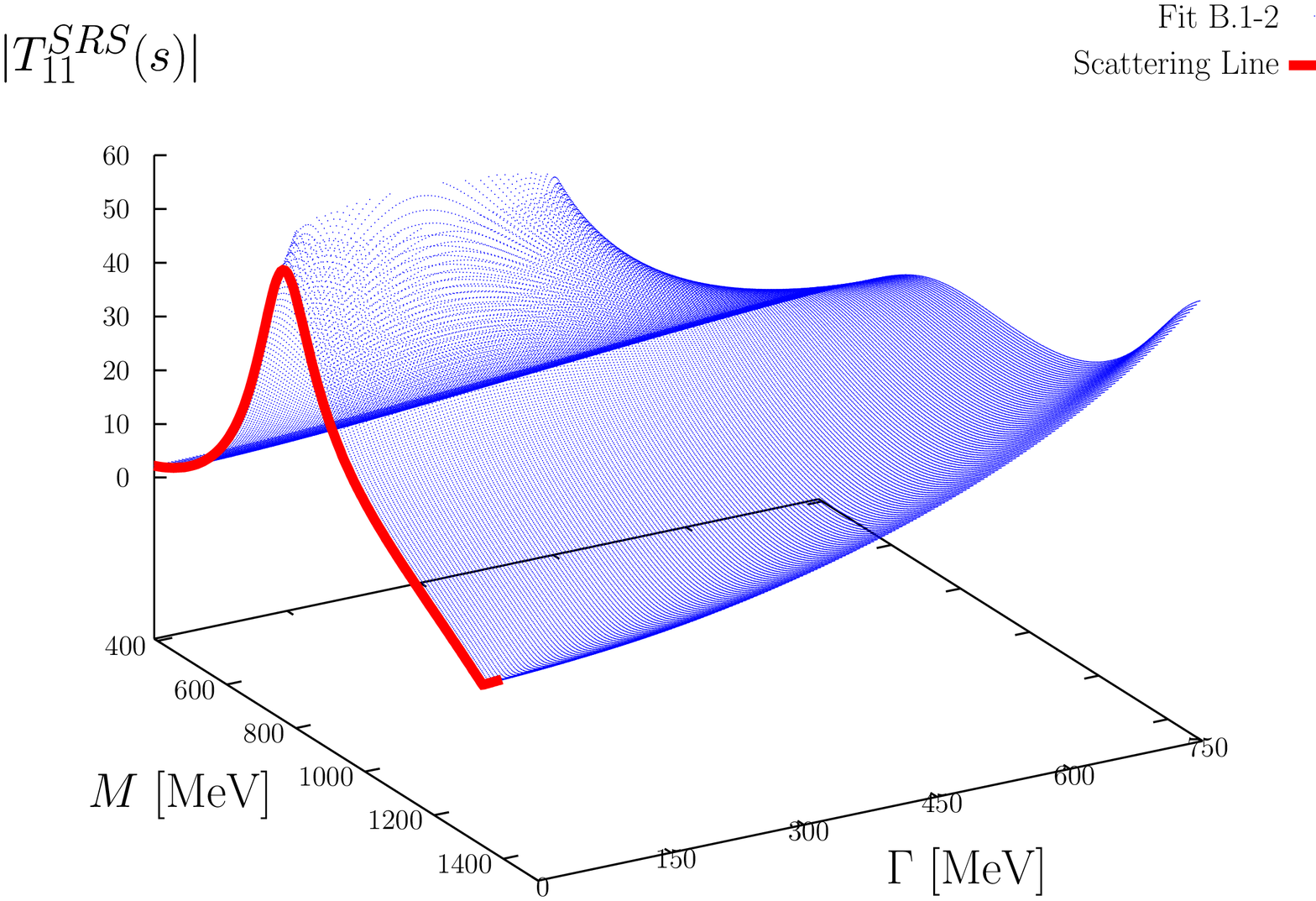}}\\\vspace{1cm}
\makebox[0pt]{\includegraphics[height=5.5cm]{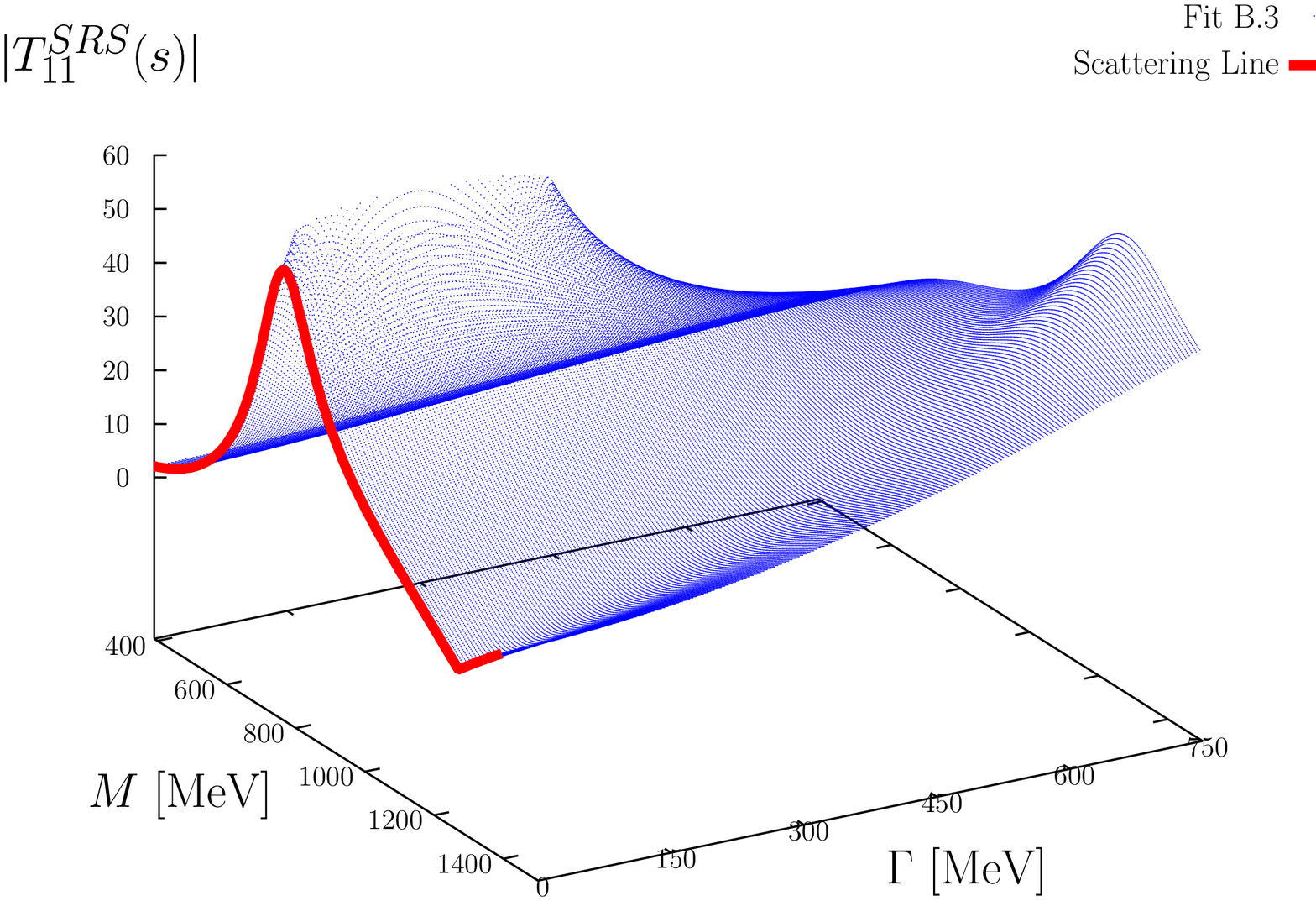}}
\end{center}
\caption{ Modulus of the SRS $T_{11}(s)$ amplitude, as a function of
complex $s$ ($s=M^2-{\rm i} M \Gamma$) taken in the fourth quadrant,
for fits {\bf A}, {\bf B.1}, {\bf B.2}, {\bf B.1-2} and {\bf B.3}. In
all plots the physical scattering line ($\Gamma=0$) is also
depicted. }
\label{fig:i1j1.SRS.nc3}
\end{figure*}

\section{Results for $\mathbf{N_C> 3}$}
\label{sec:results}

We extrapolate the amplitudes to $N_C> 3$, by means of
the $N_C$ dependence
\begin{eqnarray}
&&A^{{\rm SRA}+ {\rm ChPT}} (s,t,u)\Big|_{N_C \ge 3} =\frac{3}{N_C} A^{\rm SRA}
   (s,t,u)\Big|_{N_C=3}  +
   \left(\frac{3}{N_C}\right)^2\left (A^{\rm ChPT}(s,t,u)- A_4^{\rm
  SRA}(s,t,u)\right) \Big|_{N_C=3}\, ,
\end{eqnarray}
and the scaling law of Eq.~(\ref{eq:c's}).
$N_C> 3$ results are depicted in
Figs.~\ref{fig:i1j1.nc}, \ref{fig:i0j0-sigma.nc} and  \ref{fig:i0j0-scl.nc}.
\begin{figure*}[tbh]
\begin{center}
\makebox[0pt]{\hspace{-2cm}\includegraphics[height=5.5cm]{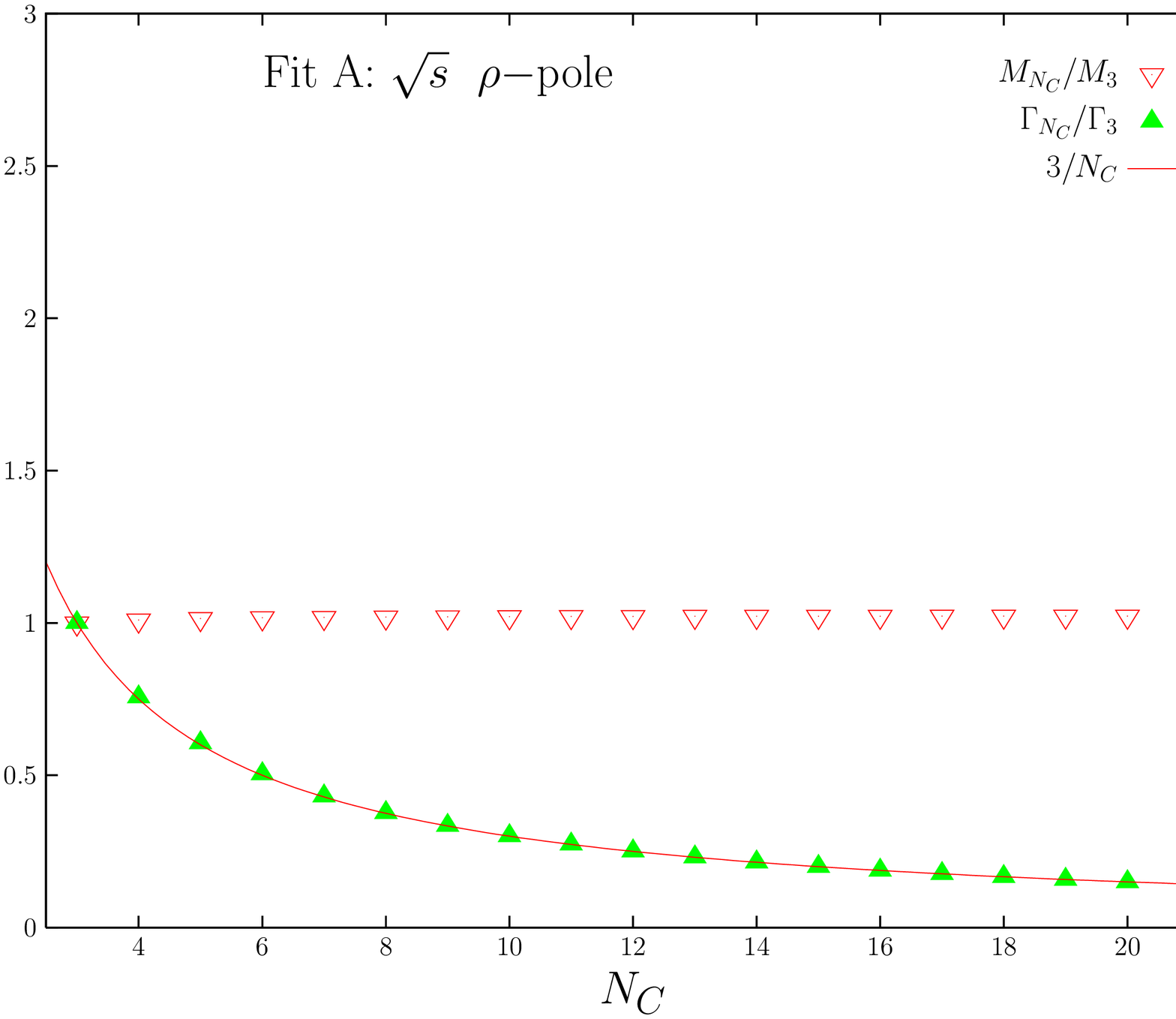}
\hspace{1cm}\includegraphics[height=5.5cm]{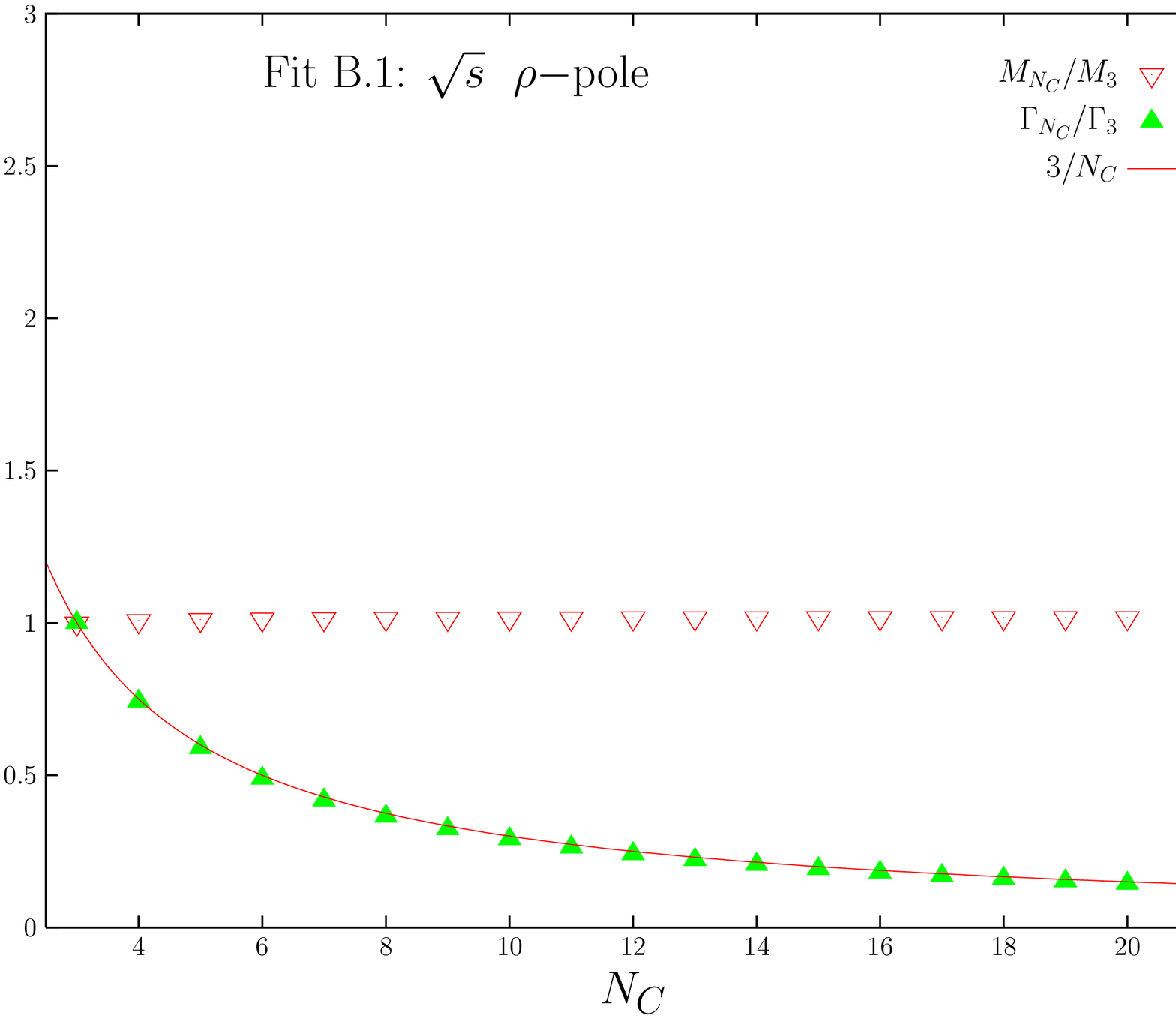}}\\\vspace{1cm}
\makebox[0pt]{\hspace{-2cm}\includegraphics[height=5.5cm]{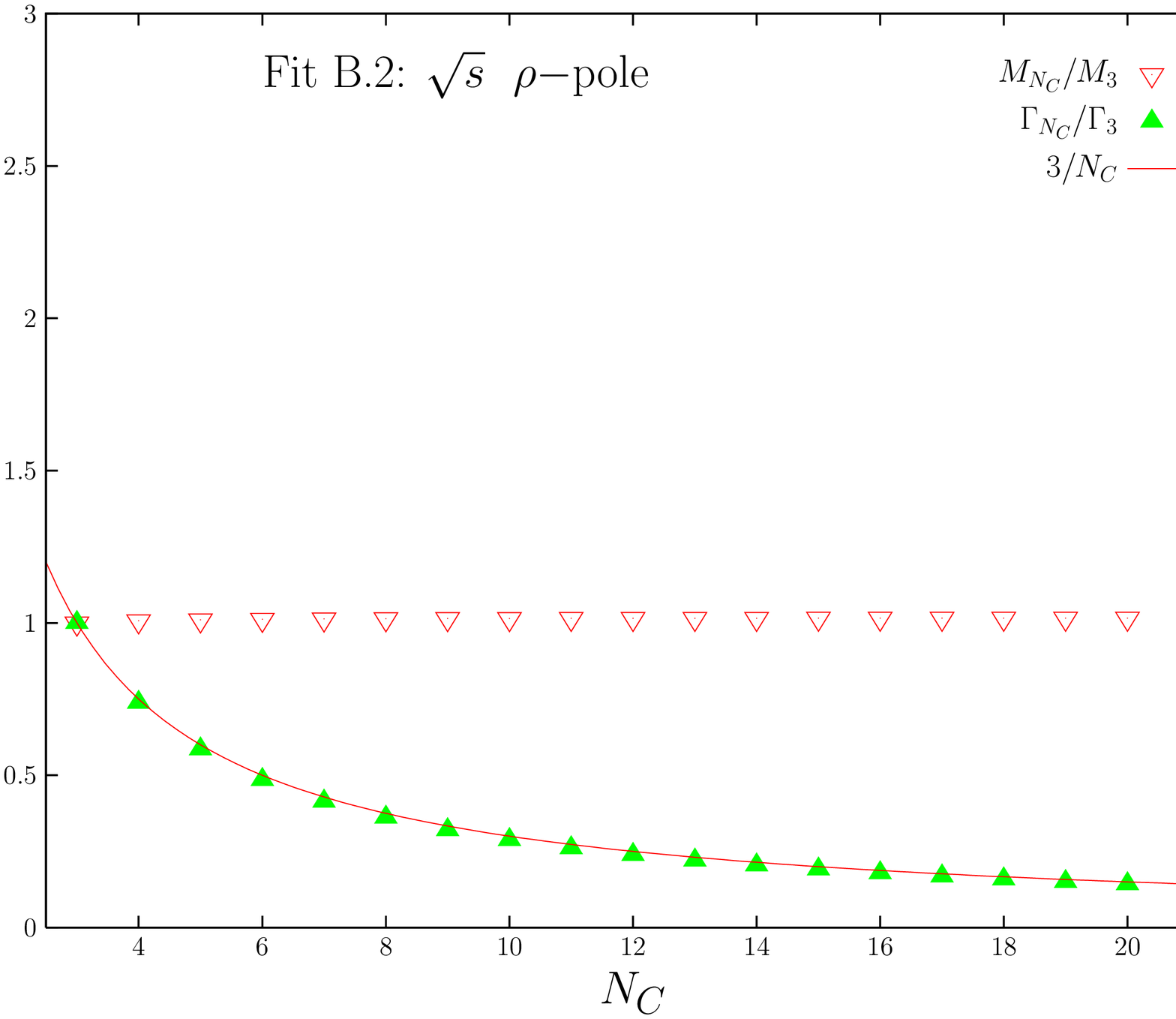}
\hspace{1cm}\includegraphics[height=5.5cm]{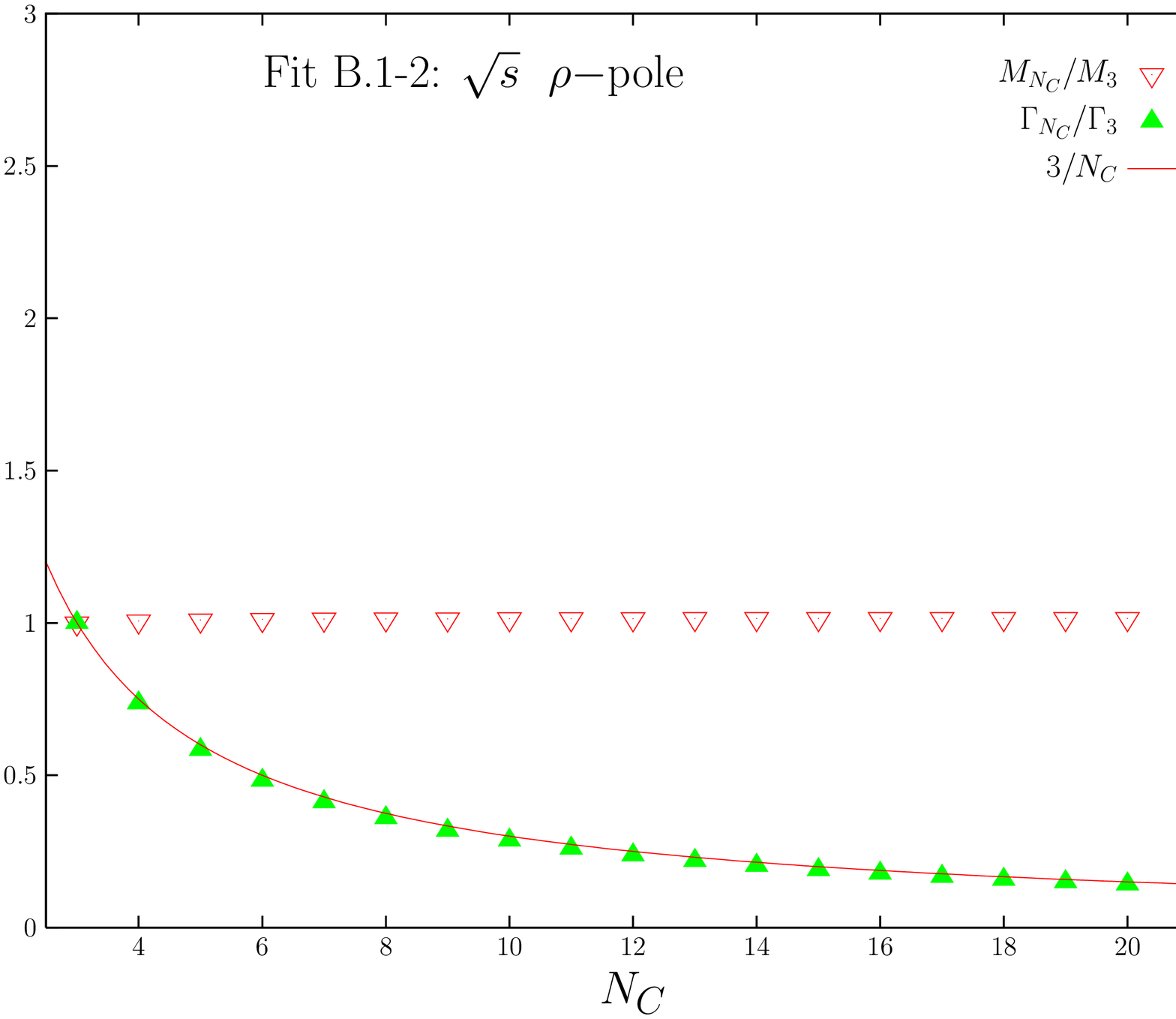}}\\\vspace{1cm}
\makebox[0pt]{\hspace{-2cm}\includegraphics[height=5.5cm]{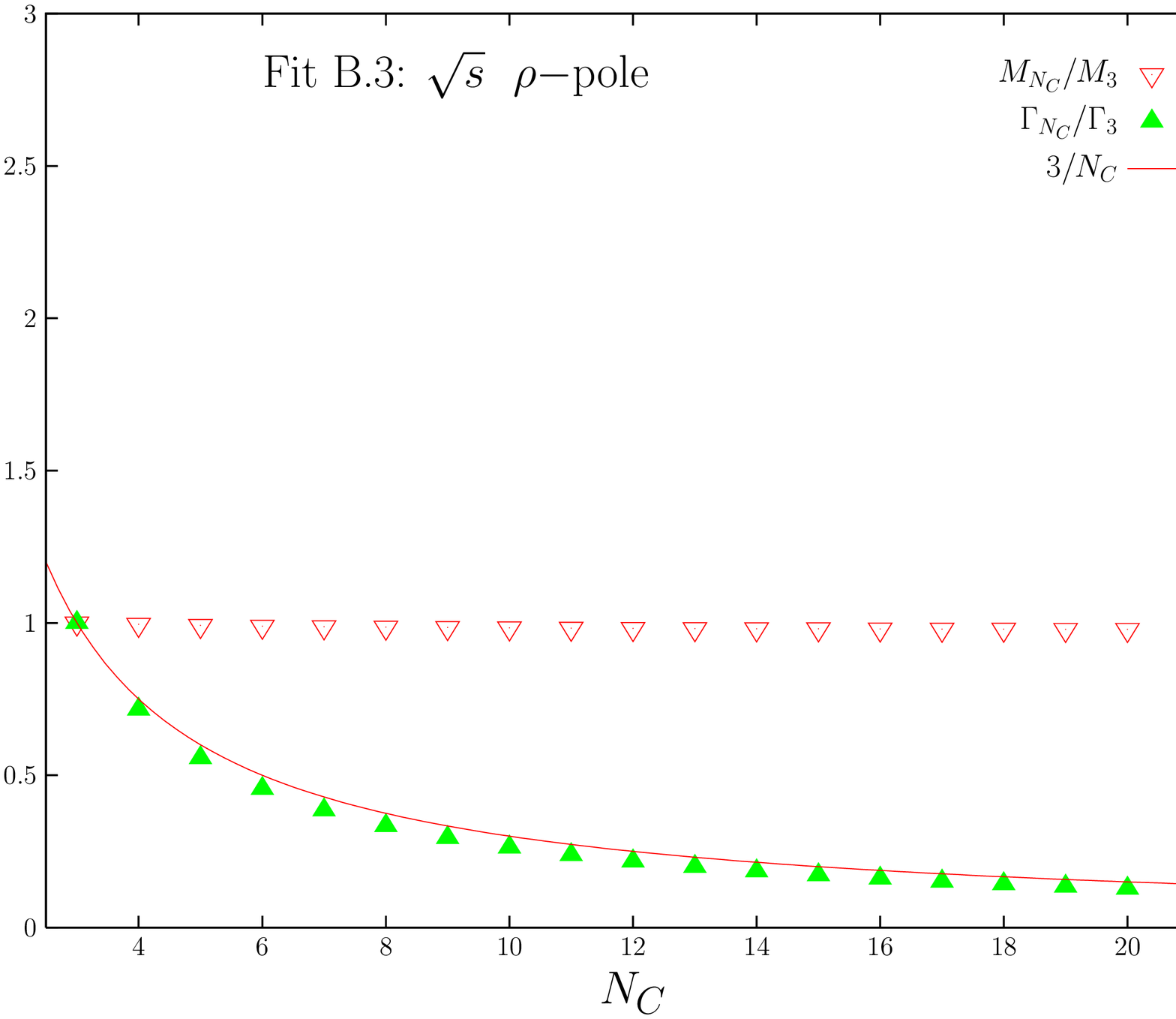}}
\end{center}
\caption{$N_C$ dependence of the $\rho$ pole position for the various
  fits described in the text. Empty [filled] triangles stand for ${\rm
  Re}\sqrt{s_\rho}$~ [$-2{\rm Im}\sqrt{s_\rho}$\,] in units of the
  $N_C=3$ values ($M_3$ and $\Gamma_3$), with $s_\rho$  the SRS pole
  position (located in the fourth quadrant) for the different $N_C$ values.}
\label{fig:i1j1.nc}
\end{figure*}
\begin{figure*}[tbh]
\begin{center}
\makebox[0pt]{\hspace{-2cm}\includegraphics[height=5.2cm]{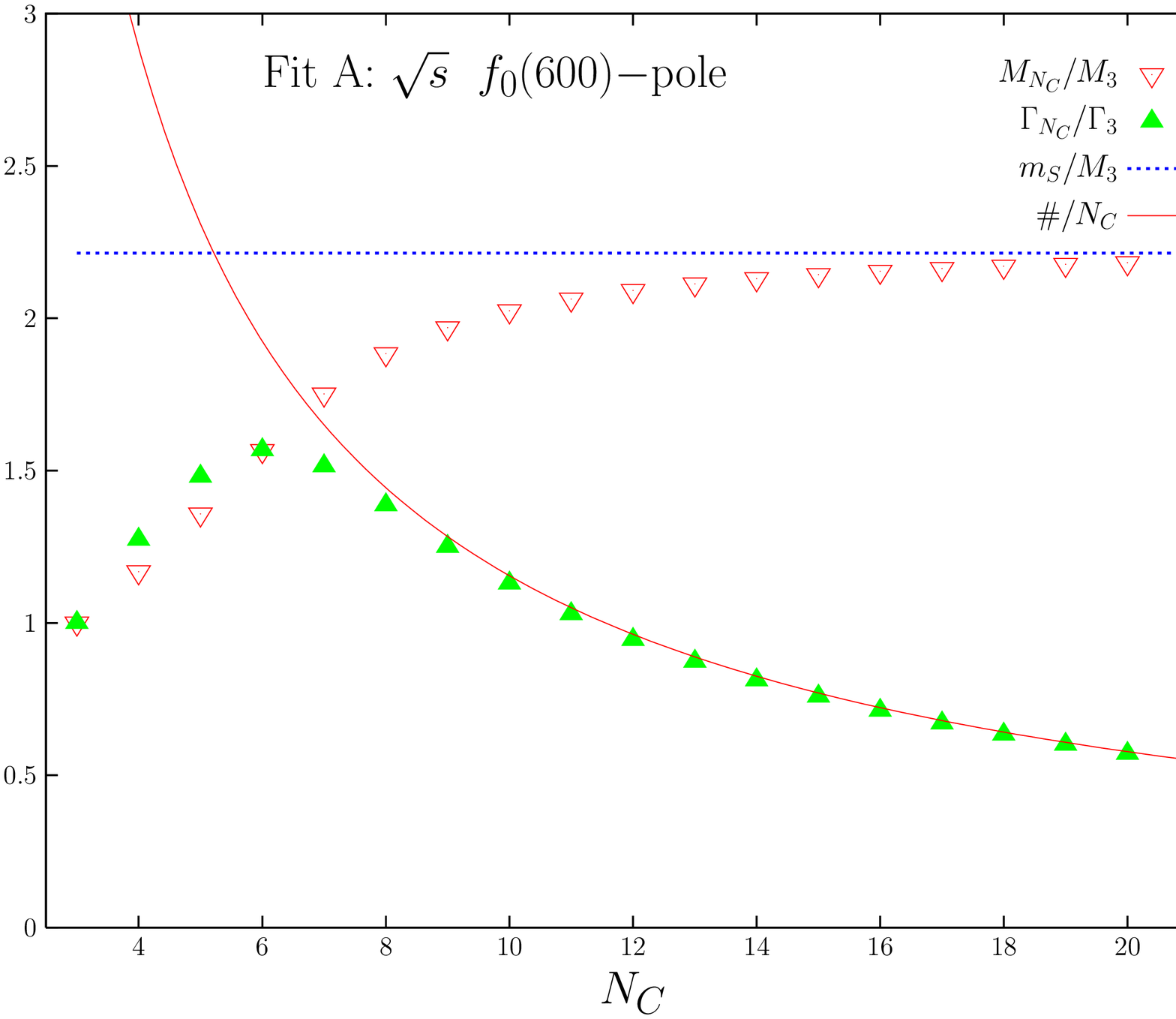}
\hspace{1cm}\includegraphics[height=5.2cm]{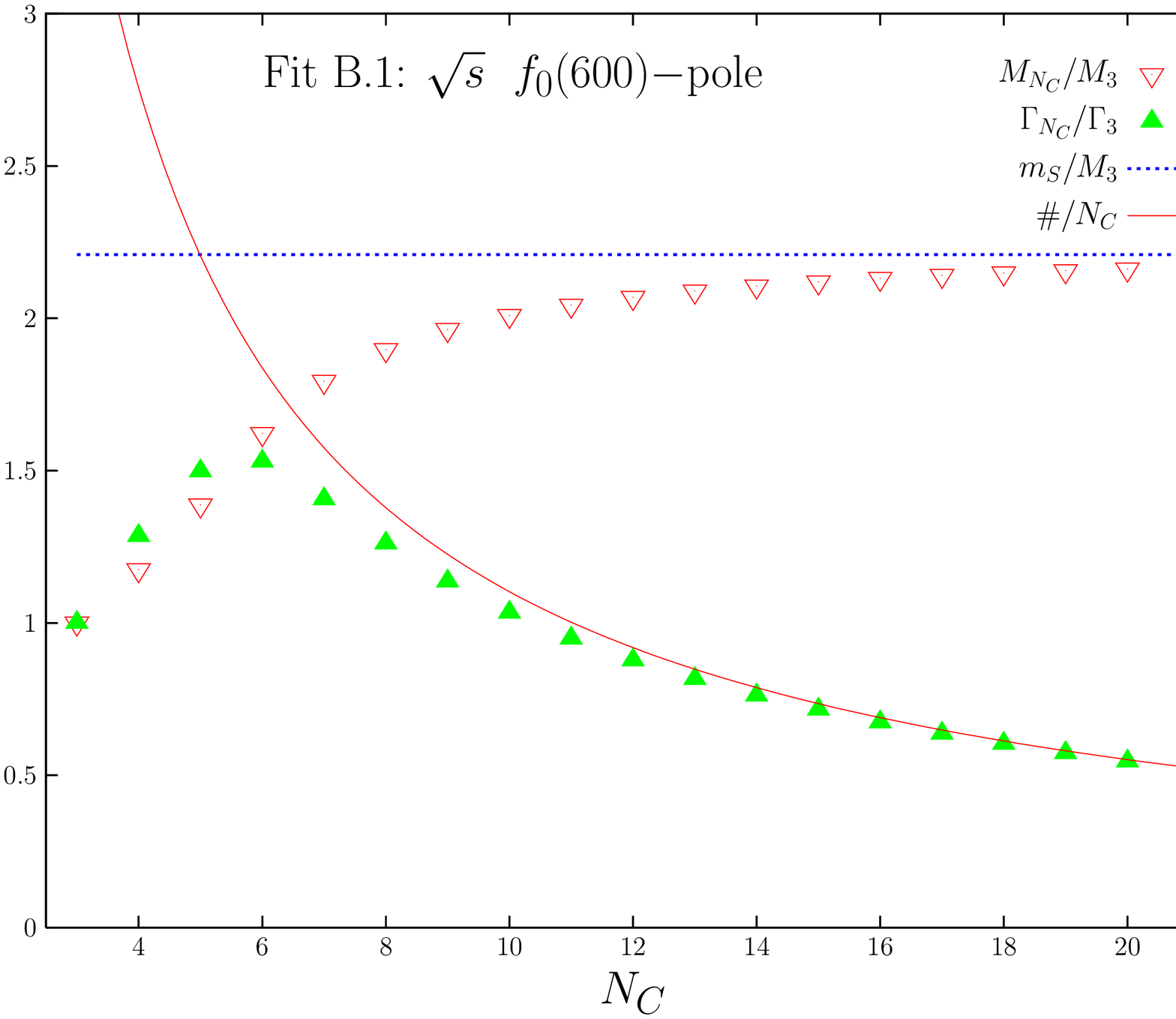}}\\ \vspace{1cm}
\makebox[0pt]{\hspace{-2cm}\includegraphics[height=5.2cm]{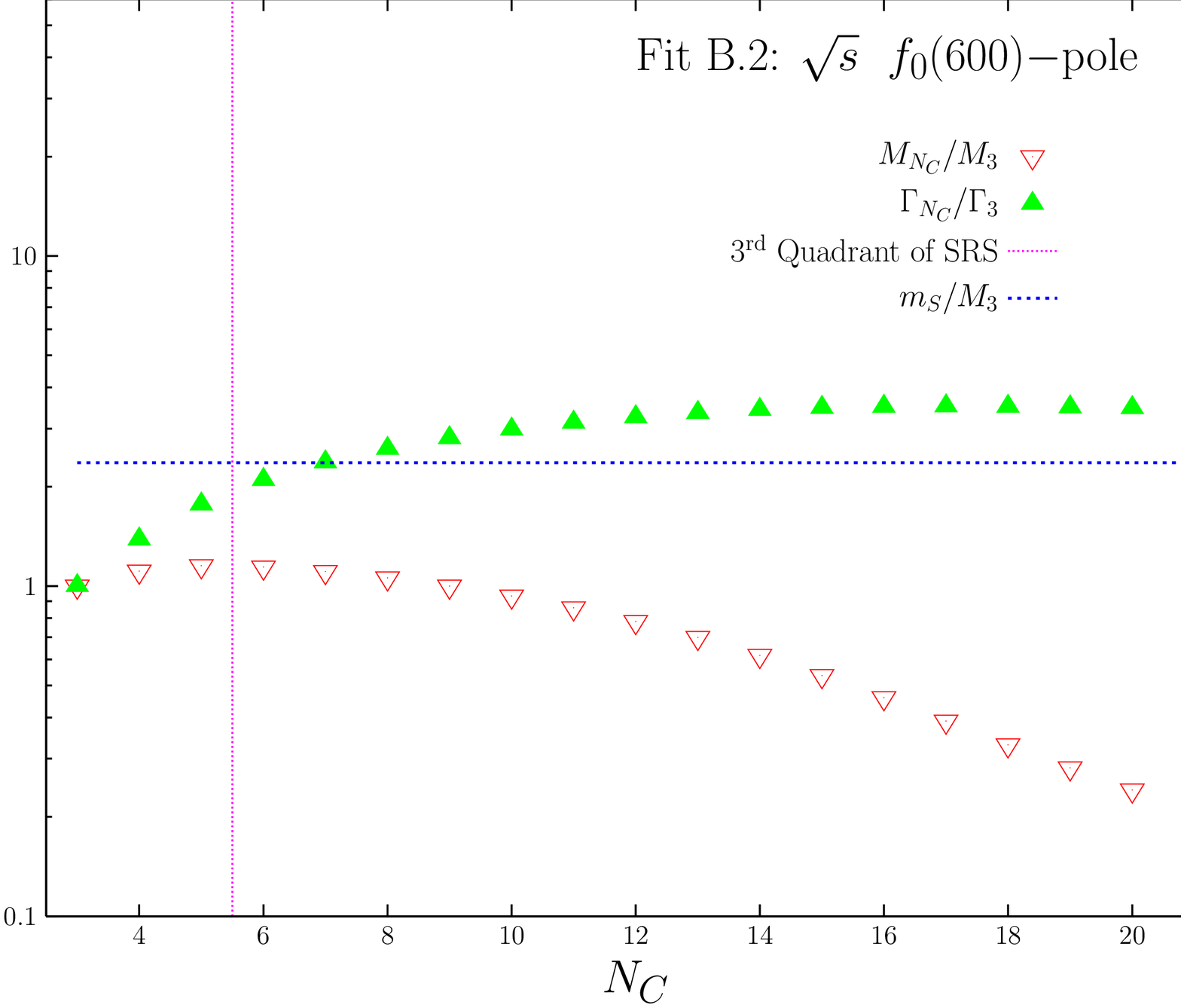}
\hspace{1cm}\includegraphics[height=5.2cm]{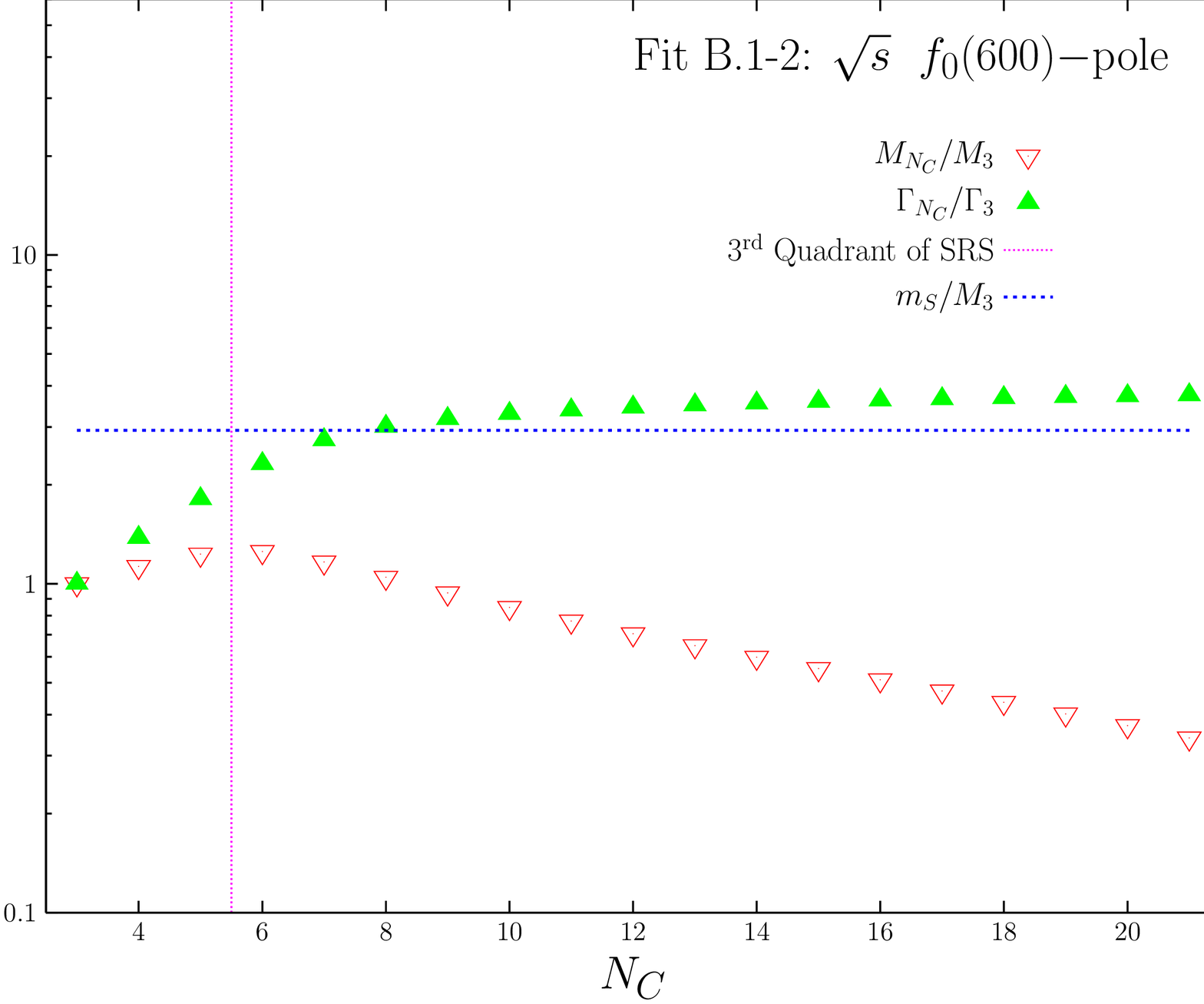}}\\ \vspace{1cm}\makebox[0pt]{\hspace{-2cm}\includegraphics[height=5.2cm]{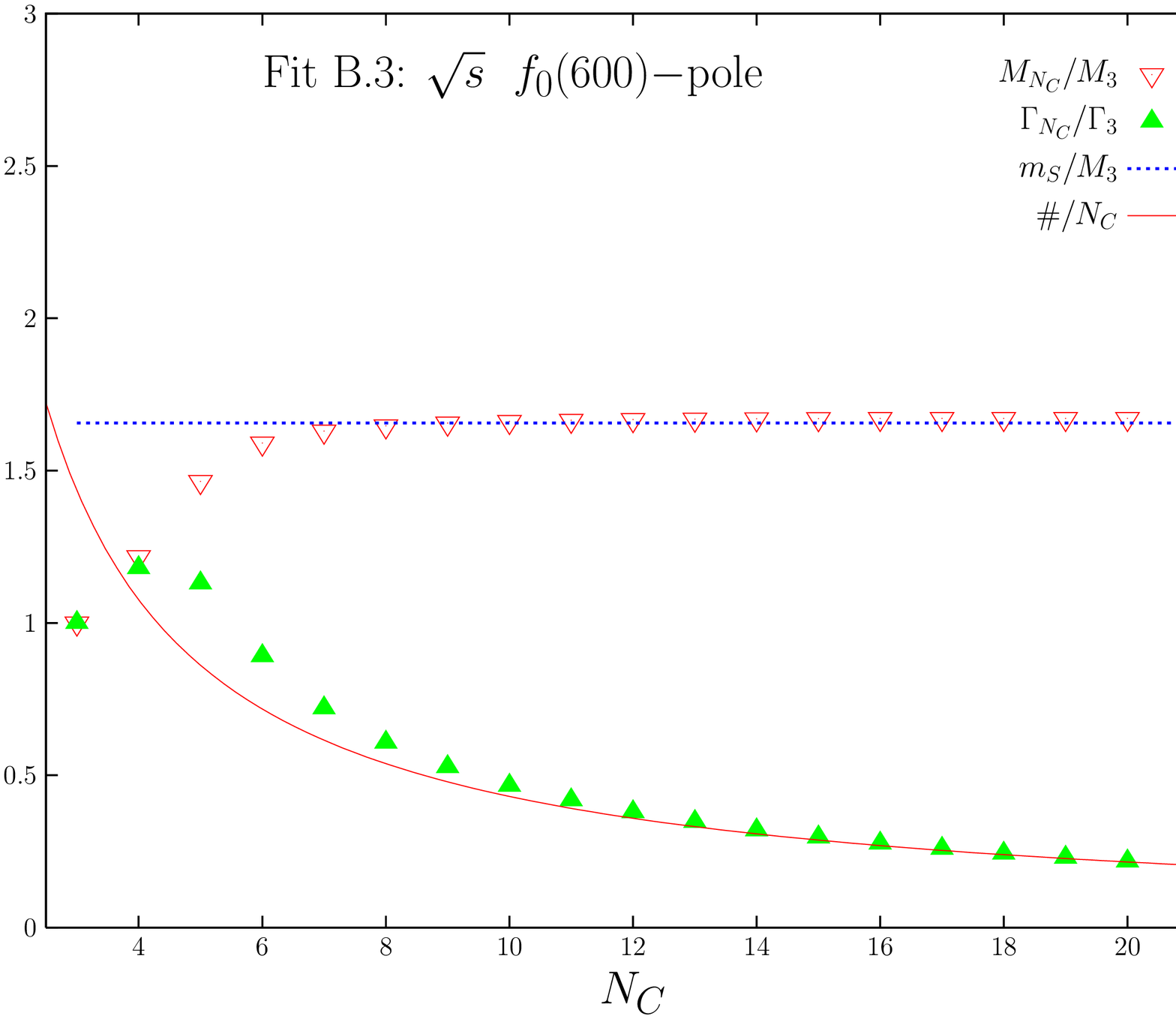}}
\end{center}
\caption{ Same as Fig.~\ref{fig:i1j1.nc}, but for the $\sigma$
  pole. In addition, the horizontal line indicates the mass of the
  scalar resonance included in the SRA amplitude (1 GeV for fits {\bf
  A}, {\bf B.1} and {\bf B.2}, 0.738 GeV for the fit {\bf B.3} and
  1.295 GeV in the case of the fit {\bf B.1-2}). For values of $N_C$
  located at the right of the vertical line in the  {\bf B.2} and
  {\bf B.1-2} panels, the pole $s_\sigma$ appears in the third
  quadrant, instead of in the fourth one, and thus, the singularity
  does not have a clear physical interpretation.}
\label{fig:i0j0-sigma.nc}
\end{figure*}
\begin{figure*}[tbh]
\begin{center}
\makebox[0pt]{\hspace{-2cm}\includegraphics[height=5.5cm]{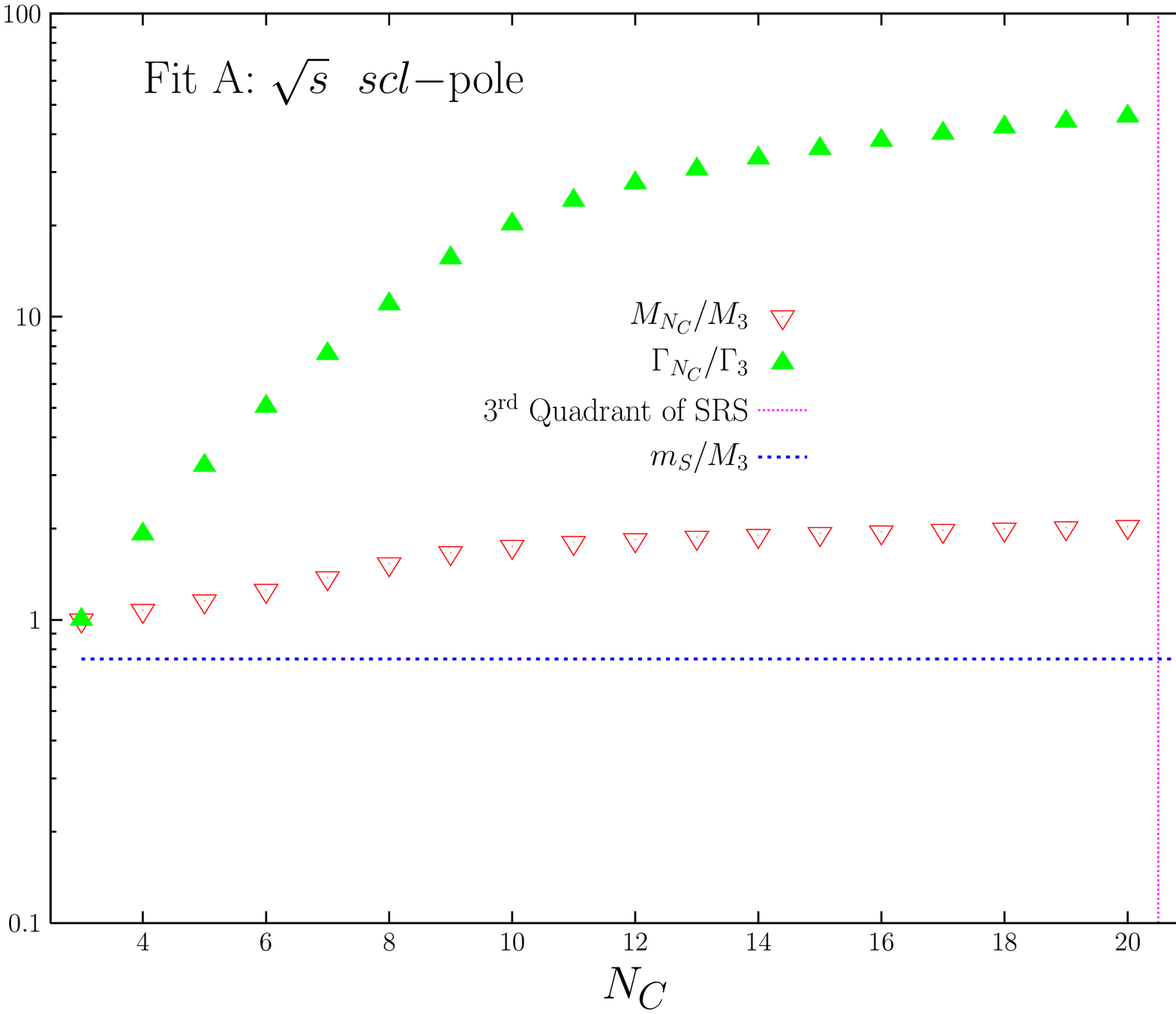}
\hspace{1cm}\includegraphics[height=5.5cm]{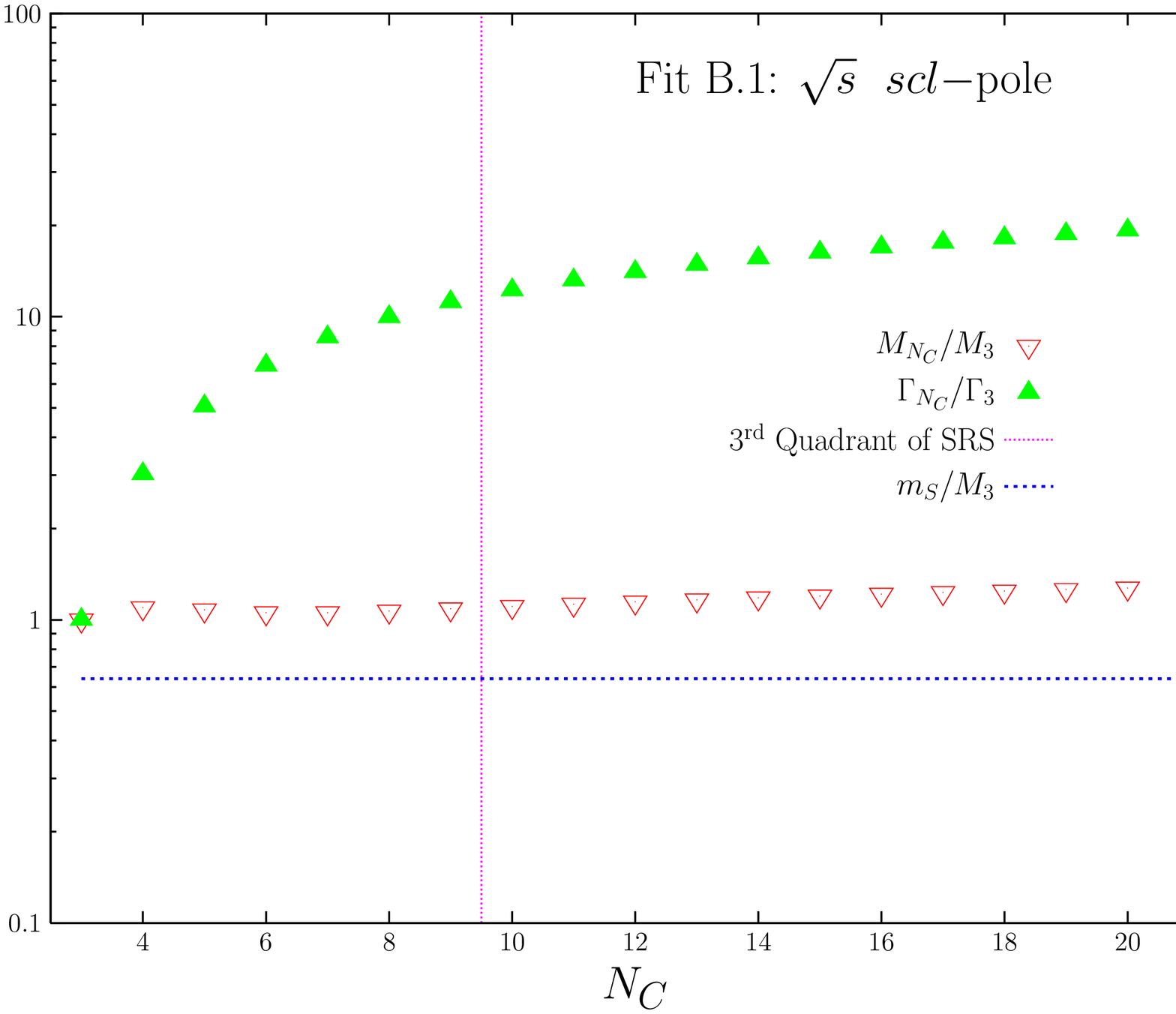}}\\\vspace{1cm}
\makebox[0pt]{\hspace{-2cm}\includegraphics[height=5.5cm]{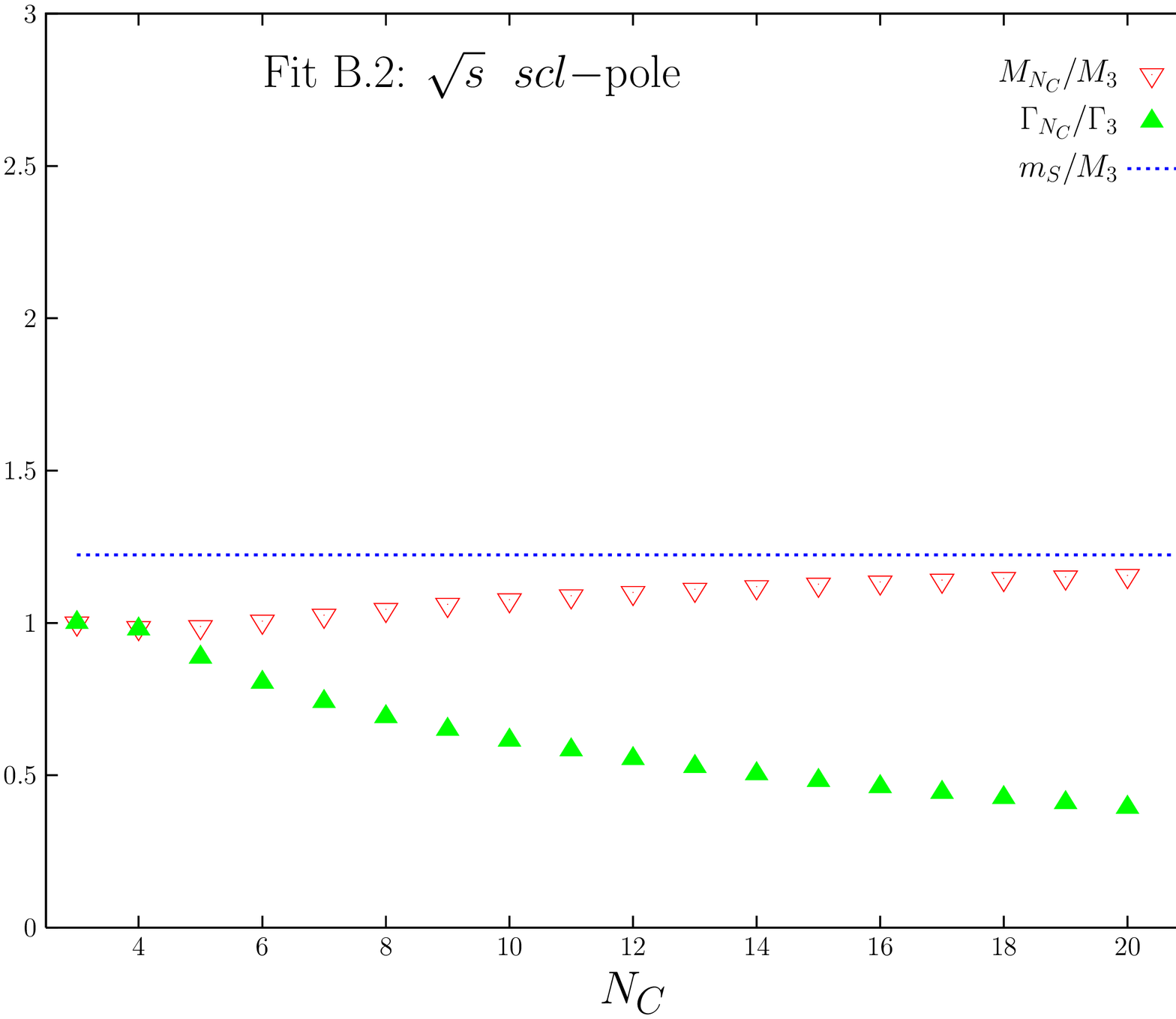}
\hspace{1cm}\includegraphics[height=5.5cm]{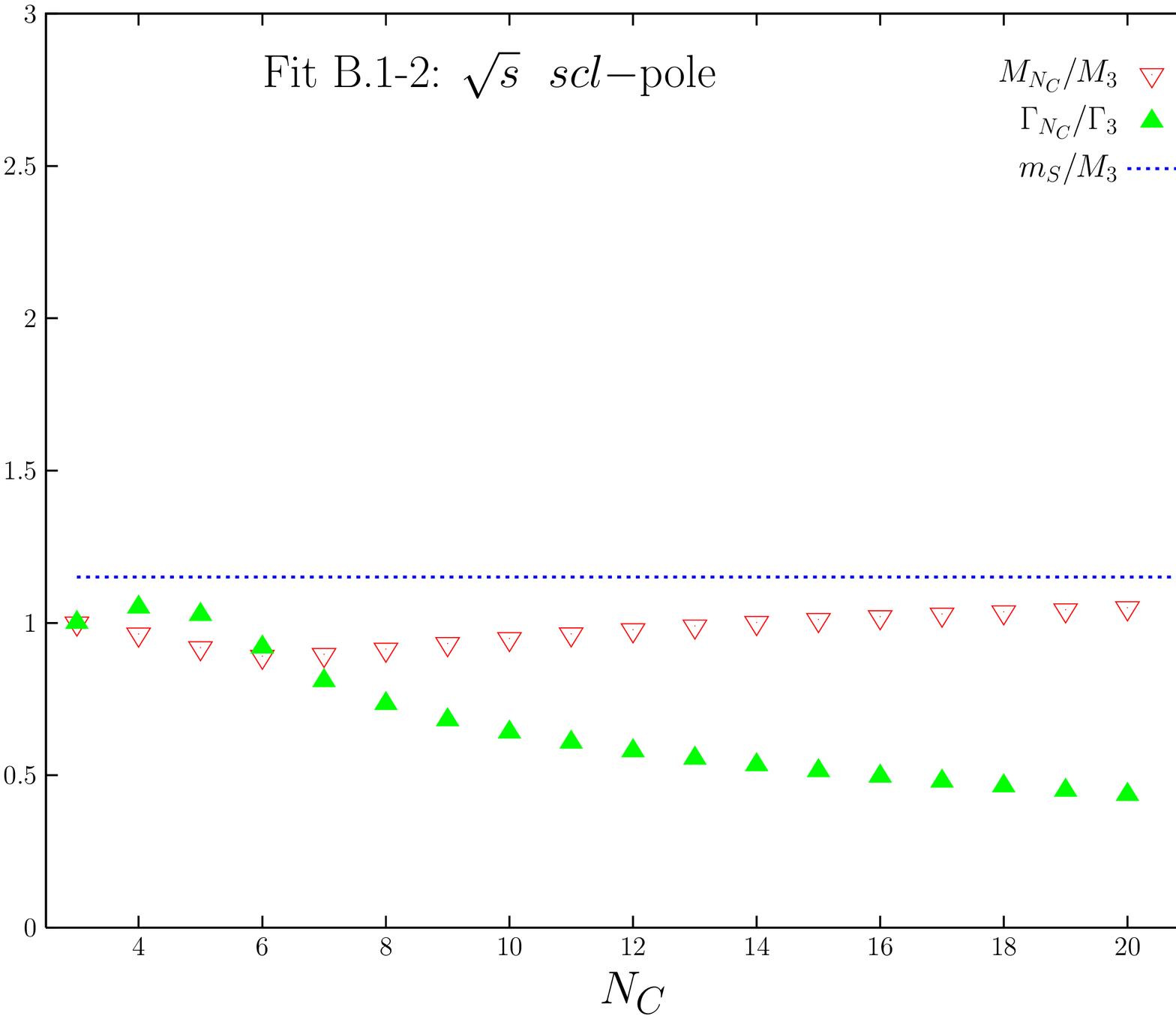}}\\\vspace{1cm}
\makebox[0pt]{\hspace{-2cm}\includegraphics[height=5.5cm]{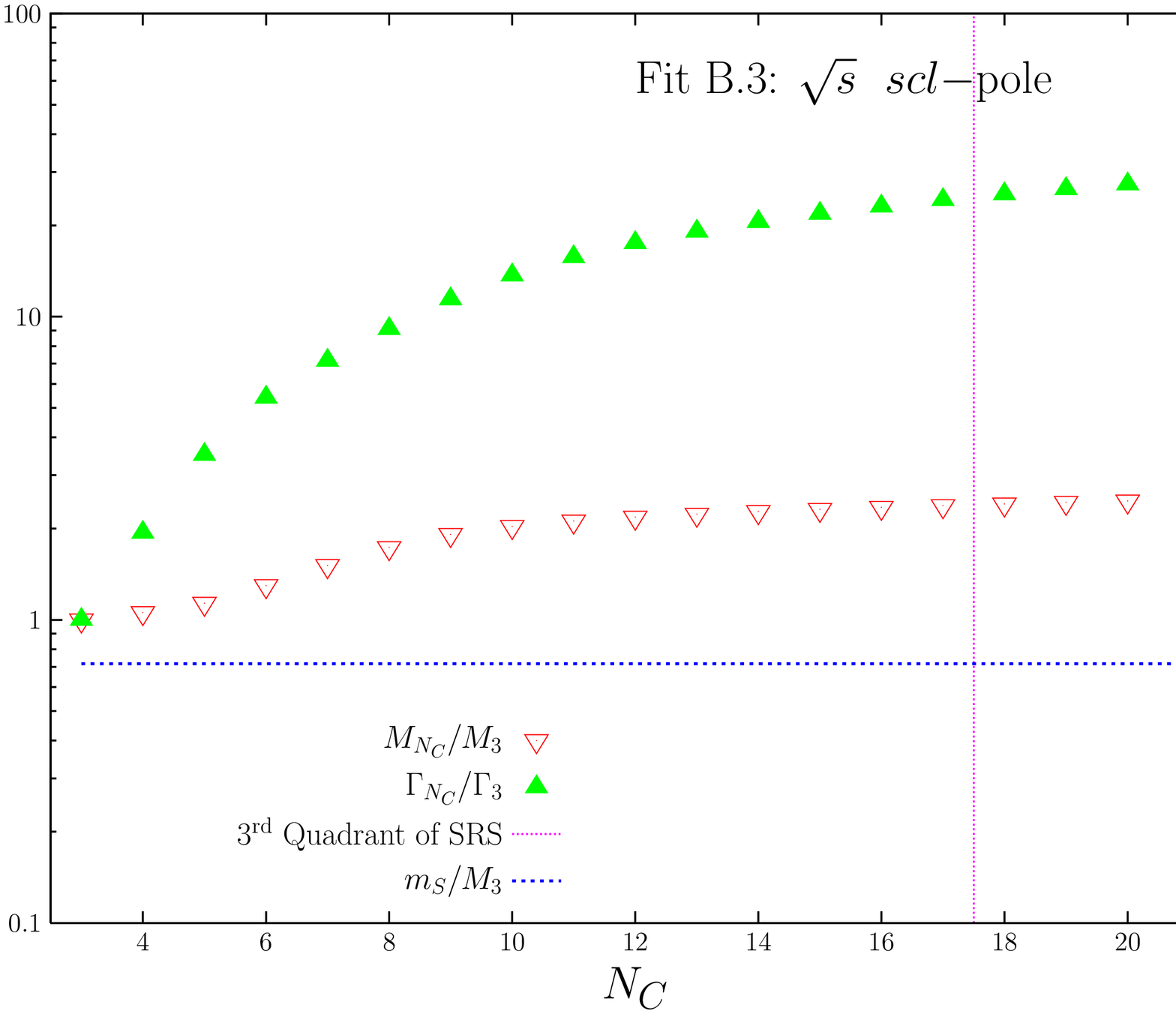}}
\end{center}
\caption{ Same as Fig.~\ref{fig:i1j1.nc}, but for the
  $scl$ pole. Now, poles appear in the third quadrant for fits {\bf
  A}, {\bf B.1}  and {\bf B.3}.}
\label{fig:i0j0-scl.nc}
\end{figure*}

In what the $\rho$ meson properties concerns, we observe that for all
five fits examined here (Fig.~\ref{fig:i1j1.nc}), both mass and width
behave as expected from a $q\bar q$ picture. Thus $m_\rho$, that did
not deviate at $N_C=3$ much from $m_V$, quickly approaches to $m_V$,
while $\Gamma_\rho$ decreases like $1/N_C$, as the number of colors
increases. This is not by any means a new result, and in the past other
authors have already reached, within an unitarized ChPT scheme, this
conclusion~\cite{Pelaez:2003dy,Pelaez:2006nj}. Indeed, in a previous
work~\cite{Nieves:2009ez}, the same result was obtained starting
from the one-loop SU(2) ChPT amplitude for massless pions and using,
as in Refs.~\cite{Pelaez:2003dy,Pelaez:2006nj}, the IAM
 to restore elastic unitarity.
However, results here are more robust, because in previous works the
leading $1/N_C$ terms appearing beyond ${\cal
O}(p^4)$~\cite{Pelaez:2003dy, Nieves:2009ez} or ${\cal
O}(p^6)$~\cite{Pelaez:2006nj} were simply ignored.  Note that the
constraint $m_V=m_S+ {\cal O}(1/N_C)$, deduced in
Ref.~\cite{Nieves:2009ez} when the one-loop unitarized $\pi\pi$
amplitude was required to be consistent with the SRA, does no longer
necessarily hold here, and we could still have both parameters to be
independent (fits {\bf A}, {\bf B.1}, {\bf B.2} and {\bf
B.1-2}). Though this could be because we keep here all $1/N_C$ terms
at all chiral orders, it might also happen that the above constraint
was just an artifact of the IAM used in \cite{Nieves:2009ez} to
unitarize the amplitudes. Nevertheless, we should note that fit {\bf
B.3}, where the constraint $m_V=m_S$ is enforced leads to
phenomenologically acceptable results as well.

Let us move on, and discuss now the scalar-isoscalar channel. For
sufficiently large $N_C$, and since unitarization corrections are
subleading, we should end up with just the unique resonance (of mass
$m_S$) included in the SRA irreducible amplitude, while the effects of
the second resonance must disappear.  We see in
Figs.~\ref{fig:i0j0-sigma.nc} and \ref{fig:i0j0-scl.nc} that this
is effectively the case. However, now it is difficult to draw robust
conclusions and the resonance that survives depends on the fit
procedure.

In the case of the fits {\bf A}, {\bf B.1} and {\bf B.3}, we see that
the resonance identified as the $f_0(600)$ for $N_C=3$ becomes the SRA
pole, with mass $m_S$ when $N_C$ is sufficiently large. There is a
first transition region for values of $N_C$ close to 3, where both the
mass and width increase with $N_C$, but above $N_C=6-8$ the resonance
width starts decreasing like $1/N_C$, and the mass approaches to the
limiting $m_S$ value.  The behavior showed in the two upper panels of
Fig.~\ref{fig:i0j0-sigma.nc} is almost identical to that of the right
upper panel of Fig. 1 in the two-loop analysis of
~\cite{Pelaez:2006nj}. The authors of this latter reference conclude
that in the case of the $\sigma$ resonance, there exists a mixing with
a $q\bar q$ sub-dominant component, arising as loop diagrams become
more suppressed at large $N_C$. The nature of the $\sigma$ resonance
in the real world ($N_C=3$) would be totally different to that of the
$\rho$ meson, being it mostly governed by chiral logarithms stemming
from unitarity and crossing symmetry~\cite{Nieves:2009ez}, justifying
the widely accepted nature of the $\sigma$ as a dynamically-generated
meson. However, within this scenario, for sufficiently large $N_C >
10$, the structure of both ($\sigma$ and $\rho$) resonances is
similar.  As pointed out recently in Ref.~\cite{RuizdeElvira:2010cs},
this solves the seeming paradox of how a distinctive nature for the
$\rho$ and $\sigma$ at $N_C = 3$ is reconciled with semi-local duality
at larger values of $N_C$.  Semi-local duality requires the
contribution of these two resonances to the $\pi^+\pi^-$ elastic cross
section to cancel\footnote{Note however, that such a statement
requires an extrapolation of the Regge behavior to somehow low
energies. For that end, the authors of Ref.~\cite{RuizdeElvira:2010cs}
use Regge trajectories for the variable $(\nu-2m_\pi^2-t/2)$ to ensure
that the imaginary part of the extrapolated Regge amplitude vanishes
at threshold.} ``on average'', since this process is purely isospin 2
in the $t$-channel, and there are no isotensor resonances at low energies.

In what respects to the second resonance found in the scalar-isoscalar
channel, we see in Fig.~\ref{fig:i0j0-scl.nc} that for fits {\bf A},
{\bf B.1} and {\bf B.3}, it follows a totally different pattern with
increasing $N_C$. Indeed, it becomes wider and wider and from one value
of $N_C$ on, the pole $s_{scl}$ turns out to be located in the third
quadrant, though $\sqrt{s_{scl}}$ still lies in the fourth quadrant,
with ${\rm Re}\ s <0$ (where the path integral for the resonance field
would not be well defined) and its effects on the scattering
disappear. We would like to mention that in Ref.~\cite{Guo:2011pa}
poles in the SRS are being searched in the variable $\sqrt{s}$. That
could be inappropriate as we pointed out in Ref.~\cite{Nieves:2009ez},
and we reiterate here. This is because, as mentioned above, there are
situations where $\sqrt{s_R}$ lies on the fourth quadrant of the SRS,
but however $s_R$ has passed to the third quadrant and thus its
meaning becomes unclear. This phenomenon, which can only happen for
broad resonances, was also illustrated in Fig.~1 of
\cite{Nieves:2009ez}, and it is precisely what happens for the
$\sigma$ meson case in Ref.~\cite{Guo:2011pa} (see the Fig. 10 in that
reference). The conclusion of Ref.~\cite{Guo:2011pa} that the $\sigma$
meson moves far away in the complex plane for large $N_C$ overlooks
the fact that it does so in the third quadrant of the complex plane.

The recent work of Ref.~\cite{Zhou:2010ra} might contradict the
findings for the $\sigma$ resonance deduced from fits {\bf A}, {\bf
  B.1} and {\bf B.3},
which in turn seem compatible with those obtained in the
two-loop analysis of Ref.~\cite{Pelaez:2006nj}.  In \cite{Zhou:2010ra},
instead, emerges a picture more consistent with that outlined in the
one-loop analysis of Ref.~\cite{Pelaez:2003dy}. Within the unitarized
quark model proposed by T\"ornqvist~\cite{Tornqvist:1995kr}, the
authors of \cite{Zhou:2010ra} find that the whole low-energy scalar
spectrum below 2 GeV, except for a possible glueball $f_0(1710)$,
could be described in one consistent picture, with the bare ``$q \bar
q$ seeds'' dressed by the hadron loops. In this model, the $\sigma$
resonance is generated as a pole of the $S$ matrix and has no
correspondence with any of the bare $q \bar q$ seeds included in the
scheme. Indeed, the $\sigma$ resonance runs away from the real axis on
the complex $s$ plane when $N_C$ increases. However, following
Ref.~\cite{RuizdeElvira:2010cs}, this scenario might not be
compatible with semi-local duality, when the number of colors is
sufficiently high. Besides, we should note that, in sharp contrast
with the work here, it is not clear whether the $\pi\pi$
amplitudes used in Ref.~\cite{Zhou:2010ra} contain or not all the leading
$1/N_C$ contributions and thus, the analysis of the behavior of the
resonance properties when $N_C$ is larger than 3 is not fully
meaningful.

The qualitative $N_C$ behavior of the two resonances found
in the scalar-isoscalar sector is substantially different when one
looks into the results of the fits {\bf B.2} and {\bf B.1-2} 
(middle panels of
Figs.~\ref{fig:i0j0-sigma.nc} and \ref{fig:i0j0-scl.nc}). There, the
role played by the $N_C=3$ $\sigma$ and $scl$ resonances is
interchanged. Thus, the $q \bar q$ component of the $\sigma$ seems to
be absent, and the $f_0(600)$ pole becomes wider and moves into the
third quadrant above $N_C > 5$. Indeed, now the behavior showed by
this pole is quite similar to that displayed for the $\sigma$ in the
right panel of Fig.4 of Ref.~\cite{Zhou:2010ra}. On the other hand, as
can be appreciated in the middle panels of
Fig.~\ref{fig:i0j0-scl.nc}, the second $I=J=0$ resonance now
becomes the scalar SRA pole included in our amplitudes.
Presumably, at high $N_C$ it would
become more delta-function-like, as the $\rho$ pole would, and it
would likely provide the needed cancellations with the contribution of
this latter resonance in the elastic $\pi^+\pi^-$
amplitude~\cite{RuizdeElvira:2010cs}.

A recent study~\cite{RuizArriola:2010fj} describes an accidental
symmetry of the full Regge tower of radially excited $0^{++}$  states,
$M^2 = a n + m_\sigma^2$.  Remarkably, the states generating doublets
with excited pion states are $f_0(600) \leftrightarrow \pi_0 (140)$,
$f_0(1370) \leftrightarrow \pi_0 (1300)$, $f_0(1710) \leftrightarrow
\pi_0 (1800)$, $f_0(2100)\leftrightarrow \pi_0 (2070) $ and $f_0(2330)
\leftrightarrow \pi_0 (2360)$, whereas the other scalar states
$f_0(980)$, $f_0(1500)$, $f_0(2020)$ and $f_0(2200)$ are not
degenerate with other mesons with light $u$ and $d$ quarks.
Arguments have been put forward as to identify the $f_0(980)$ as a
would-be glueball in the large--$N_C$ limit. As it is well known,
glueballs are more weakly coupled to mesons, ${\cal O} (1/N_C)$, than
other mesons, ${\cal O} (1/\sqrt{N_C})$. This is supported by the
rather small width ratio which yields $\Gamma_{f}/\Gamma_{\sigma} \sim
(g_{f \pi\pi}^2 m_f^3)/(g_{\sigma \pi\pi}^2 m_\sigma^3) \sim 1/N_C$,
and for $m_\sigma \sim 0.8$ MeV a ratio $g_{\sigma \pi\pi}/g_{f
  \pi\pi} \sim \sqrt{N_C}$ is obtained.

Our analysis can be improved along several lines. Firstly, one might
extend the chiral analysis to include two-loop results. Secondly, our
conclusions might be modified when coupled channels incorporating
$\bar K K$ effects are taken into account. We have given arguments why
we do not expect that this might be important as long as we remain in
the sub-threshold region, where all $\bar K K$ effects should be
included as $1/N_C$ corrections to the counter-terms.

\section{Concluding remarks}
\label{sec:conclusions}

We summarize the conclusions of this work.
First, we have constructed $\pi\pi$ amplitudes that fulfill exact
elastic unitarity, account for one-loop ChPT contributions and
include all leading terms, within the SRA, in the $1/N_C$
expansion. These amplitudes have been successfully fitted to $I=J=0$,
$I=2, J=0$ and $I=J=1$ phase shifts. Next, we have looked for poles in
the SRS of the amplitudes, and discussed their properties. Since all
leading $1/N_C$ terms are taken into account, this scheme is much more
appropriated to discuss the $N_C$ dependence of the $\sigma$ and $\rho$
masses and widths than previous ones, where the leading $1/N_C$ terms
appearing beyond ${\cal O}(p^4)$~\cite{Pelaez:2003dy, Nieves:2009ez} or
${\cal O}(p^6)$~\cite{Pelaez:2006nj} were neglected. The recent work
of Ref.~\cite{Guo:2011pa} does not identify correctly  the leading
$1/N_C$ term, and hence the conclusions of this reference
in the large--$N_C$ limit need some revision.

Robust conclusions are drawn in the case to the $\rho$ resonance, and
we confirm here that it is a stable meson in the large--$N_C$ limit,
as pointed out by other authors in the past.  In the scalar-isoscalar
sector, the overall scenario looks like somehow less predictive, since
we cannot firmly conclude whether or not the $N_C=3$ $f_0(600)$
resonance completely disappears at large $N_C$ or it has a
sub-dominant component in its structure, which would become dominant
when the number of colors gets sufficiently high. Unfortunately, this
depends on the chosen procedure ({\bf A}, {\bf B.1} and {\bf B.3} or
{\bf B.2} and {\bf B.1-2}) to fit the phase-shift data. However, it
becomes clear the predominant di-pion component of this pole for
$N_C=3$, and that the SRA delta-function-like pole, that always
appears in the $N_C \gg 1$ limit, might help to keep the whole scheme
compatible with semi-local duality. Nevertheless, this needs to be
quantitatively checked elsewhere.

\begin{acknowledgments}
   We thank J.R. Pel\'aez, J. Ruiz de Elvira and J.J. Sanz-Cillero
   for useful communications.
   This research was supported by DGI and FEDER funds, under contracts
   FIS2008-01143/FIS, FPA2007-60323 and the Spanish Consolider-Ingenio
   2010 Programme CPAN (CSD2007-00042), by Junta de Andaluc\'\i a
   contract FQM0225, by Generalitat Valenciana under contracts
   PROMETEO/2008/069 and PROMETEO/2009/0090, and it is part of the
   European Community-Research Infrastructure Integrating Activity
   ``Study of Strongly Interacting Matter'' (acronym HadronPhysics2,
   Grant Agreement n. 227431),
   under the Seventh EU Framework Programme.
   A.P. acknowledges the support of the Alexander von Humboldt Foundation.

\end{acknowledgments}


\end{document}